\documentclass[twoside]{IEEEtran}
\IEEEoverridecommandlockouts
\normalsize

\def\Figs{figs/} % call figures (eps files) if needed
\usepackage[cmex10]{amsmath} % Use the [cmex10] option to ensure
\usepackage[noadjust]{cite}       % Improves presentation of citations, like [3--7].
\usepackage{amsmath}
\usepackage{amsopn}
\usepackage{amssymb}
\usepackage{accents}
\usepackage{pifont}
\usepackage{mathrsfs}  % provides mathscr: Ralph Smith's Formal Script
\usepackage{eucal}     % replaces mathcal by eucal
\renewcommand{\mathcal}[1]{\CMcal{#1}} % reestablishing normal definition of mathcal
\usepackage{bm}         % for nice bold math symbols using \bm{...}
\usepackage{fix-cm}     % permit arbitrary sizes of Computer Modern

\usepackage{mleftright} %provides \mleft, \mright to fix issue with spacing
\mleftright                 %redefines all \left and \right to behave
\usepackage{mathdots}   % fixes \ddots and \vdots to scale properly
\usepackage{trfsigns}   % for \Fourier, \fourier, \laplace, \Laplace
\usepackage{centernot}  % for \centernot: negation of any math symbol

\usepackage{nicefrac}   % for \nicefrac: nice frac for text-mode

\usepackage[normalscripts]{moresize}   % adds a size \ssmall between \tiny and
\makeatletter
\newcommand*\if@single[3]{%
  \setbox0\hbox{${\mathaccent"0362{#1}}^H$}%
  \setbox2\hbox{${\mathaccent"0362{\kern0pt#1}}^H$}%
  \ifdim\ht0=\ht2 #3\else #2\fi
  }
\newcommand*\rel@kern[1]{\kern#1\dimexpr\macc@kerna}
\newcommand*\widebar[1]{\@ifnextchar^{{\wide@bar{#1}{0}}}{\wide@bar{#1}{1}}}
\newcommand*\wide@bar[2]{\if@single{#1}{\wide@bar@{#1}{#2}{1}}{\wide@bar@{#1}{#2}{2}}}
\newcommand*\wide@bar@[3]{%
  \begingroup
  \def\mathaccent##1##2{%
    \if#32 \let\macc@nucleus\first@char \fi
    \setbox\z@\hbox{$\macc@style{\macc@nucleus}_{}$}%
    \setbox\tw@\hbox{$\macc@style{\macc@nucleus}{}_{}$}%
    \dimen@\wd\tw@
    \advance\dimen@-\wd\z@
    \divide\dimen@ 3
    \@tempdima\wd\tw@
    \advance\@tempdima-\scriptspace
    \divide\@tempdima 10
    \advance\dimen@-\@tempdima
    \ifdim\dimen@>\z@ \dimen@0pt\fi
    \rel@kern{0.6}\kern-\dimen@
    \if#31
      \overline{\rel@kern{-0.6}\kern\dimen@\macc@nucleus\rel@kern{0.4}\kern\dimen@}%
      \advance\dimen@0.4\dimexpr\macc@kerna
      \let\final@kern#2%
      \ifdim\dimen@<\z@ \let\final@kern1\fi
      \if\final@kern1 \kern-\dimen@\fi
    \else
      \overline{\rel@kern{-0.6}\kern\dimen@#1}%
    \fi
  }%
  \macc@depth\@ne
  \let\math@bgroup\@empty \let\math@egroup\macc@set@skewchar
  \mathsurround\z@ \frozen@everymath{\mathgroup\macc@group\relax}%
  \macc@set@skewchar\relax
  \let\mathaccentV\macc@nested@a
  \if#31
    \macc@nested@a\relax111{#1}%
  \else
    \def\gobble@till@marker##1\endmarker{}%
    \futurelet\first@char\gobble@till@marker#1\endmarker
    \ifcat\noexpand\first@char A\else
      \def\first@char{}%
    \fi
    \macc@nested@a\relax111{\first@char}%
  \fi
  \endgroup
}
\makeatother

\makeatletter
\def\IEEElabelanchoreqn#1{\bgroup
\def\@currentlabel{\p@equation\theequation}\relax
\def\@currentHref{\@IEEEtheHrefequation}\label{#1}\relax
\Hy@raisedlink{\hyper@anchorstart{\@currentHref}}\relax
\Hy@raisedlink{\hyper@anchorend}\egroup}
\makeatother

\newcommand{\midk}[1]{\kern0.1em #1 \kern0.1em}
\newcommand{\middlek}[1]{\kern0.1em \middle#1 \kern0.1em}
\newcommand{\bigk}[1]{\kern-0.1em \bigm#1 \kern-0.1em}
\newcommand{\Bigk}[1]{\kern-0.1em \Bigm#1 \kern-0.1em}
\newcommand{\biggk}[1]{\kern-0.1em \biggm#1 \kern-0.1em}
\newcommand{\Biggk}[1]{\kern-0.1em \Biggm#1 \kern-0.1em}

\ifx\stefansf\undefined
  \newcommand{\tn}[1]{\textnormal{#1}}
\else
  \newcommand{\tn}[1]{\textnormal{\textsf{#1}}}
\fi

\newcommand{\vect}[1]{\mathbf{#1}} % vector
\newcommand{\vectg}[1]{\bm{#1}} % italic vector (works also for greek)
\renewcommand{\vect}[1]{\vectg{#1}} % switch by default to second version!
\newcommand{\inv}[1]{#1^{-1}} % inverse  
\newcommand{\trans}[1]{#1^{\textup{\textsf{\tiny T}}}} % transpose  
\newcommand{\Naturals}{\mathbb{N}}   % natural numbers 1,2,3,...
\newcommand{\norm}[1]{\left\|#1\right\|}
\newcommand{\enorm}[1]{\|#1\|}

\newcommand{\set}[1]{\mathcal{#1}}           % set
\newcommand{\cset}[1]{\mathcal{#1}^{\tn{c}}} % complement set
\newcommand{\card}[1]{\left|#1\right|}       % cardinality of a set.

\newcommand{\bigcard}[1]{\bigl|#1\bigr|}

\newcommand{\const}[1]{\textnormal{\usefont{U}{eur}{m}{n}\selectfont #1}} % Euler
\newcommand{\HH}{\mathop{}\!\const{H}}  % entropy
\newcommand{\Hb}{\HH_{\tn{b}}}          % binary entropy function
\newcommand{\II}{\mathop{}\!\const{I}}  % mutual information

\newcommand{\HP}[1]{\HH\left(#1\right)} 
\newcommand{\eHP}[1]{\HH(#1)} 
\newcommand{\bigHP}[1]{\HH\bigl(#1\bigr)}

\newcommand{\HPcond}[2]{\HH\left(#1 \kern0.1em\middle|\kern0.1em #2\right)}
\newcommand{\eHPcond}[2]{\HH(#1 \kern0.1em|\kern0.1em #2)} 
\newcommand{\bigHPcond}[2]{\HH\bigl(#1 \kern-0.1em \bigm| \kern-0.1em#2\bigr)}
\newcommand{\BigHPcond}[2]{\HH\Bigl(#1 \kern-0.1em \Bigm| \kern-0.1em#2\Bigr)}

\newcommand{\MinH}{\HH_\infty}  % min-entropy
\newcommand{\MHcond}[2]{\MinH\left(#1 \kern0.1em\middle|\kern0.1em #2\right)}
\newcommand{\eMHcond}[2]{\MinH(#1 \kern0.1em|\kern0.1em #2)} 
\newcommand{\bigMHcond}[2]{\MinH\bigl(#1 \kern-0.1em \bigm| \kern-0.1em#2\bigr)}
\newcommand{\BigMHcond}[2]{\MinH\Bigl(#1 \kern-0.1em \Bigm| \kern-0.1em#2\Bigr)}

\newcommand{\MI}[2]{\II\left(#1 \kern0.1em{;}\kern0.1em #2\right)} 
\newcommand{\eMI}[2]{\II(#1 \kern0.1em{;}\kern0.1em #2)} 
\newcommand{\bigMI}[2]{\II\bigl(#1 \kern0.1em{;}\kern0.1em #2\bigr)}
\newcommand{\BigMI}[2]{\II\Bigl(#1 \kern0.1em{;}\kern0.1em #2\Bigr)}
\newcommand{\MIcond}[3]{\II\left(#1 \kern0.1em{;}\kern0.1em #2 \kern0.1em\middle|\kern0.1em #3\right)}
\newcommand{\eMIcond}[3]{\II(#1 \kern0.1em{;}\kern0.1em #2 \kern0.1em|\kern0.1em #3)} 
\newcommand{\bigMIcond}[3]{\II\bigl(#1 \kern0.1em{;}\kern0.1em #2 \kern-0.1em \bigm| \kern-0.1em#3\bigr)}
\newcommand{\BigMIcond}[3]{\II\Bigl(#1 \kern0.1em{;}\kern0.1em #2 \kern-0.1em \Bigm| \kern-0.1em#3\Bigr)}

\newcommand{\SMI}{\II_\infty}  % maximal leakage
\newcommand{\SI}[2]{\SMI\left(#1 \kern0.1em{;}\kern0.1em #2\right)} 
\newcommand{\eSI}[2]{\SMI(#1 \kern0.1em{;}\kern0.1em #2)} 
\newcommand{\bigSI}[2]{\SMI\bigl(#1 \kern0.1em{;}\kern0.1em #2\bigr)}
\newcommand{\BigSI}[2]{\SMI\Bigl(#1 \kern0.1em{;}\kern0.1em #2\Bigr)}
\newcommand{\SIcond}[3]{\SMI\left(#1 \kern0.1em{;}\kern0.1em #2 \kern0.1em\middle|\kern0.1em #3\right)}
\newcommand{\eSIcond}[3]{\SMI(#1 \kern0.1em{;}\kern0.1em #2 \kern0.1em|\kern0.1em #3)} 
\newcommand{\bigSIcond}[3]{\SMI\bigl(#1 \kern0.1em{;}\kern0.1em #2 \kern-0.1em \bigm| \kern-0.1em#3\bigr)}
\newcommand{\BigSIcond}[3]{\SMI\Bigl(#1 \kern0.1em{;}\kern0.1em #2 \kern-0.1em \Bigm| \kern-0.1em#3\Bigr)}

\newcommand{\MaxL}{\mathop{}\!\mathsf{MaxL}}  % maximal leakage
\newcommand{\ML}[2]{\MaxL\left(#1 \kern0.1em{;}\kern0.1em #2\right)} 
\newcommand{\eML}[2]{\MaxL(#1 \kern0.1em{;}\kern0.1em #2)} 
\newcommand{\bigML}[2]{\MaxL\bigl(#1 \kern0.1em{;}\kern0.1em #2\bigr)}
\newcommand{\BigML}[2]{\MaxL\Bigl(#1 \kern0.1em{;}\kern0.1em #2\Bigr)}
\newcommand{\MLcond}[3]{\MaxL\left(#1 \kern0.1em{;}\kern0.1em #2 \kern0.1em\middle|\kern0.1em #3\right)}
\newcommand{\eMLcond}[3]{\MaxL(#1 \kern0.1em{;}\kern0.1em #2 \kern0.1em|\kern0.1em #3)} 
\newcommand{\bigMLcond}[3]{\MaxL\bigl(#1 \kern0.1em{;}\kern0.1em #2 \kern-0.1em \bigm| \kern-0.1em#3\bigr)}
\newcommand{\BigMLcond}[3]{\MaxL\Bigl(#1 \kern0.1em{;}\kern0.1em #2 \kern-0.1em \Bigm| \kern-0.1em#3\Bigr)}

\newcommand{\relD}{\mathop{}\!\mathscr{D}}         % relative entropy
\newcommand{\relDf}[2]{\relD\left(#1 \kern0.1em\middle\|\kern0.1em #2\right)}
\newcommand{\erelDf}[2]{\relD(#1 \kern0.1em\|\kern0.1em #2)} 
\newcommand{\bigrelDf}[2]{\relD\bigl(#1 \kern-0.1em \bigm\| \kern-0.1em#2\bigr)}
\newcommand{\BigrelDf}[2]{\relD\Bigl(#1 \kern-0.1em \Bigm\| \kern-0.1em#2\Bigr)}
\newcommand{\biggrelDf}[2]{\relD\biggl(#1 \kern-0.1em \biggm\| \kern-0.1em#2\biggr)}
\newcommand{\BiggrelDf}[2]{\relD\Biggl(#1 \kern-0.1em \Biggm\| \kern-0.1em#2\Biggr)}

\newcommand{\Prob}{\operatorname{\tn{Pr}}}

\newcommand{\Prscond}[2]{\Pr\left(#1 \kern0.1em\middle|\kern0.1em #2\right)}
\newcommand{\ePrscond}[2]{\Pr(#1 \kern0.1em|\kern0.1em #2)} 
\newcommand{\bigPrscond}[2]{\Pr\bigl(#1 \kern-0.1em \bigm| \kern-0.1em#2\bigr)}
\newcommand{\BigPrscond}[2]{\Pr\Bigl(#1 \kern-0.1em \Bigm| \kern-0.1em#2\Bigr)}
\newcommand{\biggPrscond}[2]{\Pr\biggl(#1 \kern-0.1em \biggm| \kern-0.1em#2\biggr)}
\newcommand{\BiggPrscond}[2]{\Pr\Biggl(#1 \kern-0.1em \Biggm| \kern-0.1em#2\Biggr)}

\newcommand{\Prv}[1]{\Pr\left[#1\right]}

\newcommand{\Prvcond}[2]{\Pr\left[#1 \kern0.1em\middle|\kern0.1em #2\right]}
\newcommand{\ePrvcond}[2]{\Pr[#1 \kern0.1em|\kern0.1em #2]} 
\newcommand{\bigPrvcond}[2]{\Pr\bigl[#1 \kern-0.1em \bigm| \kern-0.1em#2\bigr]}
\newcommand{\BigPrvcond}[2]{\Pr\Bigl[#1 \kern-0.1em \Bigm| \kern-0.1em#2\Bigr]}
\newcommand{\biggPrvcond}[2]{\Pr\biggl[#1 \kern-0.1em \biggm| \kern-0.1em#2\biggr]}
\newcommand{\BiggPrvcond}[2]{\Pr\Biggl[#1 \kern-0.1em \Biggm| \kern-0.1em#2\Biggr]}

\newcommand{\Exp}{\operatorname{\textnormal{\textsf{E}}}}

\newcommand{\E}[2][]{\Exp_{#1}\left[#2\right]}

\newcommand{\BigE}[2][]{\Exp_{#1}\Bigl[#2\Bigr]}

\newcommand{\Econd}[3][]{\Exp_{#1}\left[#2 \kern0.1em\middle|\kern0.1em #3\right]}
\newcommand{\eEcond}[3][]{\Exp_{#1}[#2 \kern0.1em|\kern0.1em #3]}
\newcommand{\bigEcond}[3][]{\Exp_{#1}\bigl[#2 \kern-0.1em \bigm| \kern-0.1em #3\bigr]}
\newcommand{\BigEcond}[3][]{\Exp_{#1}\Bigl[#2 \kern-0.1em \Bigm| \kern-0.1em #3\Bigr]}
\newcommand{\biggEcond}[3][]{\Exp_{#1}\biggl[#2 \kern-0.1em \biggm| \kern-0.1em #3\biggr]}
\newcommand{\BiggEcond}[3][]{\Exp_{#1}\Biggl[#2 \kern-0.1em \Biggm| \kern-0.1em #3\Biggr]}

\newcommand{\Covcond}[4][]{\mathop{}\!\mathsf{Cov}_{#1}\left[{#2},{#3}\kern0.1em\middle|\kern0.1em{#4}\right]}
\newcommand{\eCovcond}[4][]{\mathop{}\!\mathsf{Cov}_{#1}[{#2},{#3}\kern0.1em|\kern0.1em{#4}]}
\newcommand{\bigCovcond}[4][]{\mathop{}\!\mathsf{Cov}_{#1}\bigl[{#2},{#3}\kern-0.1em\bigm|\kern-0.1em{#4}\bigr]}
\newcommand{\BigCovcond}[4][]{\mathop{}\!\mathsf{Cov}_{#1}\Bigl[{#2},{#3}\kern-0.1em\Bigm|\kern-0.1em{#4}\Bigr]}
\newcommand{\biggCovcond}[4][]{\mathop{}\!\mathsf{Cov}_{#1}\biggl[{#2},{#3}\kern-0.1em\biggm|\kern-0.1em{#4}\biggr]}
\newcommand{\BiggCovcond}[4][]{\mathop{}\!\mathsf{Cov}_{#1}\Biggl[{#2},{#3}\kern-0.1em\Biggm|\kern-0.1em{#4}\Biggr]}

\newcommand{\Uniform}[1]{\mathcal{U}\left(#1\right)} % Uniform

\newcommand{\Bernoulli}[1]{\tn{Bernoulli}\left(#1\right)} % Bernoulli dist.

\newcommand{\indep}{\mathrel{\bot}\joinrel\mathrel{\mkern-5mu}\joinrel\mathrel{\bot}}  %independent
\newcommand{\markov}{\mathrel{\multimap}\joinrel\mathrel{-}\joinrel\mathrel{\mkern-6mu}\joinrel\mathrel{-}} % Markov chain
\ifx\stefanchinese\undefined 
   
\else
  \ifx\stefanxelatex\undefined 
  \else
     
  \fi
\fi

\providecommand{\abs}[1]{\left\lvert#1\right\rvert}

\newcommand{\dd}{\mathop{}\!\mathrm{d}}
\newcommand{\eqdef}{\triangleq} % definition

\newcommand{\eps}{\epsilon}

\makeatletter

\newcommand{\Rmnum}[1]{\expandafter\@slowromancap\romannumeral #1@}
\makeatother

\usepackage{lipsum} % print dummy text
\usepackage{balance} % balance the text in the final page under twocolumn formatting
\usepackage{amsbsy}
\usepackage{booktabs} % required for showing good \hline in matrix form --> \midrule
\usepackage{enumitem}
\usepackage{mathtools}
\usepackage{amsmath,amssymb,amsfonts,amsthm} % amsthm: typesetting theorems (AMS style)
\usepackage[lined,boxed,commentsnumbered,linesnumbered,ruled]{algorithm2e}
\usepackage{comment}
\usepackage{nicefrac} % displaying a nice fraction in in-line math
\usepackage{pgfplots}
\pgfplotsset{compat=newest}
\usepackage{tikz}
\usetikzlibrary{shapes, patterns,decorations.text, decorations.pathreplacing}

\newtheorem{lemma}{Lemma}
\newtheorem{theorem}{Theorem}
\newtheorem{definition}{Definition}

\newtheorem{example}{Example}

\usepackage[capitalize]{cleveref}

\crefname{equation}{\unskip}{\unskip} %\crefname{type}{singular}{plural}
\crefname{claim}{Claim}{Claims} 
\usepackage{array}
\usepackage{multirow}
\newcolumntype{C}[1]{>{\centering\arraybackslash}p{#1}}
\allowdisplaybreaks

\renewcommand{\vect}[1]{\vectg{#1}} % switch by default to second version!
\newcommand{\Hwt}[1]{w_{\mathsf{H}}\left(#1\right)} % the Hamming weight
\newcommand{\Spt}[1]{\chi\left(#1\right)} % the support function of a vector
\newcommand{\Nat}[1]{\mathbb{N}_{#1}} % natural numbers until N
\newcommand{\collect}[1]{\mathscr{#1}} % the collection notation
\DeclareMathSymbol{\mhyph}{\mathord}{operators}{`\-} % better dash in math mode
\newcommand*{\Resize}[2][4]{\resizebox{#1}{!}{\ensuremath{#2}}} % Resize based on line or column width
\makeatletter
\renewcommand*\env@matrix[1][*\c@MaxMatrixCols c]{%
  \hskip -\arraycolsep
  \let\@ifnextchar\new@ifnextchar
  \array{#1}}
\makeatother

\renewcommand{\HH}{\mathop{}\!\mathsf{H}} % entropy
\renewcommand{\II}{\mathop{}\!\mathsf{I}}  % mutual information

\renewcommand{\HH}{\mathop{}\!\mathsf{H}} % entropy
\newcommand{\WIL}{\mathop{}\!\mathsf{WIL}}  % or \II^{\mathsf{worst}}, the worst-case information leakage
\newcommand{\WL}[2]{\WIL\left(#1 \kern0.1em{;}\kern0.1em #2\right)} 
\newcommand{\eWL}[2]{\WIL(#1 \kern0.1em{;}\kern0.1em #2)} 
\newcommand{\bigWL}[2]{\WIL\bigl(#1 \kern0.1em{;}\kern0.1em #2\bigr)}
\newcommand{\BigWL}[2]{\WIL\Bigl(#1 \kern0.1em{;}\kern0.1em #2\Bigr)}
\newcommand{\ee}{\mathrm{e}}
\renewcommand{\Exp}{\operatorname{\mathbb{E}}}
\definecolor{darkgreen}{rgb}{0, 0.5, 0}
\renewcommand{\r}{\color{red}} % color red for reviewer 1 in the first round
\newcommand{\g}{\color{darkgreen}} % color green for the editor in the first round
\newcommand\scalemath[2]{\scalebox{#1}{\mbox{\ensuremath{\displaystyle #2}}}}

\begin{document}

% \title{Paper Title*\\
\title{Multi-Server Weakly-Private Information Retrieval}

\author{Hsuan-Yin Lin,~\IEEEmembership{Senior Member,~IEEE}, Siddhartha Kumar, Eirik Rosnes,~\IEEEmembership{Senior~Member,~IEEE},\\ Alexandre Graell i Amat,~\IEEEmembership{Senior~Member,~IEEE}, and Eitan Yaakobi,~\IEEEmembership{Senior~Member,~IEEE}% <-this % stops a space
  \thanks{This work was partially funded by the Swedish Research Council (grant \#2016-04253), the Israel Science Foundation (grant \#1817/18), and by the Technion Hiroshi Fujiwara Cyber Security Research Center and the Israel National Cyber Directorate. This article was presented in part at the IEEE International Symposium on Information Theory (ISIT), Paris, France, July 2019.}
\thanks{H.-Y.~Lin, S.~Kumar, and E.~Rosnes are with Simula UiB, N--5006 Bergen, Norway (e-mail: lin@simula.no; kumarsi@simula.no; eirikrosnes@simula.no).}
\thanks{A.~Graell i Amat is with the Department of Electrical Engineering, Chalmers University of Technology, SE--41296 Gothenburg, Sweden, and Simula UiB (e-mail: alexandre.graell@chalmers.se).}
\thanks{E.~Yaakobi is with the Department of Computer Science, Technion --- Israel Institute of Technology, Haifa, 3200003 Israel (email: yaakobi@gmail.com).} 
}

%\author{Hsuan-Yin Lin~\IEEEmembership{Senior~Member,~IEEE}, Siddhartha Kumar~\IEEEmembership{Member,~IEEE}, Eirik~Rosnes,~\IEEEmembership{Senior~Member,~IEEE},\\ Alexandre~Graell~i~Amat,~\IEEEmembership{Senior~Member,~IEEE}, and Eitan Yaakobi,~\IEEEmembership{Senior~Member,~IEEE}}

\maketitle
% As a general rule, do not put math, special symbols or citations in the abstract
\begin{abstract}
Private information retrieval (PIR) protocols ensure that a user can download a file from a database without revealing any information on the identity of the requested file to the servers storing the database. While existing protocols strictly impose that no information is leaked on the file’s identity, this work initiates the study of the tradeoffs that can be achieved by relaxing the perfect privacy requirement. We refer to such protocols as weakly-private information retrieval (WPIR) protocols. In particular, for the case of multiple noncolluding replicated servers, we study how the download rate, the upload cost, and the access complexity can be improved when relaxing the perfect privacy constraint. To quantify the information leakage on the requested file's identity we consider mutual information (MI), worst-case information leakage, and maximal leakage (MaxL). We present two WPIR schemes, denoted by Scheme~A and Scheme~B, based on two recent PIR protocols and show that the download rate of the former can be optimized by solving a convex optimization problem. We also show that Scheme~A achieves an improved download rate compared to the recently proposed scheme by Samy \emph{et al.} under the so-called $\epsilon$-privacy metric. Additionally, a family of schemes based on partitioning is presented. Moreover, we provide an information-theoretic converse bound for the maximum possible download rate for the MI and MaxL privacy metrics under a practical restriction on the alphabet size of queries and answers. For two servers and two files, the bound is tight under the MaxL metric, which settles the WPIR capacity in this particular case. Finally, we compare the performance of the proposed schemes and their gap to the converse bound.
\end{abstract}


\section{Introduction}
\label{sec:introduction}

Private information retrieval (PIR) was introduced in the computer science literature by Chor \emph{et al.} in~\cite{ChorGoldreichKushilevitzSudan95_1,ChorGoldreichKushilevitzSudan98_1}. A PIR scheme allows a user to  retrieve an arbitrary file from a database that is stored on either a single or multiple servers without revealing any information about the identity of the requested file. The efficiency of a PIR scheme is measured in terms of the total communication load, consisting of both the upload and download cost for the retrieval of a single file. It was already shown in the original work of Chor \emph{et al.} \cite{ChorGoldreichKushilevitzSudan98_1} that in the case that the database is stored on a single server, all files need to be downloaded in order to achieve perfect privacy, i.e., no information leakage on the identity of the requested file.  It has been extensively studied how to 
reduce the communication cost using several copies of the
database, see, e.g., \cite{BeimelIshaiKushilevitRaymond02_1,DvirGopi16_1,Yekhanin08_1,CorriganGibbsKogan20_1}. %, and  also the recent work in \cite{CorriganGibbsKogan20_1} on PIR with sublinear online time without increasing the server-side storage requirements. 

From an information-theoretic perspective and for many practical scenarios, the file size is typically much larger than the size of the queries to all servers. Therefore, rather than accounting for both the upload and the download cost, as usually done in the computer science literature, in the information theory literature efficiency is typically measured in terms of the download cost, or equivalently, in terms of the download rate. The download rate{\textemdash or the PIR rate\textemdash}is defined as the ratio between the requested file size and the average number of downloaded symbols for the retrieval of a single file. The maximum possible PIR rate of all possible schemes is called the PIR capacity. The PIR capacity for the classical PIR model of replicated servers was characterized by Sun and Jafar~\cite{SunJafar17_1}. 

To achieve a lower storage overhead, PIR protocols have also been considered jointly with coded distributed storage systems (DSSs), where the data is encoded by a linear code and then stored on several servers in a distributed manner~\cite{ShahRashmiRamchandran14_1,ChanHoYamamoto15_1,FazeilVardyYaakobi15_1}. The case of maximum distance separable (MDS) coded servers was considered in~\cite{TajeddineGnilkeElRouayheb18_1, BanawanUlukus18_1,SunTian19_1,ZhuYanQiTang20_1,ZhouTianSunLiu20_1,LiKarpukHollanti20_1}, while the case of arbitrary linear coded servers was studied in \cite{KumarLinRosnesGraellAmat19_1, LinKumarRosnesGraellAmat18_2, Freij-Hollanti-etal19_1, LavauzelleTajeddineFreij-HollantiHollanti21_1}. The concept of PIR has also been extended to several other relevant scenarios, which include colluding servers~\cite{SunJafar18_2, Freij-HollantiGnilkeHollantiKarpuk17_1, TajeddineGnilkeElRouayheb18_1, KumarLinRosnesGraellAmat19_1, DOliveiraElRouayheb20_1, Freij-Hollanti-etal19_1, HolzbaurFreij-HollantiLiHollanti21_1app}, robust PIR with Byzantine or unresponsive servers~\cite{SunJafar18_2,BanawanUlukus19_1,TajeddineGnilkeKarpukFreij-HollantiHollanti19_1}, multi-round PIR~\cite{SunJafar18_3}, multi-file PIR~\cite{BanawanUlukus18_2}, optimal download cost of PIR for an arbitrary file size~\cite{SunJafar17_3}, optimal upload cost of PIR, i.e., the minimum required amount of query information~\cite{TianSunChen19_1}, access complexity of PIR, i.e., the number of symbols accessed across all servers for the retrieval of a single file~\cite{ZhangYaakobiEtzionSchwartz19_1}, tradeoff between the storage and download cost of PIR~\cite{Tian20_1}, cache-aided PIR~\cite{Tandon17_1,WeiBanawanUlukus19_1}, PIR with side information~\cite{LiGastpar18_1,WeiBanawanUlukus19_2,KadheGarciaHeidarzadehElRouayhebSprintson20_1,ChenWangJafar20_2,HeidarzadehKazemiSprintson21_1}, PIR on graph-based replication systems~\cite{RavivTamoYaakobi20_1,JiaJafar20_1}, PIR with secure storage~\cite{YangShinLee18_1,JiaSunJafar19_1,JiaJafar20_2}, functional PIR codes~\cite{ZhangEtzionYaakobi20_1}, and private proximity retrieval codes~\cite{ZhangYaakobiEtzion21_1}.

All of the aforementioned extensions of PIR impose perfect privacy, i.e., no information leakage. However, this assumption is quite restrictive and may be relaxed for several practical applications, as leaking part of the information of the identity of the requested file is legitimate as long as there is still enough ambiguity on the file's identity to
meet the privacy requirement specified by the user. For example, the user may be willing to share with the servers that
the file is a movie (and not a book or other forms of files), or only the movie's genre, whereas keeping private  the identity of the movie. Relaxing the perfect privacy requirement of PIR has been considered briefly in the computer science literature previously. As early as in 2002,  Asonov \emph{et al.}~\cite{AsonovFreytag02_1} introduced the concept of repudiation as a relaxation of PIR. Their main motivation was to reduce the preprocessing complexity of queries, while keeping optimal communication (upload and download) cost and response time. However, the condition of repudiation can be achieved even if the server can determine the identity of the requested file almost surely. Hence, it does not provide a good level of information-theoretic privacy. More than a decade later, Toledo \emph{et al.}~\cite{ToledoDanezisGoldberg16_1} adopted a privacy metric based on \emph{differential privacy}{\g~\cite{DworkMcSherryNissimSmith06_1,Dwork06_1}} and traded off privacy for reduced communication cost. In \cite{ToledoDanezisGoldberg16_1}, several schemes that hide the query identity were proposed and studied. However, different mechanisms and security assumptions from those of information-theoretic PIR were considered. Interestingly, the authors claim that the proposed approaches can be applied to information-theoretic PIR, but they did not study any fundamental information-theoretic tradeoffs between information leakage and different costs under the considered privacy metric.  

This paper takes a first step away from the perfect privacy requirement of the information-theoretic PIR framework. Our goal is to study the tradeoffs between different parameters of PIR, such as download rate, upload cost, and access complexity, while relaxing the perfect privacy requirement on the identity of the desired file.
% user is willing to leak some information on the identity of the retrieved file.
We refer to such a scenario as \emph{weakly-private} information retrieval (WPIR).
% The practical motivation for studying this particular relaxation is that in several scenarios, leaking part of the information of the identity of the requested file is legitimate as long as there is still enough ambiguity on the file's identity to meet the privacy requirement specified by the user. 
How to properly measure information leakage has been studied extensively in the computer science literature, see, e.g., \cite{WagnerEckhoff18_1} and references therein. Mutual information (MI)~\cite{LinHewettAltman02_1,SankarRajagopalanPoor13_1,CuffYu16_1}, that captures the \emph{average} information leakage between the private data and the adversary’s observations, maximal leakage (MaxL)~\cite{Smith09_1,IssaWagnerKamath20_1}, and worst-case information leakage (WIL)~\cite{KopfBasin07_1}, are among the most popular information-theoretic privacy leakage metrics, along with (local) differential privacy~\cite{DworkMcSherryNissimSmith06_1,Dwork06_1,Kasiviswanathan-etal08_1,DuchiJordanWainwright13_1}. To the best of our knowledge, using MI as a privacy metric originates from the domain of genome privacy and was first considered in \cite{LinHewettAltman02_1}. Although, the MI privacy metric has a less clear operational meaning than MaxL, the presented results for the MI privacy metric provide fundamental insight into the tradeoff between download cost and privacy leakage, which is also valid and complements the presented results for the other considered privacy metrics.
%
%
%By relaxing   the perfect privacy requirement, the download cost can be improved beyond PIR capacity, establishing a trade-off between information leakage and download cost. We consider the mutual information (MI) and the maximum %leakage (MaxL) privacy metric~\cite{Smith09_1,IssaWagnerKamath18_1sub}, which is considered the most robust information-theoretic metric for information leakage in the literature. Our main contributions are as follows:
In this work, we consider the case of replicated noncolluding servers, mainly focusing on the MI and MaxL privacy metrics. We propose a WPIR scheme by building upon a PIR protocol recently introduced in~\cite{TianSunChen19_1} and study its tradeoffs between download rate, upload cost, and access complexity. In particular, we show that by relaxing the perfect privacy requirement, the download rate can be improved beyond PIR capacity.

The main contributions can be summarized as follows:

%By building upon a PIR protocol recently introduced in~\cite{TianSunChen18_1sub} and study its tradeoffs between
%different parameters while relaxing the privacy constraint.

\begin{itemize}
\item We introduce the concept of an $(\const{M},n)$ information retrieval (IR) scheme for a DSS with $n$ servers storing $\const{M}$ files using a global random strategy vector and a corresponding scheme, referred to as Scheme A, by building upon a PIR protocol introduced in~\cite{TianSunChen19_1}. By selecting each entry of the global random strategy according to a Bernoulli distribution, we provide for the special case of $n=2$ servers closed-form expressions for the achievable download rate, upload cost, access complexity, and privacy leakage (see \cref{thm:Scheme1_Mn2-IID-Bernoulli}). %The corresponding WPIR scheme is referred to as Scheme A.
\item In addition, we adopt the privacy metric introduced in the related works~\cite{SamyTandonLazos19_1,SamyAttiaTandonLazos21_1}, the so-called $\eps$-privacy, and compare our proposed Scheme~A to their leaky PIR scheme. By using a global random strategy for which each entry is independent and identically distributed (i.i.d.) according to a Bernoulli distribution, we show that Scheme~A performs better in terms of download rate for the case of $n=2$ servers.
\item By using a time-sharing argument (see \cref{thm:time-sharing_MI} and the discussion in \cref{sec:minimization-leakage_Scheme1-TimeSharing}), the download rate of Scheme~A can be improved. For both the MI and MaxL privacy metrics we show that optimizing the download rate for Scheme~A with time-sharing over the global random strategy can be framed as a convex optimization problem (see \cref{sec:minimization-leakage_Scheme1-TimeSharing}). 
\item We provide an information-theoretic converse result for the maximum possible download rate for an $(\const{M},n)$ IR scheme for both the MI and MaxL privacy metrics in \cref{thm:C_MI-converse,thm:C_MaxL-converse}, respectively, under a practical restriction on the alphabet size of queries and answers. The converse is derived using a known result between the entropy difference and the total variation (TV) distance of two probability distributions (see \cref{lem:entropy-diff_TV}). For the special case of $(\const{M},n)=(2,2)$ the WPIR capacity is provided in Theorems~\ref{thm:WPIRcapacity_M2n2-1stNotLeak_MI} (assuming that only one of the two servers can leak) and \ref{thm:WPIRcapacity_M2n2_MaxL} for the MI and MaxL privacy metrics, respectively. Moreover, we show that Scheme~A with time-sharing and with each entry of the global random strategy selected according to a Bernoulli distribution achieves the WPIR capacity for both privacy metrics under the above restrictions in this special case. % and $\const{M}=2$ files, we show that the achievable download rate for Scheme 1 is the maximum possible, i.e., it achieves WPIR capacity.
 \item Extensive numerical results showing the tradeoff between download rate, upload cost, access complexity, and privacy leakage are presented in \cref{sec:numerical-results} for Scheme~A (with and without time-sharing). As a comparison, we also compare with an alternative proposed constant-rate IR scheme, referred to as Scheme~B and based on the PIR scheme in~\cite[Lem.~4]{KumarLinRosnesGraellAmat19_1}.
\end{itemize}

%For the MaxL privacy metric and $\const{M}=2$ files, we show that the achievable downrate for Scheme I is the maximum possible, i.e., it achieves capacity.

%A Converse Bound with Restricted Alphabets of Queries and Download Symbols
%
%
\subsection{Related Work}

Independently, the download rate-leakage tradeoff has been studied by Samy \emph{et al.} \cite{SamyTandonLazos19_1} under the name of leaky PIR using a privacy metric related to differential privacy. The leaky PIR framework was recently also extended to symmetric PIR~\cite{SamyAttiaTandonLazos21_1} and to the consideration of latent attributes in the single server case~\cite{SamyAttiaTandonLazos20_1}. Symmetric PIR is a variant of PIR where in addition the user cannot learn anything about the remaining files in the database when the user retrieves its desired file~\cite{SunJafar19_1,WangSkoglund19_1,WangSkoglund19_2}. Moreover, Zhou \emph{et al.} \cite{ZhouGuoTian20_1} have recently studied the same problem under the MaxL privacy metric. Their scheme builds upon the same PIR protocol as our proposed Scheme~A. Moreover, by allowing for a permutation of the query strategy across the servers in addition to an arbitrary global random strategy, improved performance can be achieved. It can be shown that their scheme is equivalent to our Scheme~A with time-sharing. %A closed-form expression for the optimal tradeoff between download rate and privacy leakage of the scheme is derived. 

Our companion paper \cite{LinKumarRosnesGraellAmatYaakobi21_1} studies the corresponding problem for the single server setting under both the MI and MaxL privacy metrics. In particular, by relating the WPIR problem to rate-distortion theory, the capacity of single-server WPIR is fully characterized. Lastly, the related work in \cite{GuoZhouTian20_1} is also worth mentioning. In contrast to WPIR, where information leakage on the identity of the desired file to the servers is considered,  the information leakage of the nondesired files to the user for classical PIR was studied in \cite{GuoZhouTian20_1}.

\subsection{Organization of Paper}

The remainder of this paper is organized as follows. Section~\ref{sec:setup-preliminaries} presents the notation, basic definitions, preliminaries, and the problem formulation. In Section~\ref{sec:partition-schemes}, we present a partition scheme which first divides the database into equally-sized partitions and then uses a given IR scheme to retrieve a file from the corresponding partition. \cref{sec:Scheme1} presents an IR scheme built upon the PIR protocol introduced in~\cite{TianSunChen19_1}, referred to as Scheme~A, while \cref{sec:Scheme2} presents a constant-rate IR scheme based on the PIR scheme in \cite[Lem.~4]{KumarLinRosnesGraellAmat19_1}, referred to as Scheme~B. For both schemes and for two servers we provide closed-form expressions for the download rate, upload cost, access complexity, and information leakage under a Bernoulli global random strategy. In Section~\ref{sec:Scheme1}, using Scheme~A to retrieve files from a partition of the partition scheme is also analyzed. Based on Scheme~A, a WPIR scheme achieving a better download rate than the leaky PIR scheme under the $\eps$-privacy metric~\cite{SamyTandonLazos19_1,SamyAttiaTandonLazos21_1} is proposed in Section~\ref{sec:Scheme1_epsP}. The minimization of the information leakage for Scheme~A with time-sharing is considered in Section~\ref{sec:minimization-leakage_Scheme1-TimeSharing}. In particular, we show that the minimization problem is a convex optimization problem for both the MI and MaxL privacy metrics. Then, in \cref{sec:converse-results_MI,sec:converse-results_MaxL} we present converse results on the minimum download cost for both privacy metrics under a practical restriction on the alphabet size of queries and answers. Numerical results comparing Schemes~A (with and without time-sharing) and B in terms of download rate, upload cost, access complexity, and information leakage are presented in Section~\ref{sec:numerical-results}. Finally, some conclusions are drawn in Section~\ref{sec:conclusion}.

% In Section~\ref{sec:characterizaiton_download-leakage}, we introduce the download-leakage function of single-server WPIR, which is defined as the minimum achievable download cost for a given information leakage constraint. Moreover, we discuss some properties of the function when the leakage is measured in terms of the MI or MaxL metrics. In Section~\ref{sec:WPIR-schemes_basic-partition}, a basic solution for single-server WPIR is presented in which the files  are partitioned into several partitions. In Section~\ref{sec:capacity_single-server-WPIR}, we give a closed-form expression for the single-server WPIR capacity for both the MI and MaxL metrics. A capacity-achieving WPIR scheme is proposed in Section~\ref{sec:achievability}. The converse result on the minimum download cost for the MI metric is provided in Section~\ref{sec:converse_Thm1}, while that of the MaxL metric is given in Section~\ref{sec:converse_Thm2}. Finally, Section~\ref{sec:conclusion} concludes the paper. 

\section{Preliminaries}
\label{sec:setup-preliminaries}

\subsection{Notations}
\label{sec:notation-definitions}

We denote by $\Naturals$ the set of all positive integers, $[a]\eqdef\{1,2,\ldots,a\}$, and $[a:b]\eqdef\{a,a+1,\ldots,b\}$ for $a,b\in\{0\}\cup\Naturals$, $a \leq b$. Vectors are denoted by bold letters, random variables (RVs) (either scalar or vector) by uppercase letters, and sets by calligraphic uppercase letters, e.g., $\vect{x}$, $X$, and $\set{X}$, respectively. Moreover, $\cset{\set{X}}$ denotes the complement of a set $\set{X}$ in a universe set. The all-zero matrix of dimensions $a\times b$ is represented by $\vect{0}_{a\times b}$, or simply by  $\vect{0}$ when the dimensions are not important. % , while the all-one matrix of dimensions $a\times b$ is referred to as $\vect{1}_{a\times b}$.
For a given index set $\set{S}$, we write $X^\set{S}$ and $Y_\set{S}$ to represent $\bigl\{X^{(m)}\colon m\in\set{S}\bigr\}$ and $\bigl\{Y_l\colon l\in\set{S}\bigr\}$, respectively. $X\indep Y$ means that the two RVs $X$ and $Y$ are independent. $\E[X]{\cdot}$ denotes the expectation over the RV $X$. $X\sim\Bernoulli{p}$ denotes a Bernoulli-distributed RV with $\Prv{X=1}=p=1-\Prv{X=0}$ and $X\sim\Uniform{\set{S}}$ a uniformly-distributed RV over the set $\set{S}$. $\trans{(\cdot)}$ denotes the transpose of its argument. The Hamming weight of a binary vector $\vect{x}$ is denoted by $\Hwt{\vect{x}}$, while its support is denoted by $\Spt{\vect{x}}$. $\sigma(\cdot)$ denotes a left cyclic permutation, while $a$ left cyclic shifts are obtained through function composition and denoted by $\sigma^a(\cdot)$. The inner product of $\vect{x}$ and $\vect{y}$ is denoted by $\langle\vect{x},\vect{y}\rangle$. $\HP{X}$, $\HP{P_X}$, or $\HP{p_{1},\ldots,p_{\card{\set{X}}}}$ represents the entropy of $X$, where $P_{X}(\cdot)=(p_1,\ldots,p_{\card{\set{X}}})$ denotes the distribution of the RV $X$, while $\eMI{X}{Y}$ is the MI between $X$ and $Y$ (in bits). $\Hb(p)\eqdef -p\log_2{p}-(1-p)\log_2{(1-p)}$ is the binary entropy function. With some abuse of notation, when the marginal distribution of either $X$ or $Y$ is assumed fixed and known, the MI between $X$ and $Y$ is sometimes simply written as $\eMI{X}{Y} \equiv \II(P_{X|Y}) \equiv \II(P_{Y|X})$.
% The Galois field with $q$ elements is denoted by $\GF(q)$, where $q$ is a prime or a power of a prime.

\subsection{System Model}
\label{sec:system-model}

We consider a DSS with $n$ noncolluding replicated servers, each storing $\const{M}$ independent files $\vect{X}^{(1)},\ldots,\vect{X}^{(\const{M})}$, where each file $\vect{X}^{(m)}=\trans{\bigl(X_1^{(m)},\ldots,X_\beta^{(m)}\bigr)}$, $m\in [\const{M}]$, has length $\beta$, and can be seen as a $\beta\times 1$ vector over an alphabet $\set{X}$. Assume that each element of $\vect{X}^{(m)}$ is chosen independently and uniformly at random from $\set{X}$. Thus, we have $\bigHP{\vect{X}^{(m)}}=\beta\log_2{\card{\set{X}}}$, $\forall\,m\in [\const{M}]$.

In information retrieval (IR), a user wishes to efficiently retrieve one of the $\const{M}$ files stored in the replicated DSS. Similar to the detailed mathematical description in \cite{TianSunChen19_1}, we assume that the requested file index $M$ is a RV and $M\sim\Uniform{[\const{M}]}$.\footnote{Here, we assume for simplicity that the  requested file index $M$  is uniformly distributed. However, this assumption can be lifted, which is referred to as semantic PIR in the literature~\cite{VithanaBanawanUlukus20_1}.} We give the following definition of an IR scheme.
\begin{definition}
  \label{Def:Mn-IRscheme}
  An $(\const{M},n)$ IR scheme $\collect{C}$ for a DSS with $n$  servers storing $\const{M}$ files
  consists of:
  \begin{itemize}
  \item A global random strategy $\vect{S}$, whose alphabet is $\set{S}$.
    % is a finite set
  \item $n$ query-encoding functions $\phi_l$, $l\in [n]$, that generate $n$ queries $\vect{Q}_l=\phi_l(M,\vect{S})$ with
    alphabet $\set{Q}_l$, where query $\vect{Q}_l$ is sent to server $l$.
    % and the joint distribution $P_{\vect{Q}_1,\ldots,\vect{Q}_n|M}$ is specified by the system designer,
  \item $n$ answer-length functions $\ell_l(\vect{Q}_l)$, with range $\{0\}\cup\Nat{}$, that define the length of the answers. $\ell_l(\vect{Q}_l)$ is a function of the query $\vect{Q}_l$, which is independent of the particular realization of the files.
  \item $n$ answer functions % $\varphi_l$
    \begin{IEEEeqnarray*}{c}
      \varphi_l\colon\set{Q}_l\times\set{X}^{\beta\const{M}}\to\set{A}^{\ell_l},\quad l\in[n],
    \end{IEEEeqnarray*}
    that return the answers $\vect{A}_l=\varphi_l(\vect{Q}_l,\vect{X}^{[\const{M}]})$, where $\set{A}$ is the download symbol alphabet. % for all $l\in [n]$
  \item $n$ access-number functions $\delta_l(\vect{Q}_l)$, with range $\{0\}\cup\Nat{}$, that define the number of
    symbols accessed by $\vect{Q}_l$.
  \end{itemize}
  This scheme should satisfy the condition of \emph{perfect retrievability},
  \begin{IEEEeqnarray}{c}
    \bigHPcond{\vect{X}^{(M)}}{\vect{A}_{[n]},\vect{Q}_{[n]},M}=0.
    \label{eq:retrievability}
  \end{IEEEeqnarray}
\end{definition}

  Since a user should be able to generate the queries without any prior knowledge of the realizations of the files, it is reasonable to assume that the queries and the files are independent, i.e.,
  \begin{IEEEeqnarray}{rCl}
    \eMI{\vect{X}^{[\const{M}]}}{\vect{Q}_{[n]}}=0.
    \label{eq:independent_files-queries}
  \end{IEEEeqnarray}  
  This particular assumption is used in the converse proofs of \cref{sec:converse-results_MI,sec:converse-results_MaxL} (see Appendix~\ref{sec:proof_HPm-LB}).
% Here, when the user decides the requested file index $M$, query $\vect{Q}_l$ is sent to the $l$-th storage server, and
% in response to the received query, server $l$ sends answer $\vect{A}_l$ back to the user.

Note that a PIR scheme is an $(\const{M},n)$ IR scheme that satisfies perfect privacy for all servers, i.e., for every $m,m'\in[\const{M}]$ with $m\neq m'$, the condition
\begin{IEEEeqnarray}{c}
  \label{eq:strong-privacy}
  \Prvcond{\vect{Q}_l=\vect{q}_l}{M=m}=\Prvcond{\vect{Q}_l=\vect{q}_l}{M=m'}
\end{IEEEeqnarray}
holds for all $\vect{q}_l\in\set{Q}_l$, $l\in [n]$. The privacy constraint~\eqref{eq:strong-privacy} is equivalent to the
statement that
% all the random queries $\vect{Q}_{[n]}$ are independent of the index of the requested file,
$M\indep \vect{Q}_l$. We denote by $\vect{Q}_l^{(m)}$ the query sent to server $l$ if file $\vect{X}^{(m)}$ is requested, which
is a RV with probability mass function (PMF) $P_{\vect{Q}_l^{(m)}}(\vect{q}_l)\eqdef\Prvcond{\vect{Q}_l=\vect{q}_l}{M=m}=P_{\vect{Q}_l|M}(\vect{q}_l|m)$.

We refer to an $(\const{M},n)$ IR scheme that does not satisfy \eqref{eq:strong-privacy} as a \emph{WPIR scheme}, as
opposed to a PIR scheme that leaks no information.

% Note that under the setting of Definition~\ref{Def:Mn-IRscheme}, an $(\const{M},n)$ IR scheme can be seen as a
% hypothesis testing problem, where the user decides one of the $\const{M}$ hypotheses (files) to be retrieved by
% sending the queries $\vect{Q}_l^{(m)}$, $l\in [n]$ and each file $m\in [\const{M}]$ should be requested at least
% once. Hence, without loss of generality, we assume
% \begin{IEEEeqnarray*}{c}
%   \min_{m\in [\const{M}]}\max_{\vect{q}_l\in\set{Q}_l}\Prvcond{\vect{Q}_l=\vect{q}_l}{M=m}>0,\quad\forall\,l\in [n],
% \end{IEEEeqnarray*}
% since each hypothesis should at least be requested once during the $(\const{M},n)$ IR scheme, otherwise, we could
% exclude the $m$-th hypothesis if $\max_{\vect{q}_l\in\set{Q}_l}\Prvcond{\vect{Q}_l=\vect{q}_l}{M=m}=0$.

\subsection{Metrics of Information Leakage}
\label{sec:metrics_information-leakage}

In this paper, to measure the information leakage between $M$ and $\vect{Q}_l$ for an IR scheme, we first consider MI, which measures the \emph{average} amount of information about the requested file index $M$ from the queries $\vect{Q}_l$. Moreover, we also consider WIL~\cite{KopfBasin07_1} and MaxL, which is considered a robust information leakage quantity~\cite{IssaWagnerKamath20_1}. For the MI privacy metric, we use the following theorem to motivate the definition of information leakage for an $(\const{M},n)$ IR scheme.
% \begin{proposition}[Time-Sharing Principle for the MI Metric]
%   \label{prop:time-sharing_MI}
\begin{theorem}[Time-Sharing Principle for the MI Metric]
  \label{thm:time-sharing_MI}
  Consider an $(\const{M},n)$ IR scheme $\collect{C}$, where the leakage of the $l$-th server is defined as $\eMI{M}{\vect{Q}_l}$, $l\in[n]$. Then, there exists an $(\const{M},n)$ IR scheme $\widebar{\collect{C}}$ with leakage $\bar{\rho}\eqdef\frac{1}{n}\sum_{l\in[n]}\eMI{M}{\vect{Q}_l}$ for every server.
  % \end{proposition}
\end{theorem}
\begin{IEEEproof}
  The theorem is proven by a time-sharing argument. Assume that the IR scheme $\collect{C}$ is given by the query-encoding functions $\phi_l$, answer functions $\varphi_l$, $l\in[n]$, and a random strategy $\vect{S}$.

  Next, define query-encoding functions $\bar{\phi_l}$, answer functions $\bar{\varphi}_l$, $l\in[n]$, and a random strategy $\vect{S}_T$ for an $(\const{M},n)$ IR scheme $\widebar{\collect{C}}$ as follows. Given a requested file index $M$, the user  chooses a $T\sim\Uniform{[n]}$ and assigns the query $\widebar{\vect{Q}}_l=\bar{\phi}_l(M,\vect{S}_T)\eqdef\phi_{\sigma^{T-1}(l)}(M,\vect{S})=\vect{Q}_{\sigma^{T-1}(l)}(M,\vect{S})$ to the $l$-th server, $l\in[n]$. % , where $\sigma(\cdot)$ denotes a left cyclic permutation.
  
  The answer functions for $\widebar{\collect{C}}$ are defined as $\bar{\varphi_l}\bigl(\widebar{\vect{Q}}_l,\vect{X}^{[\const{M}]}\bigr)\eqdef\varphi_{\sigma^{T-1}(l)}\bigl(\phi_{\sigma^{T-1}(l)}(M,\vect{S}),\vect{X}^{[\const{M}]}\bigr)$, $l\in[n]$, and hence perfect retrievability is achieved due to the perfect retrievability of the IR scheme $\collect{C}$.
  
  The MI information leakage of the $l$-th server is 
  \begin{IEEEeqnarray*}{rCl}
    \eMI{M}{\widebar{\vect{Q}}_l}& = &\bigHP{\widebar{\vect{Q}}_l}-\bigHPcond{\widebar{\vect{Q}}_l}{M}
    \nonumber\\
    & = &\bigHP{\vect{Q}_{\sigma^{T-1}(l)}}-\bigHPcond{\vect{Q}_{\sigma^{T-1}(l)}}{M}
    \nonumber\\
    & \stackrel{(a)}{=} &\sum_{t=1}^n\Prv{T=t}\eMIcond{\vect{Q}_{\sigma^{t-1}(l)}}{M}{T=t}\\
    & = &\frac{1}{n}\sum_{l'=1}^n\eMI{\vect{Q}_{l'}}{M},\quad\forall\,l\in[n],\nonumber
  \end{IEEEeqnarray*}
  where $(a)$ follows from the definition of conditional mutual information.
\end{IEEEproof}
\cref{thm:time-sharing_MI} indicates that we can always obtain an $(\const{M},n)$ IR scheme with equal MI leakage at each server by cyclically shifting the servers' queries of an existing $(\const{M},n)$ IR scheme $\collect{C}$ $n$ times. Such a time-sharing scheme is denoted by $\widebar{\collect{C}}$.

Hence, to characterize the overall leakage of a given $(\const{M}, n)$ IR scheme $\collect{C}$ in terms of MI, we consider the information leakage metric
\begin{IEEEeqnarray}{c} \label{eq:MI_leakge_def}
  \rho^{(\mathsf{MI})}(\collect{C})\eqdef\frac{1}{n}\sum_{l\in [n]}\eMI{M}{\vect{Q}_l}.
\end{IEEEeqnarray}

The WIL of the $l$-th server is defined as $\WL{M}{\vect{Q}_l}\eqdef\eHP{M}-\min_{\vect{q}_l\in\set{Q}_l}\eHPcond{M}{\vect{Q}_l=\vect{q}_l}$. The overall WIL of a given $(\const{M},n)$ IR scheme $\collect{C}$ is then given as $\rho^{(\mathsf{WIL})}(\collect{C})\eqdef\max_{l\in [n]}\WL{M}{\vect{Q}_l}$.

Further, given a joint distribution $P_{M,\vect{Q}}$, the MaxL from $M$ to $\vect{Q}$ is defined as
\begin{IEEEeqnarray}{rCl}
  \ML{M}{\vect{Q}}& \eqdef &\log_2{\sum_{\vect{q}\in\set{Q}}\max_{m\in[\const{M}]}P_{\vect{Q}|M}(\vect{q}|m)}.
  % \nonumber
  % \\
  % & = &\SI{M}{\vect{Q}}.
  \label{eq:expression_MaxL}
\end{IEEEeqnarray}

Note that MaxL has a strong connection to the min-entropy (MinE) privacy metric, which is commonly-used in the computer science literature~\cite{Smith09_1,BartheKopf11_1}. MinE is a special case of the widely known R{\'e}nyi entropy~\cite{Renyi61_1}. The MinE information leakage and the MaxL privacy metric can be shown to be equivalent when $M$ is uniformly distributed~\cite{BartheKopf11_1,IssaWagnerKamath20_1}.

We will use \eqref{eq:expression_MaxL} as the MaxL privacy metric for the designed query distribution $P_{\vect{Q}_l|M}$ at the $l$-th server of a WPIR scheme, which is denoted by
\begin{IEEEeqnarray*}{c}
  \rho^{(\mathsf{MaxL})}(M,\vect{Q}_l)\eqdef\ML{M}{\vect{Q}_l}.
\end{IEEEeqnarray*}
The overall MaxL of a given $(\const{M},n)$ IR scheme $\collect{C}$ is then defined to be the worst-case MaxL over all servers:
\begin{IEEEeqnarray*}{c}
 \rho^{(\mathsf{MaxL})}(\collect{C})\eqdef\max_{l\in [n]}\ML{M}{\vect{Q}_l}. 
\end{IEEEeqnarray*}

The following lemma summarizes some useful properties for both the MI and MaxL privacy metrics.
\begin{lemma}[{\cite[Lem.~1, Cor.~1]{IssaWagnerKamath20_1}}]
  \label{lem:convex_MI-MaxL}
  For any joint distribution $P_{X,Y}$, we have the following.
  \begin{enumerate}
  \item (Data Processing Inequalities) If the RVs $X,Y$, and $Z$ form a Markov chain, then
    \begin{IEEEeqnarray*}{rCl}
      \MI{X}{Z}& \leq &\min\{\MI{X}{Y},\MI{Y}{Z}\},\textnormal{ and}
      \\
      \ML{X}{Z}& \leq &\min\{\ML{X}{Y},\ML{Y}{Z}\}.
    \end{IEEEeqnarray*}
  \item Consider a fixed distribution $P_X$. Then, both $\MI{X}{Y}$ and $2^{\ML{X}{Y}}$ are convex functions in $P_{Y|X}$.
  \end{enumerate}
\end{lemma}

%\subsubsection{$\eps$-Privacy Metric}
%\label{sec:eps-privacy-metric}

There are other privacy metrics that can be used to relax the perfect privacy requirement of PIR. The authors of \cite{SamyTandonLazos19_1,SamyAttiaTandonLazos21_1} introduced the $\eps$\emph{-privacy} metric based on the notion of (local) differential privacy. Under the setup discussed in this paper, we define the $\eps$-privacy leakage at the $l$-th server of an IR scheme $\collect{C}$ as
\begin{IEEEeqnarray*}{c}
  \rho^{(\eps\mhyph\mathsf{P})}(M;\vect{Q}_l)\eqdef\ln{\left(\max_{\vect{q}_l\in\set{Q}_l}\max_{m,m'\in[\const{M}]}\frac{P_{\vect{Q}_l|M}(\vect{q}_l|m)}{P_{\vect{Q}_l|M}(\vect{q}_l|m')}\right)}.
\end{IEEEeqnarray*}
Similar to the MaxL privacy metric, we also define
\begin{IEEEeqnarray*}{c}
  \rho^{(\eps\mhyph\mathsf{P})}{(\collect{C})}\eqdef\max_{l\in[n]}\left\{\rho^{(\eps\mhyph\mathsf{P})}(M;\vect{Q}_l)\right\}
  \IEEEeqnarraynumspace\label{eq:def_eps-privacy}
\end{IEEEeqnarray*}
as the $\eps$-privacy leakage of a given $(\const{M},n)$ IR scheme $\collect{C}$.

Note that $\eps$-privacy normally gives a \emph{stronger} privacy protection than the MI or MaxL metrics. Moreover, it is worth mentioning that there is a close relation between MaxL and differential privacy~\cite{DworkMcSherryNissimSmith06_1,Dwork06_1}, see, e.g.,~\cite[Thm.~3]{BartheKopf11_1}. In this work, although we mainly focus on the MI and MaxL privacy metrics, we will still use the $\eps$-privacy metric to show that our proposed Scheme~A outperforms the schemes from \cite{SamyTandonLazos19_1,SamyAttiaTandonLazos21_1} %in our information-theoretic WPIR framework 
(see~\cref{sec:Scheme1_epsP,sec:eps-privacy_Mn2}).

Throughout the paper, the information leakage metric of a WPIR scheme $\collect{C}$ is denoted by $\rho^{(\cdot)}(\collect{C})$, where the superscript indicates the leakage metric ($\mathsf{MI}$, $\mathsf{WIL}$, $\mathsf{MaxL}$, or $\eps\mhyph\mathsf{P}$) we are considering. Moreover, since $P_M$ is fixed, we will also simply write the leakage measure $\rho^{(\cdot)}(\cdot,\cdot)$ as a function of the designed query distribution $P_{\vect{Q}_l|M}$ of a WPIR scheme. For example, $\rho^{(\mathsf{MI})}(M,\vect{Q}_l)\equiv\II(P_{\vect{Q}_l|M})\equiv\rho^{(\mathsf{MI})}(P_{\vect{Q}_l|M})$, $l\in [n]$.

\subsection{Download Cost, IR Rate, Upload Cost, and Access Complexity of an $(\const{M},n)$ IR Scheme}
\label{sec:achievable-rate_IR}

For WPIR, in contrast to PIR, the download cost may be different for the retrieval of different files. Thus, the download cost can be defined as the expected download cost over all possible requested files. The download cost of a WPIR scheme $\collect{C}$ for the retrieval of the $m$-th file, denoted by $\const{D}^{(m)}(\collect{C})$, is defined as the expected length (in bits) of the returned answers across all servers over all random queries,
\begin{IEEEeqnarray*}{c}
  \const{D}^{(m)}(\collect{C})\eqdef\log_2\card{\set{A}}\sum_{l=1}^n\BigE[\vect{Q}_l^{(m)}]{\ell_l(\vect{Q}^{(m)}_l)},
\end{IEEEeqnarray*}
where $\vect{Q}_l^{(m)}$ is the RV with PMF $P_{\vect{Q}_l^{(m)}}(\vect{q}_l)=P_{\vect{Q}_l|M}(\vect{q}_l|m)$. The overall download cost of an IR scheme $\collect{C}$, denoted by $\const{D}(\collect{C})$, is defined as the expected download cost over all files, i.e.,
\begin{IEEEeqnarray*}{rCl}
  \const{D}(\collect{C})& \eqdef &\log_2{\card{\set{A}}}\E[M]{\sum_{l=1}^n\E[\vect{Q}_l^{(m)}]{\ell_l(\vect{Q}_l)}}
    \nonumber\\
    & = &\log_2{\card{\set{A}}}\sum_{l=1}^n\E[\vect{Q}_l]{\ell_l(\vect{Q}_l)}.
    \IEEEeqnarraynumspace%\label{eq:DLcst_C}
\end{IEEEeqnarray*}
Accordingly, the IR rate of an IR scheme $\collect{C}$ is defined as
\begin{IEEEeqnarray*}{c}
  \const{R}(\collect{C})\eqdef\frac{\beta\log_2{\card{\set{X}}}}{\const{D}(\collect{C})}.
\end{IEEEeqnarray*}
The upload cost $\const{U}(\collect{C})$ of an IR scheme $\collect{C}$ is defined as the sum of the entropies of the queries $\vect{Q}_{[n]}$,
\begin{IEEEeqnarray*}{c}
  \const{U}(\collect{C})\eqdef\sum_{l=1}^n\eHP{\vect{Q}_l}.\label{eq:def_upload}
\end{IEEEeqnarray*}
Moreover, the access complexity $\Delta(\collect{C})$ of an IR scheme $\collect{C}$ is defined as the expected number of accessed symbols across all servers for the retrieval of a single file,
\begin{IEEEeqnarray}{c}
  \Delta(\collect{C})\eqdef\sum_{l=1}^n\E[\vect{Q}_l]{\delta_l(\vect{Q}_l)}
  =\frac{1}{\const{M}}\sum_{m=1}^{\const{M}}\sum_{l=1}^n\E[\vect{Q}_l^{(m)}]{\delta_l(\vect{Q}_l)}.
  \IEEEeqnarraynumspace\label{eq:def_access}
\end{IEEEeqnarray}

An achievable $4$-tuple of an IR scheme is defined as follows.
\begin{definition}
  \label{def:tuple_IR}
  Consider a DSS with $n$ noncolluding servers storing $\const{M}$ files. A $4$-tuple $(\const{R},\const{U},\Delta,\rho)$ is said to be \emph{achievable} with information leakage metric $\rho^{(\cdot)}$ if there exists an $(\const{M},n)$ IR scheme $\collect{C}$ such that $\const{R}(\collect{C})=\const{R}$, $\const{U}(\collect{C})=\const{U}$, $\Delta(\collect{C})=\Delta$, and $\rho^{(\cdot)}(\collect{C})=\rho$.
\end{definition}

We remark that a PIR scheme corresponds to an $(\const{M},n)$ IR scheme with $\rho^{(\cdot)}=0$. It was shown in \cite{SunJafar17_1} that for $n$ noncolluding replicated servers and for a given number of files $\const{M}$, the PIR capacity, denoted by $\const{C}_{\const{M},n}$, is $\const{C}_{\const{M},n}=\inv{\bigl(1+\nicefrac{1}{n}+\cdots+\nicefrac{1}{n^{\const{M}-1}}\bigr)}$.

\section{Partition WPIR Scheme}
\label{sec:partition-schemes}

A simple approach for the construction of WPIR schemes is to first partition the database into $\eta$ equally-sized partitions, each consisting of $\nicefrac{\const{M}}{\eta}$ files where $\nicefrac{\const{M}}{\eta}\in\Naturals$,\footnote{While it is not necessary that each partition has an equal number of files, for simplicity in this paper we make this assumption.} and then use a given $(\nicefrac{\const{M}}{\eta},n)$ IR scheme to retrieve a file from the corresponding partition. Obviously, the resulting scheme is not a PIR scheme, since the servers gain the knowledge of which partition the requested file belongs to. In this section, we pursue this approach to construct an $(\const{M},n)$ IR scheme building on a given $(\nicefrac{\const{M}}{\eta},n)$ IR scheme as a subscheme.

The partition $(\const{M},n)$ WPIR scheme is formally described as follows. Assume the requested file $\vect{X}^{(m)}$
belongs to the $j$-th partition, where $j\in[\eta]$.
% A file $\vect{X}^{(M)}$ is requested such that the file index $M\sim\Uniform{[\const{M}]}$. The induced partition $J$
% is therefore uniformly-distributed as $J\sim\Uniform{[\eta]}$.
Then, the query $\vect{Q}_l$ is constructed as
\begin{IEEEeqnarray}{c}
  \vect{Q}_l=\bigl(\widetilde{\vect{Q}}_l,j\bigr)\in\tilde{\set{Q}}_l\times [\eta],\quad l\in[n],
  \label{eq:queries_partition-scheme}
\end{IEEEeqnarray}
where $\widetilde{\vect{Q}}_l$ is the query of an existing $(\nicefrac{\const{M}}{\eta},n)$ IR scheme $\tilde{\collect{C}}$.

The following theorem states the achievable $4$-tuple of the partition scheme.
\begin{theorem}
  \label{thm:partition-schemes}
  Consider a DSS with $n$ noncolluding servers storing $\const{M}$ files, and let $\tilde{\collect{C}}$ be an $(\nicefrac{\const{M}}{\eta},n)$ IR scheme with achievable 4-tuple $\bigl(\tilde{\const{R}},\tilde{\const{U}},\tilde{\Delta}, \tilde{\rho}^{(\cdot)}\bigr)$. Then, the $4$-tuple
  \begin{IEEEeqnarray}{rCl}
    \IEEEeqnarraymulticol{3}{l}{%
      \bigl(\const{R}(\collect{C}),\const{U}(\collect{C}),\Delta(\collect{C}),\rho^{(\cdot)}(\collect{C})\bigr)}
    \nonumber\\*\quad%
    & = &
    \bigl(\tilde{\const{R}},\tilde{\const{U}}+n\log_2{\eta},\widetilde{\Delta},\tilde{\rho}^{(\cdot)}+\log_2{\eta}\bigr)
    \label{eq:tuple_partition-scheme}
  \end{IEEEeqnarray}
  is achievable by the $(\const{M},n)$ partition scheme $\collect{C}$ constructed from $\tilde{\collect{C}}$ as described in \eqref{eq:queries_partition-scheme}.
\end{theorem}
\begin{IEEEproof}
  The proof is deferred to Appendix~\ref{sec:proof_partition-schemes}.
\end{IEEEproof}

Since a PIR scheme is also an IR scheme, this simple approach for the construction of WPIR schemes can also be adapted to use any of the existing $(\nicefrac{\const{M}}{\eta},n)$ PIR schemes in the literature as a subscheme. We refer to the partition scheme that uses a PIR scheme as the underlying subscheme and the query generation in~\eqref{eq:queries_partition-scheme} as a \emph{basic scheme} and denote it by $\collect{C}^{\mathsf{basic}}$ (it gives the $4$-tuple as in~\eqref{eq:tuple_partition-scheme} with $\tilde{\rho}^{(\cdot)}=0$). In Section~\ref{sec:partition-SchemeA}, we will present another partition WPIR scheme  based on our proposed IR scheme.

\section{$(\const{M},n)$ Scheme~A}
\label{sec:Scheme1}

In \cite[Sec.~III-B]{TianSunChen19_1}, a PIR scheme that achieves both the minimum upload and download cost was proposed. The queries $\vect{Q}_{[n]}$ of the scheme in \cite[Sec.~III-B]{TianSunChen19_1} are randomly generated according to a random strategy $\vect{S}=(S_1,\ldots,S_{\const{M}-1})$ with % independent and identically distributed (i.i.d.)
i.i.d.\ entries according to $\Uniform{[0:n-1]}$.\footnote{The PIR scheme in \cite[Sec.~III-B]{TianSunChen19_1} can be seen as a generalization of the canonical $(2,2)$ PIR scheme that was first introduced in \cite[Sec.~III-B]{SunJafar18_3} and further elaborated in \cite{TianSunChen18_1}, where the authors focused on the minimization of the storage overhead.} In this section, we introduce an $(\const{M},n)$ WPIR scheme, referred to as \emph{Scheme~A} and denoted by $\collect{C}_{\mathsf{A}}$, based on the PIR scheme in~\cite{TianSunChen19_1}. Scheme~A can be seen as a generalization of the PIR scheme in~\cite{TianSunChen19_1} where we lift the perfect privacy condition~\eqref{eq:strong-privacy}.

% Before we describe the detailed steps of Scheme~A, we provide the following preliminaries for the proposed Scheme~A.

For the proposed scheme, assume that $\set{X}=\{0,1\}$ and the file size to be $\beta=n-1$. We represent a query by a length-$\const{M}$ vector $\vect{q}_l=(q_{l,1},\ldots,q_{l,\const{M}})\in\set{Q}_l\subseteq[0:n-1]^{\const{M}}$. Also, the realization of $\vect{S}$ is denoted by a length-$(\const{M}-1)$ vector $\vect{s}=(s_1,\ldots,s_{\const{M}-1})$, $s_j\in [0:n-1]$, $j\in [\const{M}-1]$.

Before describing Scheme~A in detail for the general case, for simplicity we first present Scheme~A for the case of $\const{M}=2$ files and $n=2$ servers (i.e., both servers $1$ and $2$ store $\vect{X}^{(1)}$, $\vect{X}^{(2)}$) in the following example.
\begin{example}
  \label{ex:Ex1_WPIR_M2n2}
  We illustrate the $(2,2)$ Scheme~A obtained by adopting a nonuniformly-distributed random strategy $\vect{S}$ giving a joint PMF $P_{\vect{Q}_1,\vect{Q}_2}(\vect{q}_1,\vect{q}_2)$ as in Table~\ref{tab:joint-distribution_Q1-Q2}.
  % (the realization of $\vect{Q}_l$ is a length-$2$ vector $\vect{q}_l=(q_{l,1},q_{l,2})$, $l=1,2$).
  % \begin{IEEEeqnarray*}{rCl}
  %   \Scale[0.9]{ P_{\vect{Q}_1^{(1)},\vect{Q}_2^{(1)}}(\vect{q}_1,\vect{q}_2)}& = &\Scale[0.9]{
  %   \begin{cases}
  %     1-p & \textnormal{if } \vect{q}_1=(0,0),\vect{q}_2=(1,0),
  %     \\
  %     p & \textnormal{if } \vect{q}_1=(1,1),\vect{q}_2=(0,1),
  %     \\
  %     0 & \textnormal{otherwise},
  %   \end{cases}}\\
  %   %   \label{eq:M1n2-Scheme1-Bernoulli}\\
  %   \Scale[0.9]{P_{\vect{Q}_1^{(2)},\vect{Q}_2^{(2)}}(\vect{q}_1,\vect{q}_2)}& = &\Scale[0.9]{
  %   \begin{cases}
  %     1-p & \textnormal{if } \vect{q}_1=(0,0),\vect{q}_2=(0,1),
  %     \\
  %     p & \textnormal{if } \vect{q}_1=(1,1),\vect{q}_2=(1,0),
  %     \\
  %     0 & \textnormal{otherwise}.
  %   \end{cases}}
  %   \IEEEeqnarraynumspace\label{eq:M2n2-Scheme1-Bernoulli}
  % \end{IEEEeqnarray*}
  \begin{table}[t!]
    \centering
    \caption{The joint PMF of $\vect{Q}_1,\vect{Q}_2$ for $0\leq p\leq\frac{1}{2}$, given $M=1,2$.}
    \label{tab:joint-distribution_Q1-Q2}
    \vspace{-2ex}
    \Resize[0.9\columnwidth]{
    \begin{IEEEeqnarraybox}[
      \IEEEeqnarraystrutmode
      \IEEEeqnarraystrutsizeadd{4pt}{5pt}]{v/c/v/c/v/c/V/c/v}
      \IEEEeqnarrayrulerow\\
    & P_{\vect{Q}_1^{(1)},\vect{Q}_2^{(1)}}(\vect{q}_1,\vect{q}_2) && \vect{q}_2=(1,0)
    && \vect{q}_2=(0,1)&& P_{\vect{Q}_1^{(1)}}(\vect{q}_1)
    \\*\hline
    & \vect{q}_1=(0,0) && 1-p && 0 && 1-p
    \\\hline
    & \vect{q}_1=(1,1) && 0     && p && p
    \\\IEEEeqnarraydblrulerow\\% \hline
    & P_{\vect{Q}_2^{(1)}}(\vect{q}_2) && 1-p && p && 
    \\*\IEEEeqnarrayrulerow
  \end{IEEEeqnarraybox}}
  \\
  (a)
  \\[1mm]
  \Resize[0.9\columnwidth]{
  \begin{IEEEeqnarraybox}[
    \IEEEeqnarraystrutmode
    \IEEEeqnarraystrutsizeadd{4pt}{5pt}]{v/c/v/c/v/c/V/c/v}
    \IEEEeqnarrayrulerow\\
    & P_{\vect{Q}_1^{(2)},\vect{Q}_2^{(2)}}(\vect{q}_1,\vect{q}_2)
    && \vect{q}_2=(1,0) && \vect{q}_2=(0,1) && P_{\vect{Q}_1^{(2)}}(\vect{q}_1)
    \\*\hline
    & \vect{q}_1=(0,0)&& 0 && 1-p && 1-p
    \\\hline
    & \vect{q}_1=(1,1)&& p && 0   && p
    \\\IEEEeqnarraydblrulerow\\% \hline
    & P_{\vect{Q}_2^{(2)}}(\vect{q}_2)   && p && 1-p && 
    \\*\IEEEeqnarrayrulerow
  \end{IEEEeqnarraybox}}
  \\
  (b)
  \end{table}
  Files $\vect{X}^{(1)}$ and $\vect{X}^{(2)}$ are composed of one stripe each ($\beta=n-1=1$). The answers $\vect{A}_1$ and $\vect{A}_2$ are given by $\bigl(\vect{A}_1(\vect{q}_1), \vect{A}_2(\vect{q}_2)\bigr)=\bigl(X^{(1)}_{q_{1,1}}+X^{(2)}_{q_{1,2}},X^{(1)}_{q_{2,1}}+X^{(2)}_{q_{2,2}}\bigr)$, where $X^{(m)}_0 = 0$ for all $m \in [2]$.

  One can easily verify that perfect retrievability is satisfied for the above $(2,2)$ IR scheme. Its IR rate is a function of $p$ and is given by $\const{R}(p)=\inv{(p+(1-p)+p)}=\inv{(1+p)}$. Observe that $M\indep \vect{Q}_1$, which implies that $\eMI{M}{\vect{Q}_1}=\WL{M}{\vect{Q}_1}=\ML{M}{\vect{Q}_1}=0$, i.e., it does not leak any information on the identity of the retrieved file to the first server.

  The information leakage is $\rho^{(\mathsf{MI})}=\frac{1-\Hb(p)}{2}$, $\rho^{(\textnormal{WIL})}=1-\Hb(p)$, and $\rho^{(\mathsf{MaxL})}=\log_2{[2(1-p)]}$ for $0\leq p\leq\frac{1}{2}$. From this derivation, it follows that the $(2,2)$ Scheme~A achieves perfect privacy for $p=\frac{1}{2}$. The IR rate of the $(2,2)$ Scheme~A, $\const{R}(\collect{C}_{\mathsf{A}})$, is depicted in Fig.~\ref{fig:WPIRn2_M2} as a function of the information leakage $\rho^{(\cdot)}$. Interestingly, by sacrificing perfect privacy, it is possible to achieve an IR rate larger than the $2$-server PIR capacity for $2$ files. As expected, the IR rate increases with increasing information leakage.
  % Moreover, concerning upload cost and access complexity, we have $\const{U}=1+\Hb(p)$ and $\Delta=1+2p$ for this $(2,2)$ IR scheme.
  \begin{figure}[t!]
    \centering
    % \documentclass[conference, 9pt]{IEEEtran} 
% \pagestyle{empty}	
% %\documentclass[
% %%    draft,
% %	%handout,
% %	%dvips,
% %	%xcolor=dvipsnames,
% %	9pt,
% %	mathserif
% %]{beamer}

% \usepackage{epsfig,amsfonts,amsbsy,bm,mathrsfs}
% \usepackage[nolist]{acronym}
% \usepackage{mathrsfs} 
% \usepackage{graphicx,cite,amssymb,amsmath,bm}
% \usepackage{mathtools}
% \usepackage{dsfont}
% \mathtoolsset{showonlyrefs}
% \usepackage{amsmath} 
% \usepackage[utf8]{inputenc}
% %\usepackage{subcaption}
% \usepackage{tikz}
% \usetikzlibrary{fadings}
% \usetikzlibrary{shadows.blur}
% \usetikzlibrary{shapes,arrows}
% \usetikzlibrary{calc,shapes.misc}
% \usetikzlibrary{decorations.pathreplacing}
% \usepackage{ifthen}
% \usetikzlibrary{positioning}
% \usepackage{pgfplots,relsize}
% \usetikzlibrary{plotmarks}
% \usepackage{ctable}
% \usepackage{pgfplots}
% \usepackage[english]{babel}
% \usepackage{xcolor}
% \usepackage{color}
% \usepackage[utf8]{inputenc}
% \usepackage[T1]{fontenc}
% \usetikzlibrary{decorations.pathreplacing,shapes.misc}
% \usetikzlibrary{patterns}
% \usepackage{colortbl}
% \usepackage{balance}

% \newcommand{\const}[1]{\textnormal{\usefont{U}{eur}{m}{n}\selectfont #1}} % Euler

% \begin{document}
%

%\begin{frame}[label=final]
%
%\begin{figure}

%\resizebox{0.8\paperwidth}{!}{%
\begin{tikzpicture}%[thick, scale=0.9, every node/.style={transform shape}]
\pgfplotsset{every tick label/.append style={font=\small}}
  
\begin{axis}[%
width=0.5\textwidth,
height=0.27\textheight,
at={(1.387in,0.821in)},
xmin=0,
xmax=1, % xmax=180,
xlabel={$\rho^{(\cdot)}$ (in bits)},
xlabel style={
	yshift=0.5ex,
	name=label},
grid style={gray,opacity=0.5,dotted},
xmajorgrids,
ymajorgrids,
% yminorgrids,
% ymode=log,
ymin=0.65, % ymin=0.0000001,
max space between ticks=20pt,
ymax=1,
ylabel={$\const{R}(\mathscr{C}_{\mathsf{A}})$},
ylabel style={
	yshift=-1.0ex,
	name=label},
axis background/.style={fill=none},
legend cell align=left,
legend style={legend style={draw=none,fill=none}, font=\scriptsize, at={(axis cs: 1.0,0.65)}, anchor=south east},
]

\addplot [color=red, solid,line width=1pt, mark=-*, mark options={solid, line width = 0.5pt, fill=white}]table[x=avgMI_WPIR,y=R_WPIR] {\Figs/data/WPIRlkg_n2f2_v4.txt}; 
\addlegendentry{$\rho^{(\textnormal{MI})}$};

\addplot [color=red, densely dotted,line width=1pt, mark=-*, mark options={solid, line width = 0.5pt, fill=white}]table[x=maxH_WPIR,y=R_WPIR] {\Figs/data/WPIRlkg_n2f2_v4.txt}; 
\addlegendentry{$\rho^{(\textnormal{WIL})}$};

\addplot [color=red,densely dashdotted,line width=1pt, mark=-*, mark options={solid, line width = 0.5pt, fill=white}]table[x=maxL_WPIR,y=R_WPIR] {\Figs/data/WPIRlkg_n2f2_v4.txt}; 
\addlegendentry{$\rho^{(\textnormal{MaxL})}$};

\addplot [only marks, color=red, mark=triangle*, mark options={solid, mark size=3pt}]table[x=rho,y=PIRC] {\Figs/data/PIRCn2f2.txt}; 
\addlegendentry{\scriptsize{$\const{C}_{2,2}$}};

\end{axis}
\end{tikzpicture}%
%}

%\end{figure}
%%
%\end{frame}

% \end{document}
    \vspace{-1ex}
    \caption{The IR rate $\const{R}(\collect{C}_{\mathsf{A}})\in\bigl[\frac{2}{3},1\bigr]$ of the proposed $(2,2)$ Scheme~A, as a function of $\rho^{(\cdot)}$. The triangle marks the $2$-server PIR capacity for $\const{M}=2$.}
    \label{fig:WPIRn2_M2}
    \vspace{-2ex}
  \end{figure}
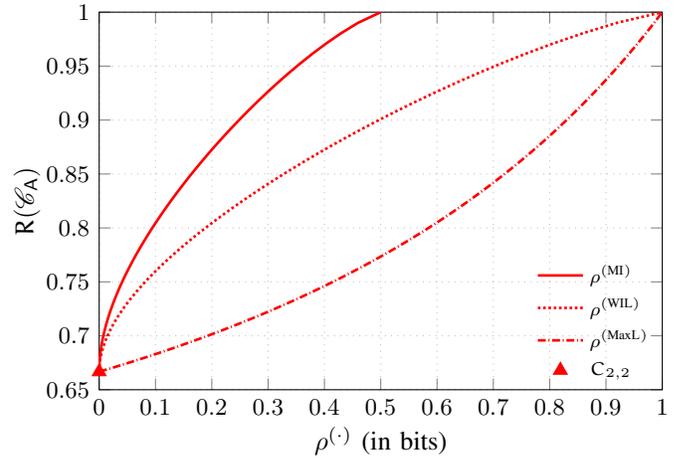  
\end{example}

% \subsection{Scheme~A}
% \label{sec:scheme1}

% We will see that Scheme~A is strongly dependent on the number of files $\const{M}$ and the number of servers $n$.

Now, we describe Scheme~A for the general case of $\const{M}$ files and $n$ servers. We assume that the user wants to download file $\vect{X}^{(m)}$ and has a random strategy $\vect{S}$ that takes on values $\vect{s}\in [0:n-1]^{\const{M}-1}$ with PMF $P_{\vect{S}}(\vect{s})$.

\subsubsection{Query Generation}
\label{sec:query-generation_Scheme1}

The query $\vect{q}_l\in\set{Q}_{l}$, $l\in[n]$, sent to the $l\textnormal{-th}$ server, resulting from the query-encoding function $\phi_l$, is defined as
\begin{IEEEeqnarray}{c}
  \vect{q}_l=\bigl(s_1,\ldots,s_{m-1},q_{l,m},s_{m},\ldots,s_{\const{M}-1}\bigr),\label{eq:Ql_Scheme1}
\end{IEEEeqnarray}
where $q_{l,m}\eqdef\bigl(l-1-\sum_{j\in[\const{M}-1]}s_j\bigr)\bmod n$. It follows that
\begin{IEEEeqnarray}{c}
  \set{Q}_l=\left\{\vect{q}_l\colon\Biggl(\sum_{m'\in[\const{M}]}q_{l,m'}\Biggr)\bmod n=l-1\right\}.\label{eq:query-set_Scheme1}
\end{IEEEeqnarray}
Note that the PMF of $\vect{Q}_l$ conditioned on the file index $M$ satisfies $P_{\bm Q^{(m)}_l}(\bm q_l)=P_{\vect{S}}(\vect{s})$.

\subsubsection{Answer Construction}
\label{sec:answer-construction_Scheme1}

The answer function $\varphi_l$ maps the query $\vect{q}_l$ into
\begin{IEEEeqnarray}{c}
  \vect{A}_l=\varphi_l(\vect{q}_l,\vect{X}^{[\const{M}]})=X^{(1)}_{q_{l,1}}+\cdots+X^{(\const{M})}_{q_{l,\const{M}}},\IEEEeqnarraynumspace\label{eq:Al_Scheme1}
\end{IEEEeqnarray}
where $X^{(m')}_0 = 0$ for all $m' \in [\const{M}]$. Further, we see that the answer-length functions satisfy
\begin{IEEEeqnarray}{c}
  \ell_l(\vect{Q}_l)=
  \begin{cases}
    0 & \textnormal{if } \vect{q}_l=\vect{0},
    \\
    1 & \textnormal{otherwise}.
  \end{cases}
  \label{eq:ell-l_Scheme1}
\end{IEEEeqnarray}

This completes the construction of the $(\const{M},n)$ Scheme~A. Note that it follows from \eqref{eq:Al_Scheme1} that $\set{A}=\{0,1\}=\set{X}$. % The perfect retrievability of Scheme~A can be verified by following the same argument as in \cite[Sec.~III-B]{TianSunChen19_1}.
% , the perfect retrievability can be verified without concerning $P_{\vect{Q}_l|M}$.
Moreover, using \eqref{eq:ell-l_Scheme1}, the IR rate of the $(\const{M},n)$ Scheme~A, $\collect{C}_{\mathsf{A}}$, can be shown to be
\begin{IEEEeqnarray}{c}
  \const{R}(\collect{C}_{\mathsf{A}})% & = &\frac{\beta}{\sum_{l=1}^n \E[\vect{Q}_l]{\ell_l(\vect{Q}_l)}}
  =\frac{n-1}{1-P_{\vect{Q}_1}(\vect{0})+n-1}.
  \IEEEeqnarraynumspace\label{eq:IRrate_Scheme1}
\end{IEEEeqnarray}
We also remark that if Scheme~A uses a random strategy $\vect{S}$ with $\{S_j\}_{j=1}^{\const{M}-1}$ i.i.d.\ according
to $\Uniform{[0:n-1]}$, then it satisfies \eqref{eq:strong-privacy} and is equivalent to the PIR capacity-achieving
scheme proposed in \cite{TianSunChen19_1}.

\subsubsection{Perfect Retrievability}
\label{sec:perfect-retrievability}

For completeness, in the following we show that Scheme~A satisfies the recovery condition in~\eqref{eq:retrievability}. % in the following lemma.
% \begin{lemma}
%   \label{lem:IR_Scheme1}
  % Consider a DSS with $n$ noncolluding servers storing $\const{M}$ files. 
  % %Consider an $n$-server uncoded DSS that stores $\const{M}$ files.
  % Then, 
  From the queries $\vect{Q}_l$ and answers $\vect{A}_l$, $l\in [n]$, designed as in~\eqref{eq:Ql_Scheme1} and~\eqref{eq:Al_Scheme1}, respectively, 
% \end{lemma}
% \begin{IEEEproof}
  given $\vect{S}=\vect{s}=(s_1,\ldots,s_{\const{M}-1})\in[0:n-1]^{\const{M}-1}$ and $M=m$, the answer from the $l$-th server can be re-written as
  \begin{IEEEeqnarray*}{rCl}
    \vect{A}_l& = &X^{(m)}_{q_{l,m}}+\left(\sum_{m'=1}^{m-1}X^{(m')}_{s_{m'}}
      +\sum_{m'=m+1}^{\const{M}}X^{(m')}_{s_{m'-1}}\right)
    \\[1mm]
    & \eqdef &X^{(m)}_{q_{l,m}}+Z,\quad l\in[n].
  \end{IEEEeqnarray*}
  Since by definition $q_{l,m}=\bigl(l-1-\sum_{j=1}^{\const{M}-1}s_j\bigr)\bmod n$, $q_{l,m}$ must range thoroughly from $0$ to $n-1$, and for $l'-1=\bigl(\sum_{j=1}^{\const{M}-1}s_j\bigr)\bmod n$, we have $\vect{A}_{l'}=X^{(m)}_{q_{l',m}}+Z=0+Z$. Thus, the user can obtain $X^{(m)}_{q_{l,m}}=\vect{A}_l-Z$, and hence retrieve $\{X^{(m)}_{1},\ldots,X^{(m)}_{n-1}\}$.
% \end{IEEEproof}

The following lemma follows immediately from the construction of Scheme~A.
\begin{lemma}
  \label{lem:PIR_Scheme1}
  Let $\{S_j\}_{j=1}^{\const{M}-1}$ be i.i.d.\ and $S_j\sim\Uniform{[0:n-1]}$ for Scheme~A. Then, it satisfies \eqref{eq:strong-privacy} and is equivalent to the PIR capacity-achieving scheme proposed in \cite{TianSunChen19_1}.
\end{lemma}
\begin{IEEEproof}
 See Appendix~\ref{sec:proof_PIR_Scheme1}.
\end{IEEEproof}

% \subsection{Analysis of the $(\const{M},2)$ Scheme~A that Uses $\{S_j\}_{j=1}^{\const{M}-1}$ being
% i.i.d.~$\sim\Bernoulli{p}$}
  
\subsection{$(\const{M},2)$ Scheme~A With $\{S_j\}_{j=1}^{\const{M}-1}$  i.i.d.\ According to $\Bernoulli{p}$}
\label{sec:achievability_Mn2-Scheme1-IID-Bernoulli}

The following result gives an achievable $4$-tuple for Scheme~A for the case of two servers and a random strategy
$\vect{S}=(S_1,\ldots,S_{\const{M}-1})$ with i.i.d.~entries according to $\Bernoulli{p}$.
\begin{theorem}
  \label{thm:Scheme1_Mn2-IID-Bernoulli} 
  Consider $0\leq p\leq\nicefrac{1}{2}$. Then, the $4$-tuple $(\const{R}_{\mathsf{A}},\const{U}_{\mathsf{A}},\Delta_{\mathsf{A}},\rho^{(\cdot)}_{\mathsf{A}}\bigr)$,
  \begin{IEEEeqnarray*}{rCl}
    \const{R}_{\mathsf{A}}& = &\inv{\bigl(1-(1-p)^{\const{M}-1}+1\bigr)},
    \\
    \const{U}_{\mathsf{A}}& = &-\sum_{w=0}^\const{M}{\const{M}\choose w} f(w,p)\log_2 f(w,p),
    \\
    \Delta_{\mathsf{A}}& = &\sum_{w=0}^\const{M}w\binom{\const{M}}{w} f(w,p),
    \\
    \rho^{(\mathsf{MI})}_{\mathsf{A}}& = &\nicefrac{\const{U}_{\mathsf{A}}}{2}-(\const{M}-1)\Hb(p),
    \\
    \rho^{(\mathsf{WIL})}_{\mathsf{A}}& = &\log_2{\const{M}}-\min\nolimits_{w\in [0:\const{M}]}\eHP{M_w}, \textnormal{ and}
    \\
    \rho^{(\mathsf{MaxL})}_{\mathsf{A}}& = &\log_2{\sum_{\substack{w\in [\const{M}]\\ w\colon\textnormal{odd}}}\binom{\const{M}}{w}(1-p)^{\const{M}-w}p^{w-1}}
  \end{IEEEeqnarray*}  
  is achievable by the $(\const{M},2)$ Scheme~A with $\{S_j\}_{j=1}^{\const{M}-1}$ i.i.d.\ according to $\Bernoulli{p}$, where 
  \begin{equation*}
    \scalemath{0.95}{f(w,p)\eqdef\frac{1}{\const{M}}\left((\const{M}-w)(1-p)^{\const{M}-w-1}p^w+ w(1-p)^{\const{M}-w}p^{w-1}\right)}
  \end{equation*} 
  and $M_w$ is a RV with PMF
  \begin{IEEEeqnarray}{c}
    P_{M_w}(m')=
    \begin{cases}
      \frac{(1-p)^{\const{M}-w-1}p^{w}}{\const{M}f(w,p)} & \textnormal{if }m'\in[\const{M}-w],
      \\[2mm]
      \frac{(1-p)^{\const{M}-w}p^{w-1}}{\const{M}f(w,p)} & \textnormal{if }m'\in[\const{M}-w+1:\const{M}].
    \end{cases}
    \nonumber\\*\label{eq:Mw-PMF_Scheme1}
  \end{IEEEeqnarray}
\end{theorem}

\begin{IEEEproof}
  See Appendix~\ref{sec:proof_Mn2-Scheme1-IID-Bernoulli}.
\end{IEEEproof}

\subsection{Partition Scheme~A: Using Scheme~A as a Subscheme}
\label{sec:partition-SchemeA}

In Section~\ref{sec:partition-schemes}, we introduced the concept of adopting an existing $(\nicefrac{\const{M}}{\eta},n)$ IR scheme to retrieve a file from a given partition. In this subsection, unlike \eqref{eq:queries_partition-scheme}, where the user sends different queries for different requested files among all partitions, we use a slightly more sophisticated way to construct a WPIR scheme by using Scheme~A as a subscheme for every partition. We refer to this scheme as \emph{partition Scheme~A} and denote it by $\collect{C}_{\mathsf{A}}^\mathsf{part}$. In the following, we present the query generation and the answer construction.

\subsubsection{Query Generation}
\label{sec:query-generation_SchemeA}

We consider the $j$-th partition, $\set{P}_j$, $j\in[\eta]$, containing all files of indices $(j-1)\nicefrac{\const{M}}{\eta}+1,\ldots,j\nicefrac{\const{M}}{\eta}$. Given a requested file with index $m=(j-1)\nicefrac{\const{M}}{\eta}+m'\in\set{P}_j$, $m'\in[\nicefrac{\const{M}}{\eta}]$, we consider an $(\nicefrac{\const{M}}{\eta},n)$ Scheme~A as a subscheme for partition $\set{P}_j$. The $l$-th query $\vect{q}_l\in\set{Q}_l$, $l\in[n]$, is defined as
\begin{IEEEeqnarray*}{rCl}
  \vect{q}_l& = &\bigl(\vect{0}_{1\times
      (j-1)\nicefrac{\const{M}}{\eta}},s_{1},\ldots,s_{m'-1},q_{l,(j-1)\nicefrac{\const{M}}{\eta}+m'},
  \nonumber\\
  && \quad\>s_{m'},\ldots,s_{\nicefrac{\const{M}}{\eta}-1},\vect{0}_{1\times(\eta-j)\nicefrac{\const{M}}{\eta}}\bigr),
\end{IEEEeqnarray*}
where $q_{l,(j-1)\nicefrac{\const{M}}{\eta}+m'}=\bigl(l-1-\sum_{j\in[\nicefrac{\const{M}}{\eta}-1]} s_j \bigr)\bmod n$. We remark that it is possible that the user sends the all-zero query $\vect{q}_l=\vect{0}$ to request different files among all partitions. In this way, since the uncertainty on the requested file is increased, it follows that the leakage of $\collect{C}_{\mathsf{A}}^\mathsf{part}$ is slightly smaller than the leakage of the basic scheme. Moreover, the query alphabet size is not exactly the same for all servers. In particular, since for every partition, an $(\nicefrac{\const{M}}{\eta},n)$ Scheme~A consists of $n^{\nicefrac{\const{M}}{\eta}-1}$ queries at each server, and only for the first server the all-zero query $\vect{0}_{1\times\const{M}}$ is sent to retrieve any one of the $\const{M}$ files, we have $\card{\set{Q}_1}=1+\eta\bigl(n^{\nicefrac{\const{M}}{\eta}-1}-1\bigr)$ and $\card{\set{Q}_l}=\eta\cdot n^{\nicefrac{\const{M}}{\eta}-1}$ for $l\in [2:n]$.

\subsubsection{Answer Construction}
\label{sec:answer-construction_SchemeA}

Similar to Scheme~A, the answer function $\varphi_l$  maps  query
$\vect{q}_l$ into $\vect{A}_l=\varphi_l(\vect{q}_l,\vect{X}^{[\const{M}]})=X^{(1)}_{q_{l,1}}+\cdots+X^{(\const{M})}_{q_{l,\const{M}}}$, %where
where $X^{(m')}_0 = 0$ for all $m' \in [\const{M}]$. 
%$X^{(m)}_{0}=\vect{0}$ is simply a dummy variable prepended to the $m$-th file beforehand. 
Further, we see that $\ell_l(\vect{Q}_l)$ satisfies~\eqref{eq:ell-l_Scheme1}.
% \begin{IEEEeqnarray*}{c}
%   \ell_l(\vect{q}_l)=
%   \begin{cases}
%     0 & \textnormal{if } \vect{q}_l=\vect{0},
%     \\
%     1 & \textnormal{otherwise}.
%   \end{cases}
%   \label{eq:\ell_l_SchemeA}
% \end{IEEEeqnarray*}
% This completes the construction of the $(\const{M},n)$ partition Scheme~A.

% Next, we derive the achievable $4$-tuples of $\collect{C}_{\mathsf{A}}^\mathsf{part}$ by using Scheme~A with
% $\{S_j\}_{j=1}^{\nicefrac{\const{M}}{\eta}-1}$ being i.i.d.~$\sim\Bernoulli{\nicefrac{1}{2}}$ as a sub-scheme for the case of $n=2$ servers.

\subsubsection{$(\const{M},n)$ Partition Scheme~A With $\{S_j\}_{j=1}^{\nicefrac{\const{M}}{\eta}-1}$ i.i.d.\ According to
  $\Uniform{[0:n-1]}$}
\label{sec:achievability_Mn-SchemeA-IID-Uniform}

We focus on a particular $(\const{M},n)$ partition Scheme~A. Since the servers can learn some information from which partition the requested file belongs to, in order to have a relatively small leakage of partition Scheme~A, it is reasonable to use Scheme~A with $\{S_j\}_{j=1}^{\nicefrac{\const{M}}{\eta}-1}$ i.i.d.\ according to $\Uniform{[0:n-1]}$ as a subscheme (i.e., a PIR subscheme, cf.~Lemma~\ref{lem:PIR_Scheme1}). Thus, this scheme works for an arbitrary number of servers $n$. We have the following result.
\begin{theorem}
  \label{thm:SchemeA_Mn-IID-Uniform}
  Let $\nicefrac{\const{M}}{\eta}$ be a positive integer with $\eta\in [\const{M}-1]$. Then, the $4$-tuple
  $\bigl(\const{R}_{\mathsf{A},\mathsf{P}},\const{U}_{\mathsf{A},\mathsf{P}},\Delta_{\mathsf{A},\mathsf{P}},\rho^{(\cdot)}_{\mathsf{A},\mathsf{P}}\bigr)$,
  \begin{IEEEeqnarray*}{rCl}
    \const{R}_{\mathsf{A},\mathsf{P}}& = &\inv{\left(1+\frac{1}{n}+\cdots+\frac{1}{n^{\nicefrac{\const{M}}{\eta}-1}}\right)},
    \\
    \const{U}_{\mathsf{A},\mathsf{P}}& = &n\bigl[(\nicefrac{\const{M}}{\eta}-1)\log_2{n}+\log_2{\eta}\bigr]-\frac{\log_2{\eta}}{{n^{\nicefrac{\const{M}}{\eta}-1}}},
    \\
    \Delta_{\mathsf{A},\mathsf{P}}& = &(n-1)\nicefrac{\const{M}}{\eta},
    \\[1mm]
    \rho^{(\mathsf{MI})}_{\mathsf{A},\mathsf{P}}& = &\log_2{\eta}-\frac{\log_2{\eta}}{n^{\nicefrac{\const{M}}{\eta}}}, \textnormal{ and}
    \\    
    \rho^{(\mathsf{WIL})}_{\mathsf{A},\mathsf{P}}& = &\log_2{\eta}=\rho^{(\mathsf{MaxL})}_{\mathsf{A},\mathsf{P}}
  \end{IEEEeqnarray*}
  is achievable by the $(\const{M},n)$ partition Scheme~A using the $(\nicefrac{\const{M}}{\eta},n)$ Scheme~A with $\{S_j\}_{j=1}^{\nicefrac{\const{M}}{\eta}-1}$ i.i.d.\ according to $\Uniform{[0:n-1]}$ as a subscheme.
\end{theorem}
\begin{IEEEproof}
  See Appendix~\ref{sec:proof_Mn-SchemeA-IID-Uniform}.
\end{IEEEproof}

Let $\bigl(\tilde{\const{R}},\tilde{\const{U}},\tilde{\Delta},0\bigr)$ be the achievable $4$-tuple of the $(\nicefrac{\const{M}}{\eta},2)$ Scheme~A with $\{S_j\}_{j=1}^{\nicefrac{\const{M}}{\eta}-1}$ i.i.d.\ according to $\Uniform{[0:n-1]}$. It follows that $\const{U}_{\mathsf{A},\mathsf{P}}=\tilde{\const{U}}+2\log_2{\eta}-\nicefrac{\log_2{\eta}}{n^{\nicefrac{\const{M}}{\eta}-1}}<\const{U}(\collect{C}^{\mathsf{basic}})$ and $\rho^{(\mathsf{MI})}_{\mathsf{A},\mathsf{P}}=\log_2{\eta}-\nicefrac{\log_2{\eta}}{n^{\nicefrac{\const{M}}{\eta}}}<\rho^{(\mathsf{MI})}(\collect{C}^{\mathsf{basic}})$, while $\const{R}_{\mathsf{A},\mathsf{P}}$, $\Delta_{\mathsf{A},\mathsf{P}}$, $\rho^{(\mathsf{WIL})}_{\mathsf{A},\mathsf{P}}$, and $\rho^{(\mathsf{MaxL})}_{\mathsf{A},\mathsf{P}}$ are identical to those of the basic scheme $\collect{C}^{\mathsf{basic}}$ in Section~\ref{sec:partition-schemes} (see the details in Appendix~\ref{sec:proof_Mn-SchemeA-IID-Uniform}).
% Since the performance of $\collect{C}_{\mathsf{A}}^\mathsf{part}$ in Thm.~\ref{thm:SchemeA_Mn-IID-Uniform} is better than that of $\collect{C}^{\mathsf{basic}}$ in Section~\ref{sec:partition-schemes}. when $(\nicefrac{\const{M}}{\eta},2)$ Scheme~A with $\{S_j\}_{j=1}^{\nicefrac{\const{M}}{\eta}-1}$ i.i.d. $\Bernoulli{\nicefrac{1}{2}}$ is used as a subscheme
Hence, in the numerical results section, the results of $\collect{C}^{\mathsf{basic}}$ are not presented.

\section{Constant-Rate $(\const{M},n)$ Scheme~B}
\label{sec:Scheme2}

We propose an alternative WPIR scheme, referred to as \emph{Scheme~B} and denoted by $\collect{C}_{\mathsf{B}}$, based on the PIR scheme in \cite[Lem.~4]{KumarLinRosnesGraellAmat19_1}. Scheme~B is constructed as follows. Assume that $\beta=n-1$ and that the user requests file $\vect{X}^{(m)}$. The random strategy $\vect{S}$ takes the form of a vector $\vect{S}=(S_1,\ldots,S_{\beta\const{M}})\in\set{X}^{\beta\const{M}}$ of length $\beta\const{M}$. The query vector $\vect{Q}_l\in\set{Q}_l=\set{X}^{\beta\const{M}}$, of length $\beta\const{M}$, is obtained as
\begin{IEEEeqnarray*}{c}
  \vect{Q}_l=\phi(m,\bm S)=\bm S+\bm v_l^{(m)},
\end{IEEEeqnarray*}%
% $\vect{Q}_l=\phi(m,\bm S)=\bm S+\bm v_l$,
where the vector $\vect{v}_l^{(m)}=(v^{(m)}_{l,1},\ldots,v^{(m)}_{l,\beta \const{M}})$ is deterministic and is completely determined by $m\in[\const{M}]$. We refer the reader to~\cite[Sec.~V]{KumarLinRosnesGraellAmat19_1} for details on the design of $\vect{v}_l^{(m)}$. Briefly, $\vect{v}_l^{(m)}$ is a binary vector, where $v_{l,i}^{(m)}=1$ denotes that the $i$-th symbol is being retrieved from the $l$-th server. The $l$-th vector has the following structure,
\begin{IEEEeqnarray*}{c}
  \vect{v}_l^{(m)}=(\vect{0}_{1\times (m-1)\beta}\mid\vect{\Delta}_l\mid\vect{0}_{1\times(\const{M}-m)\beta}),
\end{IEEEeqnarray*}
where $\vect{\Delta}_l$, $l\in[n-1]$, is the $l$-th $\beta$-dimensional unit vector, and $\vect{\Delta}_n=\vect{0}_{1 \times \beta}$. The $l$-th server responds to its corresponding query with the answer $\bm A_l\in\set{A}=\set{X}$ obtained as $\bm A_l=\varphi_l(\vect{Q}_l,\vect{X}^{[\const{M}]})\eqdef\langle\vect{Q}_l,(X^{(1)}_1,\ldots,X^{(1)}_\beta,X^{(2)}_1,\ldots,X^{(\const{M})}_\beta)\rangle$.

For the case where $\{S_j\}_{j=1}^{\beta \const{M}}$ are i.i.d.\ according to $\Uniform{\set{X}}$, Scheme~B achieves perfect privacy, and the scheme reduces to the PIR scheme in \cite[Lem.~4]{KumarLinRosnesGraellAmat19_1}. Furthermore, similar to \cite[Thm.~2]{KumarLinRosnesGraellAmat19_1}, it can be shown that the scheme achieves perfect retrievability (see \eqref{eq:retrievability}), and since its answer-lengths are constant for all possible queries of each server, the IR rate $\const{R}_{\mathsf{B}}$ of $\collect{C}_{\mathsf{B}}$ is equal to $1-\nicefrac{1}{n}$, irrespective of the information leakage $\rho^{(\cdot)}$.

We remark that for the case of multiple replicated noncolluding servers (uncoded servers), the PIR scheme in \cite[Lem.~4]{KumarLinRosnesGraellAmat19_1} is equivalent to the original PIR protocol proposed in \cite{ChorGoldreichKushilevitzSudan95_1,ChorGoldreichKushilevitzSudan98_1}. The designed queries and answers of Scheme~B result in a constant rate, while the rate of Scheme~A is dependent on the query distribution (see \eqref{eq:IRrate_Scheme1}).

In the following subsections, we consider the binary field $\set{X}=\{0,1\}$.

\subsection{$(\const{M},2)$ Scheme~B  With $\{S_j\}_{j=1}^{\const{M}}$ i.i.d.\ According to $\Bernoulli{p}$}
\label{sec:achievability_Mn2-Scheme2-IID-Bernoulli}

We have the following result.
\begin{theorem}
  \label{thm:Scheme2_n2infinity}
  Consider $0\leq p\leq\nicefrac{1}{2}$.  Then, the $4$-tuple $\bigl(\nicefrac{1}{2},\const{U}_{\mathsf{B}},\Delta_{\mathsf{B}},\rho^{(\cdot)}_{\mathsf{B}}\bigr)$,
  \begin{IEEEeqnarray*}{rCl}
    \const{U}_{\mathsf{B}}& = &-\sum_{w=0}^{\const{M}}\binom{\const{M}}{w}g(w,p)\log_2 g(w,p)+\const{M}\Hb(p),
    \\
    \Delta_{\mathsf{B}}& = &\sum_{w=0}^{\const{M}}w\binom{\const{M}}{w}\bigl(g(w,p)+h(w,p)\bigr),
    \\
    \rho_{\mathsf{B}}^{\mathsf{(MI)}}& = &\nicefrac{\const{U}_{\mathsf{B}}}{2}-\const{M}\Hb(p), 
    \\
    \rho_{\mathsf{B}}^{\mathsf{(WIL)}}& = &\log_2{\const{M}}-\min_{w\in[0:\const{M}]}\eHP{M_w'}, \textnormal{ and}
    \\
    \rho_{\mathsf{B}}^{\mathsf{(MaxL)}}& = &\log_2\biggl((1-p)^{\const{M}-1}p\nonumber\\
    &&\qquad\quad +\>\sum_{w\in [1:\const{M}]}\binom{\const{M}}{w}(1-p)^{\const{M}-(w-1)}p^{w-1}\biggr)
    \IEEEeqnarraynumspace
  \end{IEEEeqnarray*}
  is achievable by the $(\const{M},2)$ Scheme~B with $\{S_j\}_{j=1}^{\const{M}}$ i.i.d.\ according to $\Bernoulli{p}$, where $g(w,p)\eqdef\nicefrac{\bigl[(\const{M}-w)(1-p)^{\const{M}-w-1}p^{w+1}+w(1-p)^{\const{M}-w+1}p^{w-1}\bigr]}{\const{M}}$, $h(w,p)\eqdef(1-p)^{\const{M}-w}p^{w}$, and $M_w'$ is a RV with PMF
  \begin{IEEEeqnarray*}{c}
    P_{M_w'}(m')=
    \begin{cases}
      \frac{(1-p)^{\const{M}-w-1}p^{w+1}}{\const{M}g(w,p)} & \textnormal{if } m'\in[\const{M}-w],
      \\[1mm]
      \frac{(1-p)^{\const{M}-w+1}p^{w-1}}{\const{M}g(w,p)} & \textnormal{if } m'\in[\const{M}-w+1:\const{M}].
    \end{cases}
    \IEEEeqnarraynumspace
  \end{IEEEeqnarray*}
\end{theorem}

\begin{IEEEproof}
The proof is similar to the proof of \cref{thm:Scheme1_Mn2-IID-Bernoulli}, and is omitted for brevity.
\end{IEEEproof}

In the following subsection, we analyze the $(\const{M},2)$ Scheme~B with a uniformly-distributed random strategy $\vect{S}$. Note that similarly to partition Scheme~A in Section~\ref{sec:achievability_Mn-SchemeA-IID-Uniform}, we can also construct a partition scheme by using Scheme~B as a subscheme for every partition. We omit the analysis since it is almost the same as for partition Scheme~A, and the result for the $(\const{M},n)$ partition Scheme~B with $\{S_j\}_{j=1}^{\nicefrac{\const{M}}{\eta}}$ i.i.d.\ according to $\Uniform{\set{X}}$ is very close to the result in Theorem~\ref{thm:SchemeA_Mn-IID-Uniform}.

% avoid \bm{S} issue for section title
% \subsection[\texorpdfstring{Math symbols $S$}{Math symbols S}]{$(\const{M},2)$ Scheme~B With $\vect{S}$ Uniformly
% Distributed}
\subsection{$(\const{M},2)$ Scheme~B With $\vect{S}$ Uniformly
  Distributed}
\label{sec:achievability_Mn2-Scheme2-UniformSphere}

We consider the $(\const{M},2)$ Scheme~B with $\vect{S}$ uniformly distributed over all length-$\const{M}$ binary
vectors of weight $w$. In other words, $\vect{S}\sim\Uniform{\set{B}_{w,\const{M}}}$, where $\set{B}_{w,\const{M}}\eqdef\bigl\{\vect{s}\in\{0,1\}^\const{M}\colon \Hwt{\vect{s}}=w\bigr\}$.
\begin{theorem}
  \label{thm:Scheme2_Mn2-UniformSphere}
  Given any $w\in[0:\const{M}]$. Then, the $4$-tuple $\bigl(\nicefrac{1}{2},\const{U}_{\mathsf{B},\mathsf{U}},\Delta_{\mathsf{B},\mathsf{U}},\rho^{(\cdot)}_{\mathsf{B},\mathsf{U}}\bigr)$,
  \begin{IEEEeqnarray*}{rCl}
    \const{U}_{\mathsf{B},\mathsf{U}}& = &\log_2\binom{\const{M}}{w}+y(w,\const{M}),
    \\
    \Delta_{\mathsf{B},\mathsf{U}}& = &1+2w(1-\nicefrac{1}{\const{M}}),
    \\[1mm]
    \rho^{(\mathsf{MI})}_{\mathsf{B},\mathsf{U}}& = &\frac{y(w,\const{M})-\log_2{\binom{\const{M}}{w}}}{2},
    \\
    \rho^{(\mathsf{WIL})}_{\mathsf{B},\mathsf{U}}& = &\log_2{\const{M}}\nonumber\\
    && -\>\min\{\log_2{(w+1)},\log_2{(\const{M}-w+1)}\},  \textnormal{ and}
    \\
    \rho^{(\mathsf{MaxL})}_{\mathsf{B},\mathsf{U}}& = &\log_2\biggl(\frac{\const{M}-w}{w+1}+\frac{w}{\const{M}-w+1}\biggr)\IEEEeqnarraynumspace 
  \end{IEEEeqnarray*}
  is achievable by the $(\const{M},2)$ Scheme~B with $\vect{S}\sim\Uniform{\set{B}_{w,\const{M}}}$, where  $y(w,\const{M})\eqdef\log_2{\binom{\const{M}}{w}}+\log_2{\const{M}}-\nicefrac{(\const{M}-w)\log_2{(w+1)}}{\const{M}}-\nicefrac{w\log_2{(\const{M}-w+1)}}{\const{M}}$.
\end{theorem}
\begin{IEEEproof}
  See Appendix~\ref{sec:proof_Mn2-Scheme2-UniformShpere}.
\end{IEEEproof}

We remark that the analysis of the $(\const{M},2)$ Scheme~A with $\vect{S}\sim\Uniform{\set{B}_{w,\const{M}-1}}$ can also be done by following the same approach as for Theorem~\ref{thm:Scheme2_Mn2-UniformSphere}.\footnote{The scheme is not equal to that of \cite{TianSunChen19_1} because of the difference in the vector space of the random strategy. The former involves all length-($\const{M}-1$) vectors of weight $w$, while the latter consists of all vectors of length $\const{M}-1$.} However, since the resulting performance is much worse than those of the aforementioned WPIR schemes for the case of $n=2$ servers, we omit the detailed analysis in this paper.

In the rest of the paper, except for the next~\cref{sec:Scheme1_epsP} and the numerical results in~\cref{sec:numerical-results}, we consider only the MI and MaxL privacy metrics as these are more commonly used \cite{IssaWagnerKamath20_1}.

\section{$\eps$-Privacy for Scheme~A}
\label{sec:Scheme1_epsP}

In this section, we focus on the leakage analysis of the $(\const{M},n)$ Scheme~A under the $\eps$-privacy metric. Since the WPIR rate of the $(\const{M},n)$ Scheme~A is equal to~\eqref{eq:IRrate_Scheme1}, irrespective of the used privacy metric, we only need to focus on the design of the conditional query distribution $P_{\vect{Q}_l^{(m)}}(\vect{q}_l)$ at each server.

To compare our results with the works in \cite{SamyTandonLazos19_1,SamyAttiaTandonLazos21_1}, we first summarize the achievable rate for a given leakage constraint $\rho\geq 0$ as follows.
\begin{lemma}[{\cite[Eq.~(26)]{SamyAttiaTandonLazos21_1}% \footnote{Here, we adapt~\cite[Thm.~A]{SamyAttiaTandonLazos20_2sub} for the asymmetric leaky PIR scenario without concerning the database privacy leakage.}
  }]

  Consider a DSS with $n$ noncolluding servers storing $\const{M}$ files. Then, given an $\eps$-privacy leakage constraint $\rho^{(\eps\mhyph\mathsf{P})}\leq\rho$ with $\rho\geq 0$, the rate 
  \begin{IEEEeqnarray}{c}
    \const{R}^{(\eps\mhyph\mathsf{P})}_{\mathsf{LPIR}}=\inv{\biggl(1+\frac{n^{\const{M}-1}-1}{(n-1)(\ee^{\rho}+n^{\const{M}-1}-1)}\biggr)}
    \label{eq:achievable-rate_LPIR}
  \end{IEEEeqnarray}
  is achievable. Moreover, the WPIR rate under the $\eps$-privacy metric is bounded from above by
  \begin{IEEEeqnarray}{c}
    \const{R}^{(\eps\mhyph\mathsf{P})}(\collect{C})\leq\frac{1-\frac{1}{n \ee^{\rho}}}{1-\frac{1}{(n \ee^{\rho})^\const{M}}}\eqdef\const{R}^{(\eps\mhyph\mathsf{P})}_\mathsf{UB}.
    \label{eq:upper-bound_LPIR}
  \end{IEEEeqnarray}
\end{lemma}
In \cite{SamyTandonLazos19_1,SamyAttiaTandonLazos21_1}, the authors proposed the \emph{path-based approach} across databases to obtain the achievable download rate $\const{R}^{(\eps\mhyph\mathsf{P})}_{\mathsf{LPIR}}$ in \eqref{eq:achievable-rate_LPIR}. In the following subsection, we show that it is possible to achieve a better tradeoff between the $\eps$-privacy leakage and the download rate by using Scheme~A.

\subsection{$(\const{M},2)$ Scheme~A With $\{S_j\}_{j=1}^{\const{M}-1}$  i.i.d.\ According to a Bernoulli-Distributed RV}
\label{sec:achievability_Mn2-Scheme1-IID-Bernoulli_epsP}

\begin{theorem}
  \label{thm:Scheme1_Mn2-IID-Bernoulli_epsP} 
  Consider $0< p_\rho\eqdef\inv{(1+\ee^{\rho})}\leq\nicefrac{1}{2}$ for $\rho\geq 0$. Then, given an $\eps$-privacy leakage constraint $\rho^{(\eps\mhyph\mathsf{P})}\leq\rho$, the rate 
  \begin{IEEEeqnarray}{c}
    \const{R}^{(\eps\mhyph\mathsf{P})}_\mathsf{A}=\inv{\bigl(1-(1-p_\rho)^{\const{M}-1}+1\bigr)}
    \label{eq:WPIRrate_Scheme1_Mn2-IID-Bernoulli_epsP}
  \end{IEEEeqnarray}  
  is achievable by the $(\const{M},2)$ Scheme~A with $\{S_j\}_{j=1}^{\const{M}-1}$ i.i.d.\ according to $\Bernoulli{p_\rho}$.
\end{theorem}
\begin{IEEEproof}
  See Appendix~\ref{sec:proof_Mn2-Scheme1-IID-Bernoulli_epsP}.
\end{IEEEproof}

\section{Minimization of the Information Leakage for Scheme~A With Time-Sharing}
\label{sec:minimization-leakage_Scheme1-TimeSharing}

A main objective of this work is to determine the optimal WPIR scheme that leaks the smallest amount of information, subject to a given IR download cost, upload cost, or access complexity. Since both the information leakage and the IR rate can be improved based on the query generation of Scheme~A, our aim is to study the optimal tradeoff between the information leakage and the download cost for Scheme~A. In the rest of paper, we will mainly focus on the MI and MaxL privacy metrics. %the MI and MaxL. 

Following the notion of \cref{thm:time-sharing_MI} for the MI metric, we can use the time-sharing principle to construct a time-sharing scheme from Scheme~A. In particular, consider the Scheme~A $\collect{C}_\mathsf{A}$ with query-encoding functions $\phi_l$, answer functions $\varphi_l$, and a random strategy $\vect{S}$ presented in~\cref{sec:Scheme1}. We design the query-encoding functions $\bar{\phi}_l=\phi_{\sigma^{T-1}(l)}(M,\vect{S})$ and the answer functions $\bar{\varphi_l}=\varphi_{\sigma^{T-1}(l)}\bigl(\phi_{\sigma^{T-1}(l)}(M,\vect{S}),\vect{X}^{[\const{M}]}\bigr)$ of a given requested file index $M$ to construct the time-sharing scheme of Scheme~A, where $T\sim\Uniform{[n]}$. Such a scheme is referred to as time-sharing Scheme~A and denoted by $\widebar{\collect{C}}_\mathsf{A}$.\footnote{The time-sharing principle can be applied to any WPIR scheme. However, here we concentrate only on the time-sharing Scheme~A.} Recall that in Scheme~A the query realization $\vect{q}_l$ of the $l$-th server, $l\in [n]$, belongs to $\set{Q}_l$, defined in~\eqref{eq:query-set_Scheme1}, from which it follows that all of the query sets $\set{Q}_l$ are distinct. The conditional query PMF $P_{\vect{Q}_l^{(m)}}(\vect{q}_l)=P_{\vect{S}}(\vect{s})$ is independent of $M=m$, and the download cost of the $(\const{M},n)$ Scheme~A is $1-P_{\vect{S}}(\vect{0})+(n-1)$ (cf.~Section~\ref{sec:Scheme1}). Denote by $z_{\vect{s}}\eqdef P_{\vect{S}}(\vect{s})$ the PMF of the random strategy $\vect{S}$, and $\vect{q}_l\setminus\{m\}\eqdef(q_{l,1},\ldots,q_{l,m-1},q_{l,m+1},\ldots,q_{l,\const{M}})$, $m\in[\const{M}]$. By applying the time-sharing approach,  the resulting query set at the $l$-th server is $\widebar{\set{Q}}_l=[0:n-1]^{\const{M}}$, and the conditional PMF of $\widebar{\vect{Q}}_l$ given $M=m$ is
\begin{IEEEeqnarray}{rCl}
  P_{\widebar{\vect{Q}}_l^{(m)}}(\bar{\vect{q}}_l)& = &\sum_{t=1}^n\Prv{T=t}\Prvcond{\vect{Q}_{\sigma^{t-1}(l)}^{(m)}=\bar{\vect{q}}_l}{T=t}
  \nonumber\\
  & = &\frac{1}{n}z_{\vect{s}},\quad\textnormal{for }\vect{s}=\bar{\vect{q}}_l\setminus\{m\},\,\bar{\vect{q}}_l\in\widebar{\set{Q}}_l,\label{eq:use_distinct-QuerySets_Scheme1}
\end{IEEEeqnarray}
where \eqref{eq:use_distinct-QuerySets_Scheme1} follows since all query sets in Scheme~A are different. In other words, by using the time-sharing approach, we obtain a new scheme with a conditional query distribution at each server equal to
\begin{IEEEeqnarray}{c}
  \Biggl(\frac{P_{\vect{Q}^{(m)}_{\sigma^{0}(l)}}(\cdot)}{n},\cdots,\frac{P_{\vect{Q}^{(m)}_{\sigma^{n-1}(l)}}(\cdot)}{n}\Biggr)\label{eq:Ql-PMF_TimeSharingScheme1},
\end{IEEEeqnarray}
where $P_{\vect{Q}^{(m)}_l}(\cdot)=(p_1,\ldots,p_{\card{\set{Q}_l}})$ represents the conditional query distribution corresponding to $P_{\vect{Q}_l|M=m}$ for Scheme~A. Therefore, from \eqref{eq:use_distinct-QuerySets_Scheme1} or \eqref{eq:Ql-PMF_TimeSharingScheme1} it follows that every server has identical information leakage for the time-sharing Scheme~A under both the MI and MaxL privacy metrics. Note that the download cost stays the same as for Scheme~A.

For the time-sharing Scheme~A, the minimization of the information leakage $\rho^{(\cdot)}(\widebar{\collect{C}}_\mathsf{A})$ under a download cost constraint $\const{D}$ can be cast as the optimization problem
\begin{IEEEeqnarray}{rCl}
  \IEEEyesnumber\label{eq:optimization_leakage-download}
  \IEEEyessubnumber*
  \textnormal{minimize} & &\qquad\rho^{(\cdot)}(\widebar{\collect{C}}_{\mathsf{A}})
  \label{eq:objective-ft_leakage}\\
  \textnormal{subject to} & &\qquad 1-z_{\vect{0}}+(n-1) \leq\const{D},\label{eq:PMFs_download-constraint}
  \\[1mm]
  & &\sum_{\vect{s}\in [0:n-1]^{\const{M}-1}}z_{\vect{s}}=1.\label{eq:PMF_random-strategy}
\end{IEEEeqnarray}

\subsection{Optimizing the MI Leakage}
\label{sec:optimizing_MI-leakage}

In terms of the MI privacy metric, we know from~\cref{thm:time-sharing_MI} that the leakage of $\widebar{\collect{C}}_{\mathsf{A}}$ is equal to\footnote{Note that the MI leakage of Scheme~A and that of the corresponding time-sharing Scheme~A is always the same due to the definition of MI leakage in \eqref{eq:MI_leakge_def}.}
\begin{IEEEeqnarray}{rCl}
  \IEEEeqnarraymulticol{3}{l}{%
    \rho^{(\mathsf{MI})}(\widebar{\collect{C}}_{\mathsf{A}})=\eMI{M}{\widebar{\vect{Q}}_l}}\nonumber\\*%
  & = &\frac{1}{n}\sum_{l'\in[n]}\bigl[\HP{\vect{Q}_{l'}}-\eHPcond{\vect{Q}_{l'}}{M}\bigr]
  \nonumber\\
  & = &\frac{1}{n}\sum_{l'\in[n]} \eHP{P_{\vect{Q}_{l'}}}-\bigHP{(\vect{z})_{\vect{s}\in [0:n-1]^{\const{M}-1}}}
  \nonumber\\
  % & = &\eHP{P_{\bar{\vect{Q}}_1}}-\biggHP{\underbrace{\frac{(\vect{z})_{\vect{s}\in [0:n-1]^{\const{M}-1}}}{n},\cdots,\frac{(\vect{z})_{\vect{s}\in [0:n-1]^{\const{M}-1}}}{n}}_{n \textnormal{ times}}}
  % \nonumber\\
  & \stackrel{(a)}{=} &\frac{1}{n}\sum_{\bar{\vect{q}}_l\in[0:n-1]^{\const{M}}}% \scriptstyle
  \Biggl[\Biggl(-\frac{\sum_{m=1}^{\const{M}}P_{\vect{S}}(\bar{\vect{q}}_l\setminus\{m\})}{\const{M}}\Biggr)\nonumber\\
  &&\>\cdot\log_2{\left(\frac{\sum_{m=1}^{\const{M}}P_{\vect{S}}(\bar{\vect{q}}_l\setminus\{m\})}{\const{M}}\right)}\Biggr]% \nonumber\\
  % &&
  -\bigHP{(\vect{z})_{\vect{s}\in [0:n-1]^{\const{M}-1}}}\nonumber\\
  & = &\frac{1}{n}\sum_{\bar{\vect{q}}_1\in[0:n-1]^{\const{M}}}\Biggl(-\frac{\sum\limits_{m=1}^{\const{M}}z_{\bar{\vect{q}}_1\setminus\{m\}}}{\const{M}}\Biggr)\log_2{\Biggl(\frac{\sum\limits_{m=1}^{\const{M}}z_{\bar{\vect{q}}_1\setminus\{m\}}}{\const{M}}\Biggr)}\nonumber\\
  && -\>\bigHP{(\vect{z})_{\vect{s}\in [0:n-1]^{\const{M}-1}}},\IEEEeqnarraynumspace\label{eq:obj-ft_MI}
\end{IEEEeqnarray}
where~$(a)$ holds by the definition of entropy and the fact that $\bigcup_{l=1}^n\set{Q}_l=\widebar{\set{Q}}_1=[0:n-1]^{\const{M}}$. Hence, \eqref{eq:objective-ft_leakage} becomes~\eqref{eq:obj-ft_MI}.

We remark that the MI objective function $\MI{M}{\widebar{\vect{Q}}_1}$ is convex in $P_{\widebar{\vect{Q}}_1|M}$, and $P_{\widebar{\vect{Q}}_1|M}$ is subject to the following linear constraints 
\begin{IEEEeqnarray*}{rCl}
  P_{\widebar{\vect{Q}}_1|M}(\bar{\vect{q}}_1|m)=P_{\widebar{\vect{Q}}_1|M}(\bar{\vect{q}}'_1|m),\,\forall\,\bar{\vect{q}}_1\setminus\{m\}=\bar{\vect{q}}'_1\setminus\{m\}.
\end{IEEEeqnarray*}
Thus, the optimization problem~\eqref{eq:optimization_leakage-download} under MI leakage is convex. However, it is difficult to have closed-form optimal solutions for $(\const{M},n) \neq (2,2)$, %$\const{M}>2$ and $n\geq 2$
and hence instead we present numerical results of the optimized time-sharing $(\const{M},2)$ Scheme~A for several values of the number of files $\const{M}$ in Section~\ref{sec:numerical-results}. Lastly, we would like to emphasize that for any information leakage metric $\rho^{(\cdot)}$ that is convex in $P_{\widebar{\vect{Q}}_1|M}$ we end up with a convex optimization problem for the maximization of the download rate of the time-sharing Scheme~A.

\subsection{Optimizing the MaxL}
\label{sec:optimizing-MaxL}

In this subsection, we turn our attention to the minimization of the MaxL for the proposed $(\const{M},n)$ Scheme~A with time-sharing. Similar to the derivation for the MI metric, by definition~\eqref{eq:objective-ft_leakage} becomes
\begin{IEEEeqnarray*}{rCl}
  \rho^{(\mathsf{MaxL})}(\widebar{\collect{C}}_{\mathsf{A}})=\log_2{\sum_{\bar{\vect{q}}_l\in\widebar{\set{Q}}_l}\max_{m\in[\const{M}]}\frac{z_{\bar{\vect{q}}_l\setminus\{m\}}}{n}}.%\label{eq:obj-ft_MaxL}
\end{IEEEeqnarray*}

We remark that as for MI leakage, the time-sharing Scheme~A $\widebar{\collect{C}}_{\mathsf{A}}$ also has identical MaxL at each server. Using again the fact that the sets $\set{Q}_l$ are distinct and $\bigcup_{l=1}^n\set{Q}_l=\widebar{\set{Q}}_l$, we have
\begin{IEEEeqnarray*}{rCl}
  \rho^{(\mathsf{MaxL})}(\collect{C}_{\mathsf{A}})& = &\max_{l\in [n]}\Bigl\{\log_2{\sum_{\vect{q}_l\in\set{Q}_l}\max_{m\in[\const{M}]}}P_{\vect{Q}_l|M}(\vect{q}_l|m)\Bigr\}
  \\
  & = &\log_2{\biggl\{\max_{l\in[n]}\sum_{\vect{q}_l\in\set{Q}_l}\max_{m\in[\const{M}]}P_{\vect{Q}_l|M}(\vect{q}_l|m)\biggr\}}
  \\
  & \geq &\log_2{\biggl\{\frac{1}{n}\sum_{l\in[n]}\sum_{\vect{q}_l\in\set{Q}_l}\max_{m\in[\const{M}]}P_{\vect{Q}_l|M}(\vect{q}_l|m)}\biggr\}
  \\
  & = &\rho^{(\mathsf{MaxL})}(\widebar{\collect{C}}_{\mathsf{A}}).
\end{IEEEeqnarray*}

From Lemma~\ref{lem:convex_MI-MaxL} and using a similar argument as in Section~\ref{sec:optimizing_MI-leakage}, replacing the objective function with $2^{\rho^{(\mathsf{MaxL})}(\widebar{\collect{C}}_{\mathsf{A}})}$ in~\eqref{eq:optimization_leakage-download} gives a convex minimization problem. In Section~\ref{sec:numerical-results} below, we give numerical optimal values for~\eqref{eq:optimization_leakage-download} under MaxL and compare them with the converse results presented next. % QN: Should we also put the analytical solution for general (M,n) here? Which is happened to be the same as the ISIT'20 paper by Tian's group.

We remark here that the minimization of the information leakage for Scheme~A with time-sharing under a download cost constraint $\const{D}$ can also be done for the $\eps$-privacy metric. However, as this work mainly focuses on the MI and MaxL privacy metrics, we leave the analysis of this minimization of the $\eps$-privacy metric as future work. % QN: move to the conclusion for future work?

\section{Converse Results for MI Leakage}
\label{sec:converse-results_MI}

In order to present the converse results of WPIR for the MI metric, we first introduce the following measure between two PMFs.
\begin{definition}
  The TV distance between two PMFs $P_{Y_1}$ and $P_{Y_2}$ on the same finite alphabet $\set{Y}$ is defined as
  \begin{IEEEeqnarray*}{c}
    \norm{P_{Y_1}-P_{Y_2}}_\mathsf{TV}\eqdef\max_{\set{Z}\subseteq\set{Y}}\;\abs{P_{Y_1}(\set{Z})-P_{Y_2}(\set{Z})},
  \end{IEEEeqnarray*}
  where $P_Y(\set{Z})\eqdef\sum_{z\in\set{Z}}P_Y(z)$ is the probability of all realizations in the set $\set{Z}$.
\end{definition}

Next, we review a useful lemma related to the TV distance, which was presented in \cite[Lem.~2]{CuffYu16_1}.
\begin{lemma}
  \label{lem:lemma_MI-TV}
  If $\MI{X}{Y}\leq\rho$, then for any $x,x'\in\set{X}$, we have
  \begin{IEEEeqnarray*}{c}
    \norm{P_{Y|X=x}-P_{Y|X=x'}}_\mathsf{TV}\leq 1-2\inv{\Hb}(1-\rho)\eqdef\delta^{\mathsf{MI}}.
  \end{IEEEeqnarray*}
\end{lemma}

Here, since we require that $\delta^{\mathsf{MI}} \geq 0$, it is easy to see that we can specify $\inv{\Hb}(1-\rho)\in\bigl[0,\frac{1}{2}\bigr]$.

Lastly, we consider a known result between the entropy difference and the TV distance, which can be derived by using a probabilistic coupling technique. 
\begin{lemma}[{\cite[Eq.~(4)]{Zhang07_1}}]
  \label{lem:entropy-diff_TV}
  If $\norm{P_{Y|X=x}-P_{Y|X=x'}}_\mathsf{TV} \leq\delta^{\mathsf{MI}}$, then
  \begin{IEEEeqnarray}{rCl}
    \IEEEeqnarraymulticol{3}{l}{%;
      \abs{\HP{P_{Y|X=x}}-\HP{P_{Y|X=x'}}}
    }\nonumber\\*[1mm]\hspace*{3.0cm}%
    & \leq &\delta^{\mathsf{MI}}\log_2(\card{\set{Y}}-1)+\Hb(\delta^{\mathsf{MI}})
    \IEEEeqnarraynumspace\label{eq:entropy-difference_Hb}\\
    & \leq &\delta^{\mathsf{MI}}\log_2(\card{\set{Y}}-1)+1.
    \nonumber\IEEEeqnarraynumspace\label{eq:entropy-difference}% +\Hb(\delta^{\mathsf{MI}})
  \end{IEEEeqnarray}
\end{lemma}
Note that in the inequality~\eqref{eq:entropy-difference_Hb}, % if we replace the $1$ in the right-hand side of the inequality~\eqref{eq:entropy-difference} with $\Hb(\delta^{\mathsf{MI}})$, 
the upper bound becomes tight for $0\leq\delta^{\mathsf{MI}}\leq 1-\nicefrac{1}{\card{\set{Y}}}$ \cite[Thm.~6]{HoYeung10_1}.\footnote{The results shown in \cite{Zhang07_1} and \cite{HoYeung10_1} assume the variational distance as the measure between two PMFs. It can be easily shown that the TV distance is equal to the variational distance divided by $2$, i.e., $\norm{P_{Y_1}-P_{Y_2}}_\mathsf{TV}=\frac{1}{2}\sum_{y\in\set{Y}}\abs{P_{Y_1}(y)-P_{Y_2}(y)}$, see, e.g.,~\cite[Lem.~3.12]{Moser19_v47}.}

We remark that under the MI metric, we measure the overall leakage of a WPIR scheme in an average manner, i.e., $\rho^{(\mathsf{MI})}=\frac{1}{n}\sum_{l\in [n]}\eMI{M}{\vect{Q}_l}$. Here, given a leakage constraint $\rho^{(\mathsf{MI})}\leq\rho$, we assume that $\eMI{M}{\vect{Q}_l}=\rho_l$, $\forall\,l\in [n]$.

Let us define
\begin{IEEEeqnarray*}{rCl}
  %\IEEEeqnarraymulticol{3}{l}{%
    \eps^{\mathsf{MI}}(\set{Q}_l,\set{A}_l)
  %\nonumber\\*\;%
  & \eqdef &\delta_l^{\mathsf{MI}}\log_2(\card{\set{Q}_l\times\set{A}_l}-1)+1,\\
   \eps^{\mathsf{MI}}(\set{Q}_l) & \eqdef & % \Hb(\delta_l^{\mathsf{MI}})
\delta_l^{\mathsf{MI}}\log_2(\card{\set{Q}_l}-1)+1% \Hb(\delta_l^{\mathsf{MI}})
,\IEEEeqnarraynumspace\label{eq:def_eps-MI}
\end{IEEEeqnarray*}%
where $\delta_l^{\mathsf{MI}}=1-2\inv{\Hb}(1-\rho_l)$ and $\set{A}_l$ is the alphabet of the answer $\vect{A}_l$, $l\in [n]$. We give the following useful lemma.
\begin{lemma}
  \label{lem:LB_HPm}
  Given $m\neq m'$, where $m,m'\notin\set{M}\subsetneq[\const{M}-1]$, we have
  \begin{IEEEeqnarray}{rCl}
    \IEEEeqnarraymulticol{3}{l}{%
      \eHPcond{\vect{A}_{[n]}^{(m)}}{\vect{Q}_{[n]}^{(m)},\vect{X}^{\set{M}}}}\nonumber\\*\quad%
    & \geq &\beta \log_2{\card{\set{X}}}+\frac{\bigHPcond{\vect{A}^{(m')}_{[n]}}{\vect{Q}^{(m')}_{[n]},\vect{X}^{\set{M}},\vect{X}^{(m)}}}{n}
    \nonumber\\
    &&\quad -\>\frac{\sum_{l=1}^n\bigl[\eps^{\mathsf{MI}}(\set{Q}_l,\set{A}_l)+\eps^{\mathsf{MI}}(\set{Q}_l)\bigr]}{n}.
    \IEEEeqnarraynumspace\label{eq:LB_HPm}
  \end{IEEEeqnarray}
  Moreover,
  \begin{IEEEeqnarray}{c}
    \eHPcond{\vect{A}_{[n]}^{(\const{M})}}{\vect{Q}_{[n]}^{(\const{M})},\vect{X}^{[\const{M}-1]}}\geq\beta \log_2{\card{\set{X}}}.\label{eq:LB_HP-Mth}
  \end{IEEEeqnarray}
\end{lemma}
\begin{IEEEproof}
  The proof is deferred to Appendix~\ref{sec:proof_HPm-LB}.
\end{IEEEproof}

Now, we are ready to derive a general lower bound on $\const{D}$. Since $\eHPcond{\vect{A}_l^{(m)}}{\vect{Q}_l^{(m)}=\vect{q}_l}\leq\log_2{\bigcard{\set{A}}^{\ell_l(\vect{q}_l)}}$ for a given $\vect{q}_l\in\set{Q}_l$, we have
\begin{IEEEeqnarray*}{rCl}
  \const{D}^{(m)}% & = &\log_2\card{\set{A}}\sum_{l=1}^n\BigE[\vect{Q}_l^{(m)}]{\ell_l(\vect{Q}^{(m)}_l)}
  % \\
  & = &\log_2\card{\set{A}}\sum_{l=1}^n\sum_{\vect{q}_l\in\set{Q}_l}P_{\vect{Q}^{(m)}_l}(\vect{q}_l)\ell_l(\vect{q}_l)
  \\
  & \geq &\sum_{l=1}^n\sum_{\vect{q}_l\in\set{Q}_l}P_{\vect{Q}^{(m)}_l}(\vect{q}_l)\bigHPcond{\vect{A}_l^{(m)}}{\vect{Q}_l^{(m)}=\vect{q}_l}
  \\
  & = &\sum_{l=1}^n\eHPcond{\vect{A}_l^{(m)}}{\vect{Q}_l^{(m)}}.\nonumber
\end{IEEEeqnarray*} % QN: is this assumption reasonable?
Note that without loss of generality, we can assume that the conditional entropies $\eHPcond{\vect{A}_{[n]}^{(m)}}{\vect{Q}_{[n]}^{(m)}}$, $m\in[\const{M}]$, satisfy $\eHPcond{\vect{A}_{[n]}^{(1)}}{\vect{Q}_{[n]}^{(1)}}\leq\cdots\leq\eHPcond{\vect{A}_{[n]}^{(\const{M})}}{\vect{Q}_{[n]}^{(\const{M})}}$.

Hence, similar to the recursive procedure used in \cite[Sec.~V-A]{SunJafar17_1}, \cite[Sec.~VI]{BanawanUlukus18_1}, the total download cost can be bounded from below by
\begin{IEEEeqnarray}{rCl}
  \IEEEeqnarraymulticol{3}{l}{%
    \const{D}(\collect{C})
  }\nonumber\\*\quad%
  & = &\frac{1}{\const{M}}\sum_{m=1}^{\const{M}}\const{D}^{(m)}
  \geq\frac{1}{\const{M}}\sum_{m=1}^{\const{M}}\sum_{l=1}^n\bigHPcond{\vect{A}_{l}^{(m)}}{\vect{Q}_{l}^{(m)}}
  \nonumber\\[1mm]
  &  \stackrel{(a)}{\geq} &\frac{1}{\const{M}}\sum_{m=1}^{\const{M}}\eHPcond{\vect{A}_{[n]}^{(m)}}{\vect{Q}_{[n]}^{(m)}}
\nonumber\\[1mm]
  & \geq & \frac{1}{\const{M}}\sum_{m=1}^{\const{M}}\eHPcond{\vect{A}_{[n]}^{(1)}}{\vect{Q}_{[n]}^{(1)}}
  = \eHPcond{\vect{A}_{[n]}^{(1)}}{\vect{Q}_{[n]}^{(1)}}
  \nonumber\\[1mm]
  & \stackrel{(b)}{\geq} &\beta \log_2{\card{\set{X}}}+\frac{\bigHPcond{\vect{A}^{(2)}_{[n]}}{\vect{Q}^{(2)}_{[n]},\vect{X}^{(1)}}}{n}\nonumber\\[1mm]
  &&\quad -\>\frac{\sum_{l=1}^n\bigl[\eps^{\mathsf{MI}}(\set{Q}_l,\set{A}_l)+\eps^{\mathsf{MI}}(\set{Q}_l)\bigr]}{n}
  \nonumber\\
  &&\quad\vdots
  \nonumber\\[1mm]
  & \stackrel{(c)}{\geq} &\beta \log_2{\card{\set{X}}}+\sum_{m=1}^{\const{M}-1}\Biggl[\frac{\beta \log_2{\card{\set{X}}}}{n^m}\nonumber\\
  &&\hspace{2.0cm}-\>\frac{\sum_{l=1}^n\bigl[\eps^{\mathsf{MI}}(\set{Q}_l,\set{A}_l)+\eps^{\mathsf{MI}}(\set{Q}_l)\bigr]}{n^{m}}\Biggr],
  \IEEEeqnarraynumspace\label{eq:D-LB_QAalphabets}
\end{IEEEeqnarray}
where $(a)$ holds because conditioning reduces entropy, and $(b)$--$(c)$ follow by recursively applying Lemma~\ref{lem:LB_HPm} $\const{M}$ times with $\set{M}=\emptyset,\{1\},\ldots,[\const{M}-1]$, respectively. It is worth mentioning that under the assumption of perfect privacy, i.e., $\delta_l^{\mathsf{MI}}=0$, $\forall\,l\in[n]$, $\eHPcond{\vect{A}_{[n]}^{(m)}}{\vect{Q}_{[n]}^{(m)},\vect{X}^{\set{M}}}$ in \eqref{eq:LB_HPm} can be bounded from below by
  \begin{IEEEeqnarray*}{rCl}
    \IEEEeqnarraymulticol{3}{l}{%
      \eHPcond{\vect{A}_{[n]}^{(m)}}{\vect{Q}_{[n]}^{(m)},\vect{X}^{\set{M}}}}\nonumber\\*\quad%
    & \geq &\beta \log_2{\card{\set{X}}}+\frac{\bigHPcond{\vect{A}^{(m')}_{[n]}}{\vect{Q}^{(m')}_{[n]},\vect{X}^{\set{M}},\vect{X}^{(m)}}}{n}
    \IEEEeqnarraynumspace\label{eq:LB_HPm_perfect-privacy}
  \end{IEEEeqnarray*}
  by bounding \eqref{eq:sum_entropy-differences} using \eqref{eq:entropy-difference_Hb} directly in the proof of Lemma~\ref{lem:LB_HPm} (see Appendix~\ref{sec:proof_HPm-LB}). % Lemma~\ref{lem:entropy-diff_TV}.
  Following the above recursive steps, it can then be shown that the same converse results of the PIR capacity proof in \cite[Sec.~V-A]{SunJafar17_1} can be obtained.

\subsection{A Converse Bound With Restricted Alphabets of Queries and Answers} %Download Symbols
\label{sec:conv-bound_restricted-assumptions}

The expression in \eqref{eq:D-LB_QAalphabets} indicates that a general lower bound on $\const{D}$ can be arbitrarily dependent on the choice of the alphabets of the queries and answers. % download symbols.
In this subsection, a converse bound for WPIR schemes is derived from a practical design perspective. Note that in the information-theoretic PIR setup, the upload cost can be ignored as it does not scale with the file size. Moreover, to have an efficient WPIR scheme, the downloaded answer size per server should be smaller than or equal to the entire retrieved file size. Hence, we assume that
\begin{IEEEeqnarray*}{rCl}
  \IEEEyesnumber\label{eq:WPIRassumptions_converse}
  \IEEEyessubnumber*
  \card{\set{Q}_l}& \leq &\alpha<\infty, 
  \label{eq:finite-query-sizes}\\
  \card{\set{A}_l}& \leq & \card{\set{X}}^\beta,
  \label{eq:finite-download-answer}
\end{IEEEeqnarray*}
$\forall\,l\in[n]$, for some positive $\alpha\in\Naturals$, i.e., the query sizes are finite and each answer takes value on a smaller alphabet than that of the retrieved file.

We first state the following theorem, which gives an upper bound on the maximum possible WPIR rate for the MI metric.
\begin{theorem}
  \label{thm:C_MI-converse}
  Consider an $(\const{M},n)$ WPIR scheme that satisfies~\eqref{eq:WPIRassumptions_converse} and with MI leakage $\rho^{(\mathsf{MI})}\leq\rho$. Then, the maximum possible WPIR rate, denoted by $\const{R}_{\mathsf{max}}^{(\mathsf{MI})}$, is bounded from above by
  \begin{IEEEeqnarray*}{c}
    \const{R}_{\mathsf{max}}^{(\mathsf{MI})}\leq\inv{\left[\frac{1}{n^{\const{M}-1}}+2\sum_{m=1}^{\const{M}-1}\frac{1}{n^{m-1}}\inv{\Hb}(1-\rho)\right]}\eqdef\const{R}^{(\mathsf{MI})}_{\mathsf{UB}}.
  \end{IEEEeqnarray*}
\end{theorem}

\begin{IEEEproof}
Under these assumptions, \eqref{eq:D-LB_QAalphabets} becomes
\begin{IEEEeqnarray}{rCl}
  \IEEEeqnarraymulticol{3}{l}{%
    \const{D}(\collect{C})}\nonumber\\*%
  & \geq &\beta\log_2{\card{\set{X}}}+\sum_{m=1}^{\const{M}-1}\Biggl[\frac{\beta\log_2{\card{\set{X}}}}{n^m}-\sum_{l=1}^n\biggl(\frac{\delta^{\mathsf{MI}}_l\log_2{(\alpha\card{\set{X}}^\beta)}}{n^m}\nonumber\\[1mm]
  &&\hspace{3.0cm}+\>\frac{\delta^{\mathsf{MI}}_l\log_2{\alpha}}{n^m}+\frac{2}{n^m}\biggr)\Biggr].% \Hb(\delta^{\mathsf{MI}}_l)
  \IEEEeqnarraynumspace\label{eq:D-LB_fixed}
\end{IEEEeqnarray}
Dividing \eqref{eq:D-LB_fixed} by $\beta\log_2{\card{\set{X}}}$ gives
\begin{IEEEeqnarray}{rCl}
  \IEEEeqnarraymulticol{3}{l}{%
    \frac{\const{D}(\collect{C})}{\beta\log_2{\card{\set{X}}}}
  }\nonumber\\*%
  & \geq & 1+\sum_{m=1}^{\const{M}-1}\Biggl[\frac{1}{n^m}-\sum_{l=1}^n\biggl(\frac{\delta^{\mathsf{MI}}_l\log_2{\alpha}}{n^m\beta\log_2{\card{\set{X}}}}+\frac{\delta^{\mathsf{MI}}_l}{n^m}
  \nonumber\\[1mm]
  &&\hspace{1.60cm}+\>\frac{\delta^{\mathsf{MI}}_l\log_2{\alpha}}{n^m\beta\log_2{\card{\set{X}}}}+\frac{2}{n^m\beta\log_2{\card{\set{X}}}}\biggr)\Biggr]% \Hb(\delta^{\mathsf{MI}})
  \IEEEeqnarraynumspace\label{eq:normalized-D-LB_fixed}\\[1mm]
  & \xrightarrow{\:\beta \to \infty\:} &1 + \sum_{m=1}^{\const{M}-1}\frac{1}{n^m}-\sum_{m=1}^{\const{M}-1}\frac{\sum_{l=1}^n\delta^{\mathsf{MI}}_l}{n^{m}}
  \IEEEeqnarraynumspace\label{eq:D-LB_betaAsymptotic_MI}\\
  & = &1+\sum_{m=1}^{\const{M}-1}\frac{1}{n^m}-\sum_{m=1}^{\const{M}-1}\frac{\sum_{l=1}^n\bigl[1-2\inv{\Hb}(1-\rho_l)\bigr]}{n^{m}}
  \nonumber\\
  & = &1+\sum_{m=1}^{\const{M}-1}\frac{1}{n^m}-\sum_{m=1}^{\const{M}-1}\frac{n}{n^m}+2\sum_{m=1}^{\const{M}-1}\sum_{l=1}^n\frac{\inv{\Hb}(1-\rho_l)}{n^{m}}
  \nonumber\\
  & \stackrel{(a)}{\geq} &\frac{1}{n^{\const{M}-1}}+2\sum_{m=1}^{\const{M}-1}\frac{n}{n^m}\inv{\Hb}\left(1-\frac{1}{n}\sum_{l=1}^n\rho_l\right)\nonumber\\
  & \stackrel{(b)}{\geq} &\frac{1}{n^{\const{M}-1}}+2\sum_{m=1}^{\const{M}-1}\frac{n}{n^m}\inv{\Hb}(1-\rho)\nonumber
  \\
  & = &\frac{1}{n^{\const{M}-1}}+2\sum_{m=1}^{\const{M}-1}\frac{1}{n^{m-1}}\inv{\Hb}(1-\rho), \IEEEeqnarraynumspace \nonumber %\label{eq:D-converse_MI}
\end{IEEEeqnarray}
where $(a)$ and $(b)$ hold because the inverse binary entropy function is convex and increasing in $[0,1]$, respectively.
\end{IEEEproof}
We remark that in the proof of Theorem~\ref{thm:C_MI-converse} (as well as in the following converse results), the subtlety  is to allow the file size $\beta$ to go to infinity, such that we can have nontrivial converse bounds, i.e., $\const{R}^{(\cdot)}_\mathsf{UB}\leq 1$, for certain cases of $(\const{M},n)$.  We will also discuss the tightness of our proposed converse results in Section~\ref{sec:opt-DL_TimeSharing-Scheme1}. Note that the converse bound in \eqref{eq:normalized-D-LB_fixed} is in general trivial when $\beta$ decreases.

In the following, we prove the largest possible achievable WPIR rate for the special case of $(\const{M},n)=(2,2)$ under the additional constraint that only one of the two servers can leak. 
\begin{theorem}
  \label{thm:WPIRcapacity_M2n2-1stNotLeak_MI}
  Consider an $(\const{M},n)=(2,2)$ WPIR scheme that satisfies~\eqref{eq:WPIRassumptions_converse} and with MI leakage  $\rho^{(\mathsf{MI})}\leq\rho \leq 1$. Then, the maximum possible WPIR rate is % equal to
  % 
  % 
  % The maximum possible  WPIR rate under the MI privacy leakage is equal to
  \begin{IEEEeqnarray*}{c}
  % \label{thm:C_MI_M2n2}
    \const{R}^{(\mathsf{MI})}_{\mathsf{max}}(\rho)=\inv{\bigl[1+\inv{\Hb}(1-2\rho)\bigr]}
  \end{IEEEeqnarray*}
  under the assumption that only one of the two servers can leak. %, and the MI leakage is $0\leq\rho\leq 1$. 
\end{theorem}
\begin{IEEEproof}
  The achievable scheme is the Scheme~A presented in Example~\ref{ex:Ex1_WPIR_M2n2}. Thus, we only need to prove the converse. Assume without loss of generality that $\MI{M}{\vect{Q}_1}=\rho_1=0$ and $\MI{M}{\vect{Q}_2}=\rho_2=2\rho$. %Hence, from \eqref{eq:def_eps-MI}, we have
%  \begin{IEEEeqnarray*}{c}
%    \eps^{\mathsf{MI}}(\set{Q}_1,\set{A})+\eps^{\mathsf{MI}}(\set{Q}_1)=0
%  \end{IEEEeqnarray*}
%  and~\eqref{eq:D-LB_QAalphabets} becomes
%  \begin{IEEEeqnarray*}{c}
%    \const{D}(\collect{C})\geq\beta+\frac{\beta-\bigl[\eps^{\mathsf{MI}}(\set{A},\set{Q}_2)+\eps^{\mathsf{MI}}(\set{Q}_2)\bigr]}{n} 
%  \end{IEEEeqnarray*}
%  for $(\const{M},n)=(2,2)$. 
%Now we can use the exact same derivation as in Section~\ref{sec:conv-bound_restricted-assumptions} and obtain
By definition, $\delta_1^{\mathsf{MI}}=1-2\inv{\Hb}(1-\rho_1) = 0$ and we can use the exact same derivation as in Section~\ref{sec:conv-bound_restricted-assumptions} to obtain from \eqref{eq:D-LB_betaAsymptotic_MI}
  \begin{IEEEeqnarray*}{rCl}
    \frac{\const{D}(\collect{C})}{\beta\log_2{\card{\set{X}}}}& \geq &1+\frac{1}{2}-\frac{\delta_2^{\mathsf{MI}}}{2}
    \\
    & = &\frac{3}{2}-\frac{1-2\inv{\Hb}(1-2\rho)}{2}=1+\inv{\Hb}(1-2\rho),
  \end{IEEEeqnarray*}
  as $\beta\to\infty$, which completes the proof.
\end{IEEEproof}

\section{Converse Results for MaxL}
\label{sec:converse-results_MaxL}

In this section, we present the converse results for the MaxL metric. Similar to the case of the MI privacy metric, see Theorem~\ref{thm:C_MaxL-converse} presented here.% make use of the following lemma.

\begin{theorem}
  \label{thm:C_MaxL-converse}
  Consider an $(\const{M},n)$ WPIR scheme that satisfies~\eqref{eq:WPIRassumptions_converse} and with MaxL $\rho^{(\mathsf{MaxL})}\leq\rho$. Then, the maximum possible WPIR rate, denoted by $\const{R}_{\mathsf{max}}^{(\mathsf{MaxL})}$, is bounded from above by
  \begin{IEEEeqnarray*}{c}
    \const{R}_{\mathsf{max}}^{(\mathsf{MaxL})}\leq\inv{\left[1+\sum_{m=1}^{\const{M}-1}\frac{1}{n^m}-\sum_{m=1}^{\const{M}-1}\frac{2^{\rho}-1}{n^{m-1}}\right]}\eqdef\const{R}^{(\mathsf{MaxL})}_{\mathsf{UB}}.
  \end{IEEEeqnarray*}
\end{theorem}

\begin{IEEEproof}
 Similar to the case with the MI privacy metric, we first make use of the following lemma.
\begin{lemma}
  \label{lem:lemma_MaxL-TV}
  If $\ML{X}{Y}\leq \rho$, then for any $x,x'\in\set{X}$, we have
  \begin{IEEEeqnarray*}{rCl}
    \norm{P_{Y|X=x}-P_{Y|X=x'}}_\mathsf{TV}\leq 2^{\rho}-1\eqdef\delta^{\mathsf{MaxL}}.
  \end{IEEEeqnarray*}
\end{lemma}
Lemma~\ref{lem:lemma_MaxL-TV} can be proven by a similar argument as in \cite[App.~C]{CuffYu16_1}. The proof is provided in Appendix~\ref{sec:proof_MaxL-TV} for completeness.

Note that for the MaxL metric, we consider the worst-case MaxL over all servers, i.e., $\rho^{(\mathsf{MaxL})}=\max_{l\in [n]}\eML{M}{\vect{Q}_l}$. If the leakage at the $l$-th  server is $\eML{M}{\vect{Q}_l}\leq\rho_l$, then we have $\rho^{(\mathsf{MaxL})}\leq\max_{l\in[n]}\rho_l\eqdef\rho$. A lower bound on the download cost can be proven by following the same steps as in Section~\ref{sec:converse-results_MI}. Under the assumptions in \eqref{eq:WPIRassumptions_converse}, and as $\beta\to\infty$, \eqref{eq:D-LB_betaAsymptotic_MI} becomes
\begin{IEEEeqnarray}{rCl}
  \frac{\const{D}(\collect{C})}{\beta\log_2{\card{\set{X}}}}& \geq &1 + \sum_{m=1}^{\const{M}-1}\frac{1}{n^m}-\sum_{m=1}^{\const{M}-1}\frac{\sum_{l=1}^n\delta^{\mathsf{MaxL}}_l}{n^{m}}
    \nonumber \\%\label{eq:D-LB_betaAsymptotic_MaxL}\\
  & \geq &1+\sum_{m=1}^{\const{M}-1}\frac{1}{n^m}-\sum_{m=1}^{\const{M}-1}\frac{n\max_{l\in[n]}\delta_l^{\mathsf{MaxL}}}{n^{m}}
  \nonumber\\
  & = &1+\sum_{m=1}^{\const{M}-1}\frac{1}{n^m}-\sum_{m=1}^{\const{M}-1}\frac{\max_{l\in[n]}\delta_l^{\mathsf{MaxL}}}{n^{m-1}}
  \nonumber\\
  & = &1+\sum_{m=1}^{\const{M}-1}\frac{1}{n^m}-\sum_{m=1}^{\const{M}-1}\frac{2^{\rho}-1}{n^{m-1}},\IEEEeqnarraynumspace\label{eq:D-converse_MaxL}
\end{IEEEeqnarray}
where \eqref{eq:D-converse_MaxL} follows from Lemma~\ref{lem:lemma_MaxL-TV} and $\max_{l\in[n]}\delta_l^{\mathsf{MaxL}}=2^{\max_{l\in[n]}\rho_l}-1=2^\rho-1$.
\end{IEEEproof}

In the following theorem we give the maximum achievable WPIR rate under the MaxL metric for the special case of $(\const{M},n)=(2,2)$.
\begin{theorem}
  \label{thm:WPIRcapacity_M2n2_MaxL}
  Consider an $(\const{M},n)=(2,2)$ WPIR scheme that satisfies~\eqref{eq:WPIRassumptions_converse} and with MaxL  $\rho^{(\mathsf{MaxL})}\leq\rho\leq 1$. Then, the maximum possible  WPIR rate is 
    \begin{IEEEeqnarray*}{c}
    \const{R}^{(\mathsf{MaxL})}_{\mathsf{max}}(\rho)=\inv{\left[\frac{5}{2}-2^{\rho}\right]}.
  \end{IEEEeqnarray*}
  Moreover, assuming that only one of the two servers can leak information, then the maximum possible WPIR rate is
  \begin{IEEEeqnarray*}{rCl}
    \const{R}_{\mathsf{max}}^{(\mathsf{MaxL})}(\rho)=\inv{\bigl[2-2^{\rho-1}\bigr]}.
  \end{IEEEeqnarray*}
\end{theorem}
\begin{IEEEproof}
  The achievable scheme for the first assertion is Scheme~A with time-sharing. In particular, by applying the  time-sharing principle to Example~\ref{ex:Ex1_WPIR_M2n2}, we get the following conditional PMF of $\widebar{\vect{Q}}_1$ given $M=m$,
  % \Resize[0.9\columnwidth]{
  \begin{IEEEeqnarray*}{c}
    \begin{IEEEeqnarraybox}[
      \IEEEeqnarraystrutmode
      \IEEEeqnarraystrutsizeadd{7pt}{7pt}]{v/c/v/c/v/c/v}
      \IEEEeqnarrayrulerow\\
      & P_{\bar{\vect{Q}}_1|M}(\bar{\vect{q}}_1|m) && m=1 && m=2
      \\*\hline
      & \bar{\vect{q}}_1=(0,0) && \frac{1-p}{2} && \frac{1-p}{2}
      \\\hline
      & \bar{\vect{q}}_1=(0,1) && \frac{p}{2}   && \frac{1-p}{2}
      \\\hline
      & \bar{\vect{q}}_1=(1,0) && \frac{1-p}{2}   && \frac{p}{2}
      \\\hline
      & \bar{\vect{q}}_1=(1,1) && \frac{p}{2}   && \frac{p}{2}      
      \\*\IEEEeqnarrayrulerow
    \end{IEEEeqnarraybox}
  \end{IEEEeqnarray*}
  Thus, by definitions we get $2^{\rho^{(\mathsf{MaxL})}}=\nicefrac{1}{2}+(1-p)$, for $0\leq p\leq \nicefrac{1}{2}$, and hence it can be seen that $\const{D}=1+p=1+\nicefrac{3}{2}-2^{\rho^{(\mathsf{MaxL})}}=\nicefrac{5}{2}-2^{\rho^{(\mathsf{MaxL})}}$. Note that this is also the optimized time-sharing Scheme~A under the MaxL metric for $(\const{M},n)=(2,2)$, where the optimal solution  for \eqref{eq:optimization_leakage-download} is $(z^\ast_0,z^\ast_1)=(2-\const{D},\const{D}-1)$. The converse part is proven by~\eqref{eq:D-converse_MaxL}, which gives
  \begin{IEEEeqnarray*}{c}
    \frac{\const{D}(\collect{C})}{\beta\log_2{\card{\set{X}}}}\geq 1+\frac{1}{2}-\frac{2^{\rho}-1}{1}=\frac{5}{2}-2^{\rho}
  \end{IEEEeqnarray*}
  for $(\const{M},n)=(2,2)$.

  Further, the second assertion can be proven by following the same lines as in the poof of Theorem~\ref{thm:WPIRcapacity_M2n2-1stNotLeak_MI}.
\end{IEEEproof}

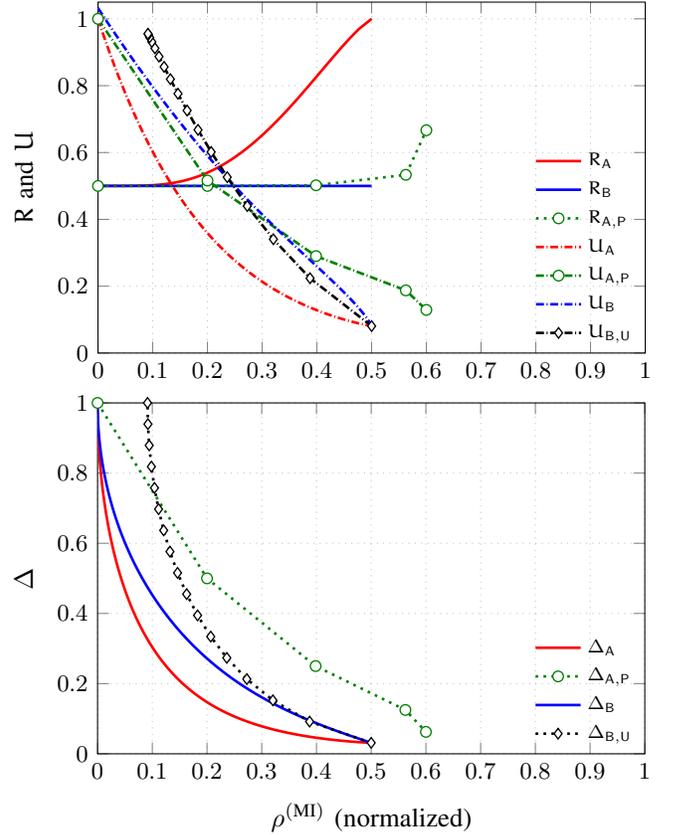
\begin{figure}[t!]
  \centering
  \begin{tikzpicture}%[thick, scale=0.9, every node/.style={transform shape}]
\pgfplotsset{every tick label/.append style={font=\small}}
\begin{axis}[%
width=\columnwidth,
height=6.25cm,
at={(1.387in,0.821in)},
xmin=0,
xmax=1, % xmax=180,
grid style={gray,opacity=0.5,dotted},
xmajorgrids,
ymajorgrids,
% yminorgrids,
% ymode=log,
ymin=0, % ymin=0.0000001,
max space between ticks=20pt,
ymax=1.05,
ylabel={$\const{R}$ and $\const{U}$},
ylabel style={
  yshift=0.5ex,
  name=label},
axis background/.style={fill=white},
legend cell align=left,
legend style={legend style={draw=none,fill=none}, font=\scriptsize,
  minimum height=0.15in, at={(axis cs: 1,0,0.0)}, anchor=south east},
]
% rates
\addplot [color=red,solid,line width=1pt, mark=-*, mark options={solid, line width = 0.5pt, fill=white}]table[x=avgMI_1,y=R_1] {\Figs/data/R1-MI_WPIRlkg_n2f32_v4.txt}; 
\addlegendentry{$\const{R}_{\mathsf{A}}$};

\addplot [color=blue,solid,line width=1pt, mark=-*, mark options={solid, line width = 0.5pt, fill=white}]table[x=avgMI_2B,y=R_2B] {\Figs/data/R2B-MI_WPIRlkg_n2f32_v4.txt}; 
\addlegendentry{$\const{R}_{\mathsf{B}}$};

\addplot [color=black!50!green,dotted,line width=1pt, mark=*, mark options={solid, line width = 0.5pt, fill=white}]table[x=avgMI_A,y=R_A] {\Figs/data/RA-MI_WPIRlkg_n2f32_v4.txt}; 
\addlegendentry{$\const{R}_{\mathsf{A},\mathsf{P}}$};

% upload costs
\addplot [color=red,densely dashdotted,line width=1pt, mark=-*, mark options={solid, line width = 0.5pt, fill=white}]table[x=avgMI_1,y=U_1] {\Figs/data/U1-MI_WPIRlkg_n2f32_v4.txt}; 
\addlegendentry{$\const{U}_{\mathsf{A}}$};

\addplot [color=black!50!green, densely dashdotted,line width=1pt, mark=*, mark options={solid, line width = 0.5pt, fill=white}]table[x=avgMI_A,y=U_A] {\Figs/data/UA-MI_WPIRlkg_n2f32_v4.txt}; 
\addlegendentry{$\const{U}_{\mathsf{A},\mathsf{P}}$};

\addplot [color=blue, densely dashdotted, line width=1pt, mark=-*, mark options={solid, line width = 0.5pt, fill=white}]table[x=avgMI_2B,y=U_2B] {\Figs/data/U2B-MI_WPIRlkg_n2f32_v4.txt}; 
\addlegendentry{$\const{U}_{\mathsf{B}}$};

\addplot [color=black, densely dashdotted, line width=1pt, mark=diamond*, mark options={solid, line width = 0.5pt, fill=white}]table[x=avgMI_2W,y=U_2W] {\Figs/data/U2W-MI_WPIRlkg_n2f32_v4.txt}; 
\addlegendentry{$\const{U}_{\mathsf{B},\mathsf{U}}$};

\end{axis}
\end{tikzpicture}%
% }
\\
% plot the access complexity
\begin{tikzpicture}%[thick, scale=0.9, every node/.style={transform shape}]
\pgfplotsset{every tick label/.append style={font=\small}}
\begin{axis}[%
width=\columnwidth,
height=6.25cm,
at={(1.387in,0.821in)},
xmin=0,
xmax=1, % xmax=180,
xlabel={$\rho^{(\textnormal{MI})}$ (normalized)},
xlabel style={
  yshift=-0.5ex,
  name=label},
grid style={gray,opacity=0.5,dotted},
xmajorgrids,
ymajorgrids,
% yminorgrids,
% ymode=log,
ymin=0, % ymin=0.0000001,
max space between ticks=20pt,
ymax=1,
ylabel={$\Delta$},
ylabel style={
  yshift=0.5ex,
  name=label},
axis background/.style={fill=white},
legend cell align=left,
legend style={legend style={draw=none,fill=none}, font=\scriptsize,
  minimum height=0.15in, at={(axis cs: 1.0,0.0)}, anchor=south east},
]

\addplot [color=red,solid,line width=1pt, mark=-*, mark options={solid, line width = 0.5pt, fill=white}]table[x=avgMI_1,y=S_1] {\Figs/data/S1-MI_WPIRlkg_n2f32_v4.txt}; 
\addlegendentry{$\Delta_{\mathsf{A}}$};

\addplot [color=black!50!green,dotted,line width=1pt, mark=*, mark options={solid, line width = 0.5pt, fill=white}]table[x=avgMI_A,y=S_A] {\Figs/data/SA-MI_WPIRlkg_n2f32_v4.txt}; 
\addlegendentry{$\Delta_{\mathsf{A},\mathsf{P}}$};

\addplot [color=blue,solid,line width=1pt, mark=-*, mark options={solid, line width = 0.5pt, fill=white}]table[x=avgMI_2B,y=S_2B] {\Figs/data/S2B-MI_WPIRlkg_n2f32_v4.txt}; 
\addlegendentry{$\Delta_{\mathsf{B}}$};

\addplot [color=black,dotted,line width=1pt, mark=diamond*, mark options={solid, line width = 0.5pt, fill=white}]table[x=avgMI_2W,y=S_2W] {\Figs/data/S2W-MI_WPIRlkg_n2f32_v4.txt}; 
\addlegendentry{$\Delta_{\mathsf{B},\mathsf{U}}$};

\end{axis}
\end{tikzpicture}%
%\end{figure}
%%
%\end{frame}

% \end{document}
  \caption{$\const{R}$, $\const{U}$, and $\Delta$ of different WPIR schemes for $\const{M}=32$, as a function of $\rho^{(\mathsf{MI})}$. For $\const{M}=32$, $\const{C}_{\const{M},2}$ is almost equal to $\nicefrac{1}{2}$.}
  \label{fig:tuple-MI_n2M32}
\end{figure}

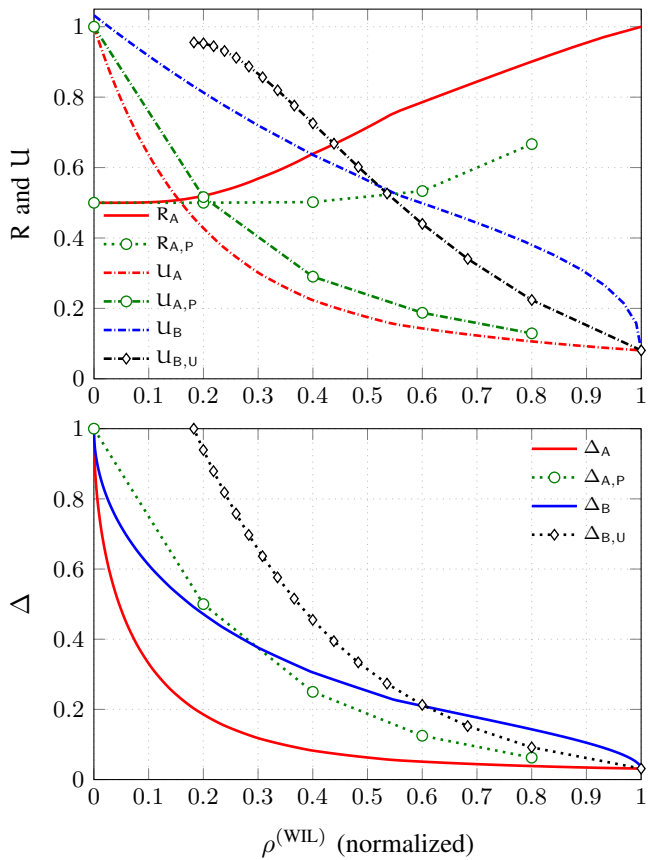
\begin{figure}[t!]
  \centering
  % \documentclass[conference, 9pt]{IEEEtran} 
% \pagestyle{empty}	
% %\documentclass[
% %%    draft,
% %	%handout,
% %	%dvips,
% %	%xcolor=dvipsnames,
% %	9pt,
% %	mathserif
% %]{beamer}

% \usepackage{epsfig,amsfonts,amsbsy,bm,mathrsfs}
% \usepackage[nolist]{acronym}
% \usepackage{mathrsfs} 
% \usepackage{graphicx,cite,amssymb,amsmath,bm}
% \usepackage{mathtools}
% \usepackage{dsfont}
% \mathtoolsset{showonlyrefs}
% \usepackage{amsmath} 
% \usepackage[utf8]{inputenc}
% \usepackage[T1]{fontenc}
% %\usepackage{subcaption}
% \usepackage{ctable}
% \usepackage{pgfplots}
% \usepackage[english]{babel}
% \usepackage{xcolor}
% \usepackage{color}
% \usepackage{tikz}
% \usetikzlibrary{fadings}
% \usetikzlibrary{shadows.blur}
% \usetikzlibrary{shapes,arrows}
% \usetikzlibrary{calc,shapes.misc}
% \usetikzlibrary{decorations.pathreplacing}
% \usepackage{ifthen}
% \usetikzlibrary{positioning}
% \usepackage{pgfplots,relsize}
% \usetikzlibrary{plotmarks}
% \usetikzlibrary{decorations.pathreplacing,shapes.misc}
% \usetikzlibrary{patterns}
% \usepackage{colortbl}
% \usepackage{balance}

% \newcommand{\tn}[1]{\textnormal{#1}}
% \newcommand{\vmat}[1]{\bm{#1}} % notation for random matrices
% \newcommand{\set}[1]{\mathcal{#1}}           % set
% \newcommand{\const}[1]{\textnormal{\usefont{U}{eur}{m}{n}\selectfont #1}} % Euler

% \newcommand{\Bernoulli}[1]{\tn{Bernoulli}\left(#1\right)} % Bernoulli dist.
% \newcommand{\Uniform}[1]{\tn{Uniform}\left(#1\right)} % Uniform dist.

% \begin{document}

\definecolor{mycolor5}{rgb}{0.00000,0.44700,0.74100}%
\definecolor{mycolor3}{rgb}{0.85000,0.32500,0.09800}%
\definecolor{mycolor1}{rgb}{0.92900,0.69400,0.12500}%
\definecolor{mycolor4}{rgb}{0.49400,0.18400,0.55600}%
\definecolor{mycolor2}{rgb}{0.46600,0.67400,0.18800}%
\noindent
%\begin{frame}[label=final]
%
%\begin{figure}
% plot the IR rate and upload cost
%\resizebox{0.8\paperwidth}{!}{%
\begin{tikzpicture}%[thick, scale=0.9, every node/.style={transform shape}]
\pgfplotsset{every tick label/.append style={font=\small}}
\begin{axis}[%
width=\columnwidth,
height=6.5cm,
at={(1.387in,0.821in)},
xmin=0,
xmax=1, % xmax=180,
grid style={gray,opacity=0.5,dotted},
xmajorgrids,
ymajorgrids,
% yminorgrids,
% ymode=log,
ymin=0, % ymin=0.0000001,
max space between ticks=20pt,
ymax=1.05,
ylabel={$\const{R}$ and $\const{U}$},
ylabel style={
  yshift=0.5ex,
  name=label},
axis background/.style={fill=white},
legend cell align=left,
legend style={legend style={draw=none,fill=none}, font=\scriptsize,
  minimum height=0.15in, at={(axis cs: 0,0.0)}, anchor=south west},
]
% rates
\addplot [color=red,solid,line width=1pt, mark=-*, mark options={solid, line width = 0.5pt, fill=white}]table[x=WIL_1,y=R_1] {\Figs/data/R1-WIL_WPIRlkg_n2f32_v4.txt}; 
\addlegendentry{$\const{R}_{\mathsf{A}}$};

% \addplot [color=blue,solid,line width=1pt, mark=-*, mark options={solid, line width = 0.5pt, fill=white}]table[x=WIL_2B,y=R_2B] {\Figs/data/R2B-WIL_WPIRlkg_n2f32_v4.txt}; 
% \addlegendentry{$\const{R}_2$};

\addplot [color=black!50!green,dotted,line width=1pt, mark=*, mark options={solid, line width = 0.5pt, fill=white}]table[x=WIL_A,y=R_A] {\Figs/data/RA-WIL_WPIRlkg_n2f32_v4.txt}; 
\addlegendentry{$\const{R}_{\mathsf{A},\mathsf{P}}$};

% upload cost
\addplot [color=red,densely dashdotted,line width=1pt, mark=-*, mark options={solid, line width = 0.5pt, fill=white}]table[x=WIL_1,y=U_1] {\Figs/data/U1-WIL_WPIRlkg_n2f32_v4.txt}; 
\addlegendentry{$\const{U}_{\mathsf{A}}$};

\addplot [color=black!50!green,densely dashdotted,line width=1pt, mark=*, mark options={solid, line width = 0.5pt, fill=white}]table[x=WIL_A,y=U_A] {\Figs/data/UA-WIL_WPIRlkg_n2f32_v4.txt}; 
\addlegendentry{$\const{U}_{\mathsf{A},\mathsf{P}}$};

\addplot [color=blue,densely dashdotted,line width=1pt, mark=-*, mark options={solid, line width = 0.5pt, fill=white}]table[x=WIL_2B,y=U_2B] {\Figs/data/U2B-WIL_WPIRlkg_n2f32_v4.txt}; 
\addlegendentry{$\const{U}_{\mathsf{B}}$};

\addplot [color=black,densely dashdotted,line width=1pt, mark=diamond*, mark options={solid, line width = 0.5pt, fill=white}]table[x=WIL_2W,y=U_2W] {\Figs/data/U2W-WIL_WPIRlkg_n2f32_v4.txt}; 
\addlegendentry{$\const{U}_{\mathsf{B},\mathsf{U}}$};

\end{axis}
\end{tikzpicture}%
% }
\\
% plot the access complexity
\begin{tikzpicture}%[thick, scale=0.9, every node/.style={transform shape}]
\pgfplotsset{every tick label/.append style={font=\small}}
\begin{axis}[%
width=\columnwidth,
height=6.25cm,
at={(1.387in,0.821in)},
xmin=0,
xmax=1, % xmax=180,
xlabel={$\rho^{(\textnormal{WIL})}$ (normalized)},
xlabel style={
  yshift=-0.5ex,
  name=label},
grid style={gray,opacity=0.5,dotted},
xmajorgrids,
ymajorgrids,
% yminorgrids,
% ymode=log,
ymin=0, % ymin=0.0000001,
max space between ticks=20pt,
ymax=1,
ylabel={$\Delta$},
ylabel style={
  yshift=0.5ex,
  name=label},
axis background/.style={fill=white},
legend cell align=left,
legend style={legend style={draw=none,fill=none}, font=\scriptsize, 
  minimum height=0.15in, at={(axis cs: 1.0,1.0)}, anchor=north east},
]

\addplot [color=red,solid,line width=1pt, mark=-*, mark options={solid, line width = 0.5pt, fill=white}]table[x=WIL_1,y=S_1] {\Figs/data/S1-WIL_WPIRlkg_n2f32_v4.txt}; 
\addlegendentry{$\Delta_{\mathsf{A}}$};

\addplot [color=black!50!green,dotted,line width=1pt, mark=*, mark options={solid, line width = 0.5pt, fill=white}]table[x=WIL_A,y=S_A] {\Figs/data/SA-WIL_WPIRlkg_n2f32_v4.txt}; 
\addlegendentry{$\Delta_{\mathsf{A},\mathsf{P}}$};

\addplot [color=blue,solid,line width=1pt, mark=-*, mark options={solid, line width = 0.5pt, fill=white}]table[x=WIL_2B,y=S_2B] {\Figs/data/S2B-WIL_WPIRlkg_n2f32_v4.txt}; 
\addlegendentry{$\Delta_{\mathsf{B}}$};

\addplot [color=black,dotted,line width=1pt, mark=diamond*, mark options={solid, line width = 0.5pt, fill=white}]table[x=WIL_2W,y=S_2W] {\Figs/data/S2W-WIL_WPIRlkg_n2f32_v4.txt}; 
\addlegendentry{$\Delta_{\mathsf{B},\mathsf{U}}$};

\end{axis}
\end{tikzpicture}%
%\end{figure}
%%
%\end{frame}

% \end{document}
  \caption{$\const{R}$, $\const{U}$, and $\Delta$ of different WPIR schemes for $\const{M}=32$, as a function of $\rho^{(\mathsf{WIL})}$. Here, $\const{R}_{\mathsf{B}}=\nicefrac{1}{2}$ is not plotted.}
  \label{fig:tuple-WIL_n2M32}
\end{figure}

\begin{figure}[t!]
  \centering
  % \documentclass[conference, 9pt]{IEEEtran} 
% \pagestyle{empty}	
% %\documentclass[
% %%    draft,
% %	%handout,
% %	%dvips,
% %	%xcolor=dvipsnames,
% %	9pt,
% %	mathserif
% %]{beamer}

% \usepackage{epsfig,amsfonts,amsbsy,bm,mathrsfs}
% \usepackage[nolist]{acronym}
% \usepackage{mathrsfs} 
% \usepackage{graphicx,cite,amssymb,amsmath,bm}
% \usepackage{mathtools}
% \usepackage{dsfont}
% \mathtoolsset{showonlyrefs}
% \usepackage{amsmath} 
% \usepackage[utf8]{inputenc}
% \usepackage[T1]{fontenc}
% %\usepackage{subcaption}
% \usepackage{ctable}
% \usepackage{pgfplots}
% \usepackage[english]{babel}
% \usepackage{xcolor}
% \usepackage{color}
% \usepackage{tikz}
% \usetikzlibrary{fadings}
% \usetikzlibrary{shadows.blur}
% \usetikzlibrary{shapes,arrows}
% \usetikzlibrary{calc,shapes.misc}
% \usetikzlibrary{decorations.pathreplacing}
% \usepackage{ifthen}
% \usetikzlibrary{positioning}
% \usepackage{pgfplots,relsize}
% \usetikzlibrary{plotmarks}
% \usetikzlibrary{decorations.pathreplacing,shapes.misc}
% \usetikzlibrary{patterns}
% \usepackage{colortbl}
% \usepackage{balance}

% \newcommand{\tn}[1]{\textnormal{#1}}
% \newcommand{\vmat}[1]{\bm{#1}} % notation for random matrices
% \newcommand{\set}[1]{\mathcal{#1}}           % set
% \newcommand{\const}[1]{\textnormal{\usefont{U}{eur}{m}{n}\selectfont #1}} % Euler

% \newcommand{\Bernoulli}[1]{\tn{Bernoulli}\left(#1\right)} % Bernoulli dist.
% \newcommand{\Uniform}[1]{\tn{Uniform}\left(#1\right)} % Uniform dist.

% \begin{document}

\definecolor{mycolor5}{rgb}{0.00000,0.44700,0.74100}%
\definecolor{mycolor3}{rgb}{0.85000,0.32500,0.09800}%
\definecolor{mycolor1}{rgb}{0.92900,0.69400,0.12500}%
\definecolor{mycolor4}{rgb}{0.49400,0.18400,0.55600}%
\definecolor{mycolor2}{rgb}{0.46600,0.67400,0.18800}%
\noindent
%\begin{frame}[label=final]
%
%\begin{figure}
% plot the IR rate and upload cost
%\resizebox{0.8\paperwidth}{!}{%
\begin{tikzpicture}%[thick, scale=0.9, every node/.style={transform shape}]
\pgfplotsset{every tick label/.append style={font=\small}}
\begin{axis}[%
width=\columnwidth,
height=6.5cm,
at={(1.387in,0.821in)},
xmin=0,
xmax=1, % xmax=180,
grid style={gray,opacity=0.5,dotted},
xmajorgrids,
ymajorgrids,
% yminorgrids,
% ymode=log,
ymin=0, % ymin=0.0000001,
max space between ticks=20pt,
ymax=1.05,
ylabel={$\const{R}$ and $\const{U}$},
ylabel style={
  yshift=0.5ex,
  name=label},
axis background/.style={fill=white},
legend cell align=left,
legend style={legend style={draw=none,fill=none}, font=\scriptsize,
  minimum height=0.15in, at={(axis cs: 0,0.0)}, anchor=south west},
]
% rates
\addplot [color=red,solid,line width=1pt, mark=-*, mark options={solid, line width = 0.5pt, fill=white}]table[x=MaxL_1,y=R_1] {\Figs/data/R1-MaxL_WPIRlkg_n2f32_v4.txt}; 
\addlegendentry{$\const{R}_{\mathsf{A}}$};

% \addplot [color=blue,solid,line width=1pt, mark=-*, mark options={solid, line width = 0.5pt, fill=white}]table[x=maxH_2B,y=R_2B] {\Figs/data/R2B-MaxL_WPIRlkg_n2f32_v4.txt}; 
% \addlegendentry{$\const{R}_2$};

\addplot [color=black!50!green,dotted,line width=1pt, mark=*, mark options={solid, line width = 0.5pt, fill=white}]table[x=MaxL_A,y=R_A] {\Figs/data/RA-MaxL_WPIRlkg_n2f32_v4.txt}; 
\addlegendentry{$\const{R}_{\mathsf{A},\mathsf{P}}$};

% upload cost
\addplot [color=red,densely dashdotted,line width=1pt, mark=-*, mark options={solid, line width = 0.5pt, fill=white}]table[x=MaxL_1,y=U_1] {\Figs/data/U1-MaxL_WPIRlkg_n2f32_v4.txt}; 
\addlegendentry{$\const{U}_{\mathsf{A}}$};

\addplot [color=black!50!green,densely dashdotted,line width=1pt, mark=*, mark options={solid, line width = 0.5pt, fill=white}]table[x=MaxL_A,y=U_A] {\Figs/data/UA-MaxL_WPIRlkg_n2f32_v4.txt}; 
\addlegendentry{$\const{U}_{\mathsf{A},\mathsf{P}}$};

\addplot [color=blue,densely dashdotted,line width=1pt, mark=-*, mark options={solid, line width = 0.5pt, fill=white}]table[x=MaxL_2B,y=U_2B] {\Figs/data/U2B-MaxL_WPIRlkg_n2f32_v4.txt}; 
\addlegendentry{$\const{U}_{\mathsf{B}}$};

\addplot [color=black,densely dashdotted,line width=1pt, mark=diamond*, mark options={solid, line width = 0.5pt, fill=white}]table[x=MaxL_2W,y=U_2W] {\Figs/data/U2W-MaxL_WPIRlkg_n2f32_v4.txt}; 
\addlegendentry{$\const{U}_{\mathsf{B},\mathsf{U}}$};

\end{axis}
\end{tikzpicture}%
% }
\\
% plot the access complexity
\begin{tikzpicture}%[thick, scale=0.9, every node/.style={transform shape}]
\pgfplotsset{every tick label/.append style={font=\small}}
\begin{axis}[%
width=\columnwidth,
height=6.25cm,
at={(1.387in,0.821in)},
xmin=0,
xmax=1, % xmax=180,
xlabel={$\rho^{(\textnormal{MaxL})}$ (normalized)},
xlabel style={
  yshift=-0.5ex,
  name=label},
grid style={gray,opacity=0.5,dotted},
xmajorgrids,
ymajorgrids,
% yminorgrids,
% ymode=log,
ymin=0, % ymin=0.0000001,
max space between ticks=20pt,
ymax=1,
ylabel={$\Delta$},
ylabel style={
  yshift=0.5ex,
  name=label},
axis background/.style={fill=white},
legend cell align=left,
legend style={legend style={draw=none,fill=none}, font=\scriptsize, 
  minimum height=0.15in, at={(axis cs: 1.0,1.0)}, anchor=north east},
]

\addplot [color=red,solid,line width=1pt, mark=-*, mark options={solid, line width = 0.5pt, fill=white}]table[x=MaxL_1,y=S_1] {\Figs/data/S1-MaxL_WPIRlkg_n2f32_v4.txt}; 
\addlegendentry{$\Delta_{\mathsf{A}}$};

\addplot [color=black!50!green,dotted,line width=1pt, mark=*, mark options={solid, line width = 0.5pt, fill=white}]table[x=MaxL_A,y=S_A] {\Figs/data/SA-MaxL_WPIRlkg_n2f32_v4.txt}; 
\addlegendentry{$\Delta_{\mathsf{A},\mathsf{P}}$};

\addplot [color=blue,solid,line width=1pt, mark=-*, mark options={solid, line width = 0.5pt, fill=white}]table[x=MaxL_2B,y=S_2B] {\Figs/data/S2B-MaxL_WPIRlkg_n2f32_v4.txt}; 
\addlegendentry{$\Delta_{\mathsf{B}}$};

\addplot [color=black,dotted,line width=1pt, mark=diamond*, mark options={solid, line width = 0.5pt, fill=white}]table[x=MaxL_2W,y=S_2W] {\Figs/data/S2W-MaxL_WPIRlkg_n2f32_v4.txt}; 
\addlegendentry{$\Delta_{\mathsf{B},\mathsf{U}}$};

\end{axis}
\end{tikzpicture}%
%\end{figure}
%%
%\end{frame}

% \end{document}
  \caption{$\const{R}$, $\const{U}$, and $\Delta$ of different WPIR schemes for $\const{M}=32$, as a function of $\rho^{(\mathsf{MaxL})}$. Here, $\const{R}_{\mathsf{B}}=\nicefrac{1}{2}$ is not plotted.}
  \label{fig:tuple-MaxL_n2M32}
\end{figure}
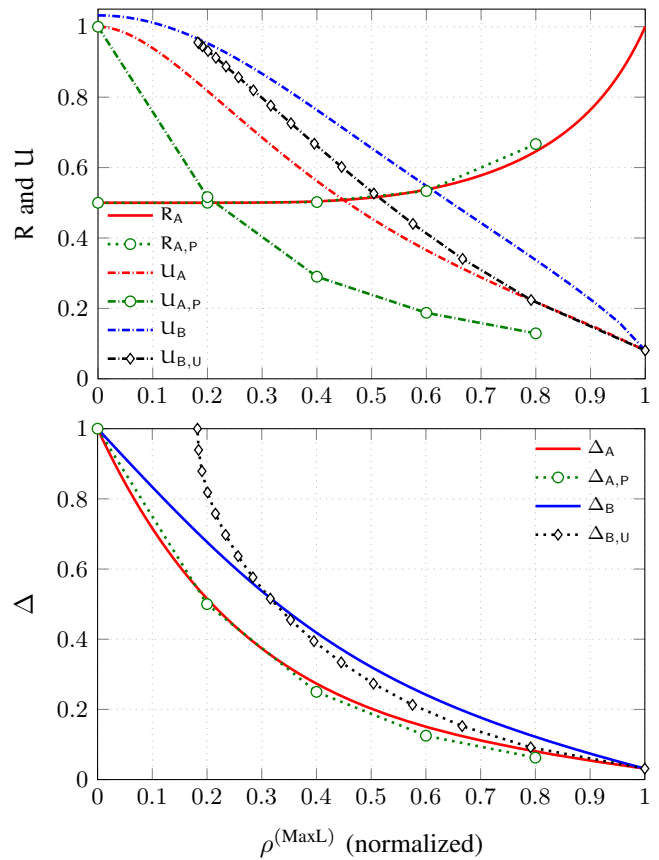

\section{Numerical Results}
\label{sec:numerical-results}

This section consists of three subsections. \cref{sec:WPIRcurves_Mn2} considers the case of two servers and compares the achievable $4$-tuples $\bigl(\const{R},\const{U},\Delta,\rho^{(\cdot)}\bigr)$ for the $(\const{M},2)$ WPIR schemes proposed in \cref{sec:achievability_Mn2-Scheme1-IID-Bernoulli,sec:achievability_Mn-SchemeA-IID-Uniform,sec:achievability_Mn2-Scheme2-IID-Bernoulli,sec:achievability_Mn2-Scheme2-UniformSphere}. Note that as the performance of $\collect{C}^{\mathsf{basic}}$ in Section~\ref{sec:partition-schemes} is inferior to the scheme in~\cref{sec:achievability_Mn-SchemeA-IID-Uniform}, the results of $\collect{C}^{\mathsf{basic}}$ are not presented. \cref{sec:eps-privacy_Mn2} also focuses on the case of two servers and compares our $(\const{M},2)$ WPIR scheme proposed in \cref{sec:achievability_Mn2-Scheme1-IID-Bernoulli_epsP} with the scheme proposed in~\cite{SamyTandonLazos19_1,SamyAttiaTandonLazos21_1} under the $\eps$-privacy metric. \cref{sec:opt-DL_TimeSharing-Scheme1} presents optimized values for the download rate for the time-sharing Scheme~A by numerically solving the convex optimization problem in \eqref{eq:optimization_leakage-download} for both the MI and MaxL privacy metrics and comparisons with the converse bounds from \cref{thm:C_MI-converse,thm:C_MaxL-converse}.

\begin{figure}[htbp!]
  \centering
\definecolor{mycolor1}{rgb}{1.00000,0.00000,1.00000}%
\begin{tikzpicture}
% \pgfplotsset{every tick label/.append style={font=\small}}
\begin{axis}[%
legend cell align=left, 
legend style={fill opacity=0.8, draw opacity=1, text opacity=1, at={(0.97,0.03)}, anchor=south east, draw=white!80!black},
width=0.85\columnwidth,
height=4.725cm, % 5.25cm,
at={(1.011in,0.642in)},
scale only axis,
xmin=0,
xmax=10,
xlabel style={font=\color{white!15!black}},
xlabel={$\rho^{(\eps\mhyph\mathsf{P})}$},
ymin=0.55,
max space between ticks=30pt,
ymax=1,
ylabel style={font=\color{white!15!black}},
ylabel={$\const{R}$},
axis background/.style={fill=white},
grid style={gray,opacity=0.5,dotted},
xmajorgrids,
ymajorgrids,
% axis background/.style={fill=white},
% legend style={draw=none,fill=none, at={(0.593,0.11)}, anchor=south west, legend cell align=left, align=left, draw=white!15!black}
legend style={legend style={draw=none,fill=none}, font=\scriptsize, at={(1.0,0.00)}, anchor=south east},
]
\addplot [color=blue,solid,line width=1pt, mark=-*, mark options={solid, line width = 0.5pt, fill=white}]
  table[row sep=crcr]{%
0	0.571428571428571\\
0.1	0.577772296459757\\
0.2	0.584568247961242\\
0.3	0.591828893786681\\
0.4	0.599563508080412\\
0.5	0.607777575646263\\
0.6	0.616472201898771\\
0.7	0.625643551840154\\
0.8	0.635282344823269\\
0.9	0.645373434122051\\
1	0.655895501082895\\
1.1	0.666820892460051\\
1.2	0.678115626137537\\
1.3	0.689739584674605\\
1.4	0.701646908078209\\
1.5	0.713786587269637\\
1.6	0.72610324851059\\
1.7	0.738538107455241\\
1.8	0.751030060517529\\
1.9	0.763516871935993\\
2	0.775936408225252\\
2.1	0.788227868328861\\
2.2	0.800332958095979\\
2.3	0.812196961660734\\
2.4	0.823769669483567\\
2.5	0.835006132465454\\
2.6	0.845867222698688\\
2.7	0.856319993018576\\
2.8	0.866337838568115\\
2.9	0.875900473247253\\
3	0.884993741595816\\
3.1	0.893609292034828\\
3.2	0.901744140409169\\
3.3	0.90940015360333\\
3.4	0.916583481971013\\
3.5	0.923303966852585\\
3.6	0.929574546006913\\
3.7	0.935410675789848\\
3.8	0.940829784748\\
3.9	0.945850769263185\\
4	0.950493538196147\\
4.1	0.954778610273627\\
4.2	0.958726765307347\\
4.3	0.962358748238326\\
4.4	0.965695023436608\\
4.5	0.96875557560025\\
4.6	0.971559752919254\\
4.7	0.974126147826168\\
4.8	0.976472510573034\\
4.9	0.978615690988057\\
5	0.9805716040172\\
5.1	0.982355214997628\\
5.2	0.983980541002974\\
5.3	0.985460665014805\\
5.4	0.986807760088412\\
5.5	0.988033121078511\\
5.6	0.989147201861307\\
5.7	0.990159656327689\\
5.8	0.991079381725013\\
5.9	0.99191456319153\\
6	0.992672718558927\\
6.1	0.993360742696862\\
6.2	0.993984950841361\\
6.3	0.994551120489758\\
6.4	0.995064531561363\\
6.5	0.995530004618561\\
6.6	0.995951937020306\\
6.7	0.996334336941717\\
6.8	0.996680855242126\\
6.9	0.996994815201563\\
7	0.997279240174243\\
7.1	0.997536879228653\\
7.2	0.99777023085885\\
7.3	0.997981564861626\\
7.4	0.998172942480333\\
7.5	0.998346234919265\\
7.6	0.998503140333074\\
7.7	0.998645199394589\\
7.8	0.998773809541791\\
7.9	0.998890238001178\\
8	0.998995633680482\\
8.1	0.999091038018981\\
8.2	0.999177394878683\\
8.3	0.999255559554541\\
8.4	0.999326306976748\\
8.5	0.999390339173133\\
8.6	0.999448292054772\\
8.7	0.999500741583247\\
8.8	0.999548209373456\\
8.9	0.999591167781678\\
9	0.999630044524536\\
9.1	0.999665226870803\\
9.2	0.999697065444446\\
9.3	0.999725877674103\\
9.4	0.9997519509211\\
9.5	0.9997755453154\\
9.6	0.999796896326259\\
9.7	0.999816217091994\\
9.8	0.999833700531133\\
9.9	0.999849521255171\\
10	0.999863837301374\\
};
\addlegendentry{$\const{R}^{(\eps\mhyph\mathsf{P})}_{\mathsf{LPIR}}$}

\addplot [color=red,solid,line width=1pt, mark=-*, mark options={solid, line width = 0.5pt, fill=white}]
  table[row sep=crcr]{%
0	0.571428571428571\\
0.1	0.579912911825456\\
0.2	0.589038265745732\\
0.3	0.598796731783248\\
0.4	0.609171780869436\\
0.5	0.620137970639929\\
0.6	0.631660890050452\\
0.7	0.643697355952522\\
0.8	0.656195873616795\\
0.9	0.669097361457164\\
1	0.682336127069972\\
1.1	0.695841068041557\\
1.2	0.709537057881429\\
1.3	0.723346466084937\\
1.4	0.7371907528352\\
1.5	0.75099207411722\\
1.6	0.764674832578934\\
1.7	0.778167113430532\\
1.8	0.791401952653384\\
1.9	0.804318396004766\\
2	0.816862320660357\\
2.1	0.828987005586507\\
2.2	0.84065345064199\\
2.3	0.851830456895444\\
2.4	0.86249449089559\\
2.5	0.872629363154484\\
2.6	0.882225755734207\\
2.7	0.891280635688851\\
2.8	0.899796590552308\\
2.9	0.907781119565881\\
3	0.915245910457508\\
3.1	0.922206126859808\\
3.2	0.928679726374149\\
3.3	0.934686824251904\\
3.4	0.940249112970404\\
3.5	0.945389343828241\\
3.6	0.950130873181289\\
3.7	0.954497273122765\\
3.8	0.958512004258149\\
3.9	0.962198146681965\\
4	0.965578184248931\\
4.1	0.968673836658276\\
4.2	0.971505933647447\\
4.3	0.974094325636801\\
4.4	0.976457825405715\\
4.5	0.978614175750415\\
4.6	0.980580038523206\\
4.7	0.982371000941837\\
4.8	0.984001595555957\\
4.9	0.985485330743434\\
5	0.986834729067894\\
5.1	0.988061371250848\\
5.2	0.989175943892521\\
5.3	0.990188289412932\\
5.4	0.991107456979635\\
5.5	0.991941753442626\\
5.6	0.992698793513439\\
5.7	0.99338554860771\\
5.8	0.994008393922337\\
5.9	0.994573153443443\\
6	0.995085142683256\\
6.1	0.995549209026041\\
6.2	0.995969769628374\\
6.3	0.996350846870007\\
6.4	0.996696101390761\\
6.5	0.997008862778272\\
6.6	0.997292157992847\\
6.7	0.99754873763068\\
6.8	0.997781100136394\\
6.9	0.997991514081567\\
7	0.998182038628302\\
7.1	0.998354542296791\\
7.2	0.998510720153889\\
7.3	0.998652109536294\\
7.4	0.998780104417554\\
7.5	0.998895968523052\\
7.6	0.999000847291581\\
7.7	0.999095778776392\\
7.8	0.999181703572721\\
7.9	0.999259473853023\\
8	0.999329861585423\\
8.1	0.999393566005391\\
8.2	0.999451220405365\\
8.3	0.999503398302025\\
8.4	0.999550619036154\\
8.5	0.999593352855596\\
8.6	0.999632025527596\\
8.7	0.999667022522965\\
8.8	0.999698692810877\\
8.9	0.999727352299749\\
9	0.999753286956595\\
9.1	0.999776755634385\\
9.2	0.999797992634303\\
9.3	0.99981721002744\\
9.4	0.999834599758201\\
9.5	0.999850335549737\\
9.6	0.999864574629836\\
9.7	0.999877459294038\\
9.8	0.999889118321203\\
9.9	0.999899668255348\\
10	0.999909214566305\\
};
\addlegendentry{$\const{R}^{(\eps\mhyph\mathsf{P})}_{\mathsf{A}}$}

\addplot [color=orange,solid,line width=1pt, mark=-*, mark options={solid, line width = 0.5pt, fill=white}]
  table[row sep=crcr]{%
0	0.571428571428571\\
0.1	0.603463373838746\\
0.2	0.634137369359315\\
0.3	0.663300628514236\\
0.4	0.690849978848276\\
0.5	0.716725041813976\\
0.6	0.740902986955971\\
0.7	0.763392640112139\\
0.8	0.78422846813394\\
0.9	0.803464828349398\\
1	0.821170739885324\\
1.1	0.83742532043592\\
1.2	0.85231394274509\\
1.3	0.865925100884344\\
1.4	0.878347934770567\\
1.5	0.889670338013932\\
1.6	0.899977564465841\\
1.7	0.909351248486265\\
1.8	0.917868759457555\\
1.9	0.925602819771469\\
2	0.932621325524275\\
2.1	0.938987319235994\\
2.2	0.94475907332465\\
2.3	0.949990251430239\\
2.4	0.954730121850269\\
2.5	0.959023803322578\\
2.6	0.962912528260899\\
2.7	0.966433912444724\\
2.8	0.969622223231741\\
2.9	0.972508640739373\\
3	0.975121508260355\\
3.1	0.977486569548021\\
3.2	0.979627191624807\\
3.3	0.981564572510673\\
3.4	0.983317933799958\\
3.5	0.984904698386133\\
3.6	0.986340653883795\\
3.7	0.987640102456961\\
3.8	0.988815997856232\\
3.9	0.989880070512885\\
4	0.990842941549313\\
4.1	0.991714226552837\\
4.2	0.99250262993151\\
4.3	0.993216030631853\\
4.4	0.99386155995364\\
4.5	0.99444567214892\\
4.6	0.994974208443573\\
4.7	0.995452455071244\\
4.8	0.995885195862592\\
4.9	0.996276759887978\\
5	0.996631064609427\\
5.1	0.996951654958142\\
5.2	0.997241738717081\\
5.3	0.99750421855407\\
5.4	0.997741721019686\\
5.5	0.997956622795338\\
5.6	0.998151074450736\\
5.7	0.998327021945894\\
5.8	0.998486226090915\\
5.9	0.998630280156886\\
6	0.998760625813055\\
6.1	0.998878567549006\\
6.2	0.998985285725591\\
6.3	0.999081848384781\\
6.4	0.999169221936334\\
6.5	0.999248280827975\\
6.6	0.999319816295699\\
6.7	0.999384544281646\\
6.8	0.999443112598686\\
6.9	0.999496107413354\\
7	0.999544059111962\\
7.1	0.999587447608554\\
7.2	0.99962670714781\\
7.3	0.999662230650926\\
7.4	0.999694373647974\\
7.5	0.999723457836069\\
7.6	0.999749774298943\\
7.7	0.999773586420161\\
7.8	0.999795132519107\\
7.9	0.999814628236139\\
8	0.999832268690767\\
8.1	0.999848230434456\\
8.2	0.999862673217603\\
8.3	0.999875741588364\\
8.4	0.999887566339332\\
8.5	0.999898265816548\\
8.6	0.999907947103946\\
8.7	0.999916707095084\\
8.8	0.99992463346288\\
8.9	0.999931805537076\\
9	0.999938295098192\\
9.1	0.999944167095929\\
9.2	0.99994948029921\\
9.3	0.999954287884356\\
9.4	0.999958637967292\\
9.5	0.999962574085109\\
9.6	0.999966135631793\\
9.7	0.999969358252502\\
9.8	0.999972274200305\\
9.9	0.999974912658988\\
10	0.99997730003513\\
};
\addlegendentry{$\const{R}^{(\eps\mhyph\mathsf{P})}_{\mathsf{UB}}$}

\end{axis}
\end{tikzpicture}%
  \caption{The rates $\const{R}^{(\eps\mhyph\mathsf{P})}_{\mathsf{LPIR}}$, $\const{R}^{(\eps\mhyph\mathsf{P})}_{\mathsf{A}}$, and $\const{R}^{(\eps\mhyph\mathsf{P})}_{\mathsf{UB}}$ for $(\const{M},n)=(3,2)$, as a function of $\rho^{(\eps\mhyph\mathsf{P})}$.}
  \label{fig:cmprsn_epsPLkg_rate_WPIRM3n2}
\end{figure}
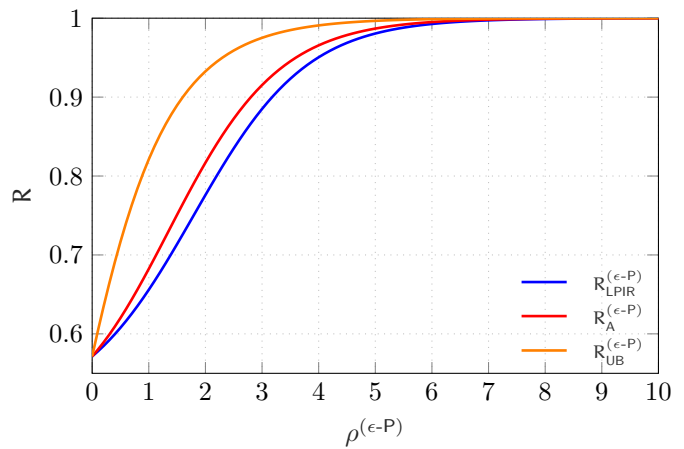

\begin{figure}[htbp!]
  \centering
  \input{\Figs/WPIR_download_rate_scheme1_under_eps_privacyM10n2_v1.tex}
  \caption{The rates $\const{R}^{(\eps\mhyph\mathsf{P})}_{\mathsf{LPIR}}$, $\const{R}^{(\eps\mhyph\mathsf{P})}_{\mathsf{A}}$, and $\const{R}^{(\eps\mhyph\mathsf{P})}_{\mathsf{UB}}$ for $(\const{M},n)=(10,2)$, as a function of $\rho^{(\eps\mhyph\mathsf{P})}$.}
  \label{fig:cmprsn_epsPLkg_rate_WPIRM10n2}
\end{figure}

\subsection{$(\const{M},2)$ WPIR Schemes}
\label{sec:WPIRcurves_Mn2}

Four $(\const{M},2)$ WPIR schemes presented in \cref{sec:achievability_Mn2-Scheme1-IID-Bernoulli,sec:achievability_Mn-SchemeA-IID-Uniform,sec:achievability_Mn2-Scheme2-IID-Bernoulli,sec:achievability_Mn2-Scheme2-UniformSphere} are illustrated. More specifically, the $4$-tuples $\bigl(\const{R}_{\mathsf{A}},\const{U}_{\mathsf{A}},\Delta_{\mathsf{A}},\rho^{(\cdot)}_{\mathsf{A}}\bigr)$ achieved by the $(\const{M},2)$ Scheme~A with $\{S_j\}_{j=1}^{\const{M}-1}$ i.i.d.~according to $\Bernoulli{p}$, $\bigl(\const{R}_{\mathsf{A},\mathsf{P}},\const{U}_{\mathsf{A},\mathsf{P}},\Delta_{\mathsf{A},\mathsf{P}},\rho^{(\cdot)}_{\mathsf{A},\mathsf{P}}\bigr)$ achieved by the $(\const{M},n)$ partition Scheme~A using the $(\nicefrac{\const{M}}{\eta},n)$ Scheme~A with $\{S_j\}_{j=1}^{\nicefrac{\const{M}}{\eta}-1}$ i.i.d.~according to $\Uniform{[0:n-1]}$ as a subscheme, $\bigl(\nicefrac{1}{2},\const{U}_{\mathsf{B}},\Delta_{\mathsf{B}},\rho^{(\cdot)}_{\mathsf{B}}\bigr)$ achieved by the $(\const{M},2)$ Scheme~B with $\{S_j\}_{j=1}^{\const{M}}$ i.i.d.~according to $\Bernoulli{p}$, and $\bigl(\nicefrac{1}{2},\const{U}_{\mathsf{B},\mathsf{U}},\Delta_{\mathsf{B},\mathsf{U}},\rho^{(\cdot)}_{\mathsf{B},\mathsf{U}}\bigr)$ achieved by the $(\const{M},2)$ Scheme~B with $\vect{S}\sim\Uniform{\set{B}_{w,\const{M}}}$, are presented for comparison. For the sake of illustration, the information leakage $\rho^{(\mathsf{MI})}$ is normalized by $\log_2{\const{M}}$ bits, while the upload cost and access complexity are normalized by $2(\const{M}-1)$ and $\const{M}$, respectively. $2(\const{M}-1)$ and $\const{M}$ are the upload cost and access complexity of the PIR capacity-achieving scheme presented in \cite{TianSunChen19_1} for the case of two servers. The upload cost $2(\const{M}-1)$ is optimal among all so-called \emph{decomposable} PIR capacity-achieving schemes \cite{TianSunChen19_1}.\footnote{Based on \cite[Def.~2]{TianSunChen19_1}, all existing PIR schemes in the literature are decomposable.}

Fig.~\ref{fig:tuple-MI_n2M32} presents the results of the four WPIR schemes for the case of $\const{M}=32$ files and leakage metric $\rho^{(\mathsf{MI})}$. We can see that Scheme~A yields the best performance in terms of download rate, upload cost, and access complexity  for all values of the information leakage. Note that the IR rate of Scheme~B with different $\vect{S}$ is always equal to $\nicefrac{1}{2}$. The results of the four  WPIR schemes for WIL and MaxL  are provided in \cref{fig:tuple-WIL_n2M32,fig:tuple-MaxL_n2M32}, respectively. For WIL, Scheme~A performs best among the four schemes for all values of the information leakage in terms of download rate, upload cost, and access complexity. However, for MaxL the partition Scheme~A (from \cref{thm:SchemeA_Mn-IID-Uniform}) has a comparable performance to Scheme~A with $\{S_j\}_{j=1}^{\const{M}-1}$ i.i.d.~according to $\Bernoulli{p}$ (from \cref{thm:Scheme1_Mn2-IID-Bernoulli}) for both download rate and access complexity. In particular, for $\rho^{(\textnormal{MaxL})}=0.8$ it exhibits a slightly higher download rate, whereas for  $0.2\le \rho^{(\textnormal{MaxL})}\le0.8$ it achieves a lower access complexity. On the other hand, it yields a significantly lower upload cost for all values of the information leakage.
% On the other hand, we also present the results in Fig.~\ref{fig:tuple-MaxL_n2M32} for different WPIR schemes under the
% $\rho^{(\mathsf{MaxL})}$ metric.  One can see that under this metric, the partition scheme presented in
% Section~\ref{sec:achievability_Mn-SchemeA-IID-Uniform} behaves rather well compared to Scheme~A with
% $\{S_j\}_{j=1}^{\const{M}-1}$ i.i.d.~according to $\Bernoulli{p}$.

\begin{figure}[t!]
  \centering
  \input{\Figs/cmprsn_MILkg_rate_WPIRn2M2.tex}
  \caption{The optimized rate $\bar{\const{R}}^{(\mathsf{MI})}_{\mathsf{opt}}$ for the time-sharing Scheme~A and $\const{R}^{(\mathsf{MI})}_{\mathsf{UB}}$ for $(\const{M},n)=(2,2)$, as a function of $\rho^{(\mathsf{MI})}$.}
  \label{fig:cmprsn_MILkg_rate_WPIRn2M2}
\end{figure}

\begin{figure}[t!]
  \centering
  \input{\Figs/cmprsn_MILkg_rate_WPIRn2M6.tex}
  \caption{The optimized rate $\bar{\const{R}}^{(\mathsf{MI})}_{\mathsf{opt}}$, $\const{R}_{\mathsf{A}}$, and $\const{R}_{\mathsf{A},\mathsf{P}}$ for $(\const{M},n)=(6,2)$, as a function of $\rho^{(\mathsf{MI})}$.}
  \label{fig:cmprsn_MILkg_rate_WPIRn2M6}
\end{figure}

\begin{figure}[t!]
  \centering
  \input{\Figs/cmprsn_MILkg_rate_WPIRn3M6.tex}
  \caption{The optimized rate $\bar{\const{R}}^{(\mathsf{MI})}_{\mathsf{opt}}$ and $\const{R}_{\mathsf{A},\mathsf{P}}$ for $(\const{M},n)=(6,3)$, as a function of $\rho^{(\mathsf{MI})}$.}
  \label{fig:cmprsn_MILkg_rate_WPIRn3M6}
\end{figure}

\subsection{$\eps$-Privacy for $(\const{M},2)$ WPIR Schemes}
\label{sec:eps-privacy_Mn2}

In this subsection, the achievable rates and the upper bound under the $\eps$-privacy metric presented in \cref{sec:Scheme1_epsP} are demonstrated. We evaluate the rate $\const{R}^{(\eps\mhyph\mathsf{P})}_{\mathsf{LPIR}}$, the upper bound $\const{R}^{(\eps\mhyph\mathsf{P})}_\mathsf{UB}$, and the rate $\const{R}^{(\eps\mhyph\mathsf{P})}_{\mathsf{A}}$ in \eqref{eq:achievable-rate_LPIR}, \eqref{eq:upper-bound_LPIR}, and \eqref{eq:WPIRrate_Scheme1_Mn2-IID-Bernoulli_epsP}, respectively. Fig.~\ref{fig:cmprsn_epsPLkg_rate_WPIRM3n2} presents the results for the case of $\const{M}=3$ files. In Fig.~\ref{fig:cmprsn_epsPLkg_rate_WPIRM10n2}, the corresponding curves for $(\const{M},n)=(10,2)$ are provided. It can be seen that our proposed $(\const{M},2)$ Scheme~A outperforms the leaky PIR scheme presented in~\cite{SamyTandonLazos19_1,SamyAttiaTandonLazos21_1}, in terms of download rate. Note that for the special case of $\const{M}=2$ files, we have $\const{R}^{(\eps\mhyph\mathsf{P})}_{\mathsf{LPIR}}=\const{R}^{(\eps\mhyph\mathsf{P})}_{\mathsf{A}}$. Moreover, the converse bound $\const{R}^{(\eps\mhyph\mathsf{P})}_{\mathsf{UB}}$ is in general not tight.

\subsection{Optimized Rates for the Time-Sharing Scheme~A}
\label{sec:opt-DL_TimeSharing-Scheme1}

In this subsection, we give the maximum download rate under a leakage constraint for the time-sharing Scheme~A described in Section~\ref{sec:minimization-leakage_Scheme1-TimeSharing} with both the MI and MaxL privacy metrics.  Since for both metrics optimizing the download rate is a convex problem (see \eqref{eq:optimization_leakage-download}), the optimal solutions can be obtained by using the CVXPY Python-embedded modeling language for convex optimization problems~\cite{DiamondBoyd16_1,Agrawal-etal18_1}. The optimal corresponding rate obtained from~\eqref{eq:optimization_leakage-download} is denoted by $\bar{\const{R}}^{(\cdot)}_{\mathsf{opt}}$. %, and the rate upper bound derived from~\eqref{eq:D-converse_MI} or \eqref{eq:D-converse_MaxL} is denoted by $\const{R}^{(\cdot)}_{\mathsf{UB}}$.
Unless specified otherwise, all solutions are numerically computed. Moreover, if the converse bound $\const{R}^{(\cdot)}_{\mathsf{UB}}$ (see \cref{thm:C_MI-converse,thm:C_MaxL-converse}) is trivial for all leakage constraints, i.e., $\const{R}^{(\cdot)}_{\mathsf{UB}}\geq 1$, we do not include it in the figures.

Under the MI privacy metric, Fig.~\ref{fig:cmprsn_MILkg_rate_WPIRn2M2} compares the optimal rate-leakage tradeoff curve for the canonical case of $(\const{M},n)=(2,2)$ to the converse bound $\const{R}^{(\mathsf{MI})}_{\mathsf{UB}}$ from  \cref{thm:C_MI-converse}, which shows that in general it is not tight. We remark that the optimal curve is equal to the curve presented in Example~\ref{ex:Ex1_WPIR_M2n2}, and it can also be shown that the analytical optimal solution can be derived directly from~\eqref{eq:optimization_leakage-download}. In Fig.~\ref{fig:cmprsn_MILkg_rate_WPIRn2M6},  for $(\const{M},n)=(6,2)$, the download rate from \cref{thm:Scheme1_Mn2-IID-Bernoulli,thm:SchemeA_Mn-IID-Uniform} is plotted as a function of the information leakage, together with the optimal download rate $\bar{\const{R}}^{(\mathsf{MI})}_{\mathsf{opt}}$ for the time-sharing Scheme~A obtained from  \eqref{eq:optimization_leakage-download}. The comparisons show that Scheme~A with $\{S_j\}_{j=1}^{\const{M}-1}$ i.i.d.\ according to $\Bernoulli{p}$ (from \cref{thm:Scheme1_Mn2-IID-Bernoulli}) exhibits a download rate that is close to being optimal. On the other hand, partition Scheme~A (from \cref{thm:SchemeA_Mn-IID-Uniform}) performs quite far from the  optimal tradeoff curve.  In Fig.~\ref{fig:cmprsn_MILkg_rate_WPIRn3M6}, the corresponding curves for $(\const{M},n)=(6,3)$ (excluding  the curve from \cref{thm:Scheme1_Mn2-IID-Bernoulli}, which assumes $n=2$) are presented. Again, partition Scheme~A performs far from the optimal tradeoff curve. Note that for both $(\const{M},n)=(6,2)$ and $(6,3)$ the converse bound from \cref{thm:C_MI-converse} is trivial.% shows that the partition scheme behaves very bad.

In \cref{fig:cmprsn_MaxLkg_rate_WPIRn2M3,fig:cmprsn_MaxLkg_rate_WPIRn2M6,fig:cmprsn_MaxLkg_rate_WPIRn3M6}, the corresponding curves for the MaxL privacy metric are depicted. In particular, the figures show results for $(\const{M},n)=(3,2)$, $(6,2)$, and $(6,3)$, respectively. Note that with the MaxL privacy metric the converse bound from \cref{thm:C_MaxL-converse} is tight for the canonical case of $(\const{M},n)=(2,2)$  (see Theorem~\ref{thm:WPIRcapacity_M2n2_MaxL}). %\cref{thm:WPIRcapacity_M2n2_MaxL} gives the optimal download rate for all leakages. 
Hence, in contrast to \cref{fig:cmprsn_MILkg_rate_WPIRn2M2}, where $(\const{M},n)=(2,2)$, we use $(\const{M},n)=(3,2)$ in \cref{fig:cmprsn_MaxLkg_rate_WPIRn2M3}.
%
%, since Theorem~\ref{thm:WPIRcapacity_M2n2_MaxL} shows that for the canonical case of $(\const{M},n)=(2,2)$, the converse bound can be achieved, we plot in Fig.~\ref{fig:cmprsn_MaxLkg_rate_WPIRn2M3} the optimal curve $\bar{\const{R}}^{(\mathsf{MaxL})}_{\mathsf{opt}}$ and the converse bound $\const{R}^{(\mathsf{MaxL})}_{\mathsf{UB}}$ for the case of $(\const{M},n)=(3,2)$.
%As for the MI privacy metric, the converse bound from \cref{thm:C_MaxL-converse} is  in general not tight. In Fig.~\ref{fig:cmprsn_MaxLkg_rate_WPIRn2M6}, we depict the curves from Theorem~\ref{thm:Scheme1_Mn2-IID-Bernoulli} and Theorem~\ref{thm:SchemeA_Mn-IID-Uniform}, and the optimal curve $\bar{\const{R}}^{(\mathsf{MaxL})}_{\mathsf{opt}}$ for $(\const{M},n)=(6,2)$. 
From \cref{fig:cmprsn_MaxLkg_rate_WPIRn2M6}, for $(\const{M},n)=(6,2)$, Scheme~A with $\{S_j\}_{j=1}^{\const{M}-1}$ i.i.d.\ according to $\Bernoulli{p}$ (from \cref{thm:Scheme1_Mn2-IID-Bernoulli}) exhibits a lower download rate than partition Scheme~A (from \cref{thm:SchemeA_Mn-IID-Uniform}), which is in contrast to the case of MI leakage where partition Scheme~A performs significantly worse (see \cref{fig:cmprsn_MILkg_rate_WPIRn2M6}). Moreover, the gap to the optimized rate $\bar{\const{R}}^{(\mathsf{MaxL})}_{\mathsf{opt}}$ is higher than with the MI privacy metric. For  $(\const{M},n)=(6,3)$, \cref{fig:cmprsn_MaxLkg_rate_WPIRn3M6} shows that the gap in download rate between the optimized rate $\bar{\const{R}}^{(\mathsf{MaxL})}_{\mathsf{opt}}$ and the rate from partition Scheme~A is smaller than with the MI privacy metric, which indicates that partition Scheme~A performs better with the MaxL privacy metric than with the MI privacy metric.  %For both privacy metric the converse bound is trivial for $(\const{M},n)=(6,3)$.}
For both $(\const{M},n)=(6,2)$ and $(6,3)$ (as for the MI privacy metric) the converse bound from \cref{thm:C_MaxL-converse} is trivial.
%
% Finally, we compare the optimal curve $\bar{\const{R}}^{(\mathsf{MaxL})}_{\mathsf{opt}}$ with $\const{R}_{\mathsf{A},\mathsf{P}}$ in Fig.~\ref{fig:cmprsn_MaxLkg_rate_WPIRn3M6}  for the case of $(\const{M},n)=(6,3)$.
%
% plots for MaxL
\begin{figure}[t!]
  \centering
  \input{\Figs/cmprsn_MaxLkg_rate_WPIRn2M3.tex}
  \caption{The optimized rate $\bar{\const{R}}^{(\mathsf{MaxL})}_{\mathsf{opt}}$ for the time-sharing Scheme~A and $\const{R}^{(\mathsf{MaxL})}_{\mathsf{UB}}$ for $(\const{M},n)=(3,2)$, as a function of $\rho^{(\mathsf{MaxL})}$.}
  \label{fig:cmprsn_MaxLkg_rate_WPIRn2M3}
\end{figure}

\begin{figure}[t!]
  \centering
  \input{\Figs/cmprsn_MaxLkg_rate_WPIRn2M6.tex}
  \caption{The optimized rate $\bar{\const{R}}^{(\mathsf{MaxL})}_{\mathsf{opt}}$, $\const{R}_{\mathsf{A}}$, and $\const{R}_{\mathsf{A},\mathsf{P}}$ for $(\const{M},n)=(6,2)$, as a function of $\rho^{(\mathsf{MaxL})}$.}
  \label{fig:cmprsn_MaxLkg_rate_WPIRn2M6}
\end{figure}

\begin{figure}[t!]
  \centering
  \input{\Figs/cmprsn_MaxLkg_rate_WPIRn3M6.tex}
  \caption{The optimized rate $\bar{\const{R}}^{(\mathsf{MaxL})}_{\mathsf{opt}}$ and $\const{R}_{\mathsf{A},\mathsf{P}}$ for $(\const{M},n)=(6,3)$, as a function of $\rho^{(\mathsf{MaxL})}$.}
  \label{fig:cmprsn_MaxLkg_rate_WPIRn3M6}
\end{figure}

\section{Conclusion}
\label{sec:conclusion}

We presented the first study of the tradeoffs that can be achieved by relaxing the perfect privacy requirement of PIR, referred to as WPIR, for the case of multiple replicated noncolluding servers. Two WPIR schemes based on two different PIR protocols, named Scheme~A and Scheme~B, and a family of schemes based on partitioning were proposed. The proposed model shows that by relaxing the perfect privacy requirement, the download rate, the upload cost, and the access complexity can be improved. In addition, we showed that Scheme~A achieves an improved download rate compared to the leaky PIR scheme proposed by Samy \emph{et al.} under the $\epsilon$-privacy metric. Under the MI and MaxL privacy metrics and with a practical restriction on the alphabet size of queries and answers, we provided an information-theoretic converse bound on the download rate. For the MaxL privacy metric and for two servers and two files, the converse bound is tight, giving the WPIR capacity in this special case. Numerous numerical results were presented, comparing the performance of the proposed schemes and their gap to the new converse bound.

Many interesting directions can be studied as future work. First of all, the derivation of a better converse bound on the download rate, as well as on the upload cost and the access complexity, is worth further investigation for general cases of $(\const{M},n)$. On the other hand, %similar to the current research line of PIR, 
practical variants of WPIR include colluding, Byzantine, and unresponsive servers, are all important topics for future research, as well as WPIR for coded DSSs and WPIR with secure storage.

\section*{Acknowledgment}

The authors would like to thank the two anonymous reviewers and the Associate Editor Prof.~Amos Beimel for their thoughtful and valuable comments.

%%%%%%%%%%%%%%%%%%%%%%%%%%%%%%%%%%%%%%%%%%%%%%%%%%%%%%%%%%%%%%%%%%%%% 
\appendices

%%%%%%%%%%%%%%%%%%%%%%%%%%%%%%%%%%%%%%%%%%%%%%%%%%%%%%%%%%%%%%%%%%%%%
\section{Proof of Theorem~\ref{thm:partition-schemes}}
\label{sec:proof_partition-schemes}

Without loss of generality, denote the requested file index $M$ by $M\equiv(M_J,J)$, where $M_J$ denotes the requested file index in the $J$-th partition. The MI based leakage at the $l$-th server, $l\in[n]$, is given as
\begin{IEEEeqnarray}{rCl}    
  \eMI{M}{\vect{Q}_l}& = &\eHP{M}-\eHPcond{M}{\vect{Q}_l}
  \nonumber\\
  & = &\eHP{M_J,J}-\eHPcond{M_J,J}{\widetilde{\vect{Q}}_l,J}
  \nonumber\\
  & = &\eHP{J}+\eHPcond{M_J}{J}-\eHPcond{M_J}{\widetilde{\vect{Q}}_l,J}
  \nonumber\\
  & \stackrel{(a)}{=}  &\eHP{J}+\eHP{M_J}-\eHPcond{M_J}{\widetilde{\vect{Q}}_l}\nonumber\\
  & = &\log_2\eta+\eMI{M_J}{\widetilde{\vect{Q}}_l},\label{eq:use_uniformM}
\end{IEEEeqnarray}
where $(a)$ follows since $M_J$ and $J$ are assumed to be uniform RVs, and hence knowing $J$ does not reveal any information about the requested file index $M_J$. Using \eqref{eq:use_uniformM} in \eqref{eq:MI_leakge_def} gives  $\rho^{(\mathsf{MI})}(\collect{C})$. Using a similar argument as above, the expressions for $\const{R}(\collect{C})$, $\const{U}(\collect{C})$, $\Delta(\collect{C})$, $\rho^{(\mathsf{WIL})}(\collect{C})$, and $\rho^{(\mathsf{MaxL})}(\collect{C})$ can be derived accordingly.

\section{Proof of Lemma~\ref{lem:PIR_Scheme1}}
\label{sec:proof_PIR_Scheme1}

Observe that if $\{S_j\}_{j=1}^{\const{M}-1}$ are i.i.d.\ according to~$\Uniform{[0:n-1]}$, then  for every
$m,m'\in [\const{M}]$ with $m\neq m'$, it holds that
\begin{IEEEeqnarray*}{c}
  \Prvcond{\vect{Q}_l=\vect{q}_l}{M=m}=\Prvcond{\vect{Q}_l=\vect{q}_l}{M=m'}=\Bigl(\frac{1}{n}\Bigr)^{\const{M}-1}
\end{IEEEeqnarray*}
for all $\vect{q}_l\in\set{Q}_l$, $l\in[n]$. Moreover, from the answer construction of \eqref{eq:Al_Scheme1}, the IR rate is
\begin{IEEEeqnarray}{rCl}
  \const{R}& = &\frac{\beta\log_2{2}}{\log_2{2}\sum_{l=1}^n\E[\vect{Q}_l]{\ell_l(\vect{Q}_l)}} \nonumber\\
  &=&\frac{n-1}{\bigl[1-P_{\vect{Q}_1}(\vect{0})\bigr]+\sum_{l=2}^n\E[\vect{Q}_l]{\ell_l(\vect{Q}_l)}}
  \nonumber\\
  & = &\frac{n-1}{\bigl(1-\frac{1}{n^{\const{M}-1}}\bigr)+(n-1)}=\frac{1-\frac{1}{n}}{1-\frac{1}{n^\const{M}}},
  \nonumber
\end{IEEEeqnarray}
which is equal to the $n$-server PIR capacity for $\const{M}$ files.

\section{Proof of Theorem~\ref{thm:Scheme1_Mn2-IID-Bernoulli}}
\label{sec:proof_Mn2-Scheme1-IID-Bernoulli}

%Consider Scheme~A in Section~\ref{sec:Scheme1} for $\const{M}\geq 2$ and $n=2$, we select that $\{S_j\}_{j=1}^{\const{M}-1}$ is i.i.d. and $S_j\sim\Bernoulli{p}$, $0\leq p\leq \frac{1}{2}$. In the following, we show that the achievable $4$-tuple with $\rho^{(\cdot)}$ is given as Theorem~\ref{thm:Scheme1_Mn2-IID-Bernoulli}.
From the theorem statement, the entries $\{S_j\}_{j=1}^{\const{M}-1}$ of the random strategy $\vect{S}$ are assumed to be i.i.d.\ according to $\Bernoulli{p}$, $0\leq p\leq \frac{1}{2}$.
Hence, $P_{\vect{Q}_1}(\vect{0})=P_{\vect{S}}(\vect{0})=(1-p)^{\const{M}-1}$, and we have % the IR rate $\const{R}(\collect{C}_{\mathsf{A}})$ becomes 
\begin{IEEEeqnarray*}{c}
  \const{R}(\collect{C}_{\mathsf{A}})=\frac{1}{\bigl[1-(1-p)^{\const{M}-1}\bigr]+1}
\end{IEEEeqnarray*}
from the general formula in~\eqref{eq:IRrate_Scheme1}. % and we have
%with $\{S_j\}_{j=1}^{\const{M}-1}$ being i.i.d.~$\sim\Bernoulli{p}$.

For the upload cost, access complexity, and the information leakage metrics, we first derive the PMF of $\vect{Q}_l$, $l=1,2$. Let us consider a query $\vect{q}_l$ to the $l$-th server that has $\Hwt{\vect{q}_l}=w$. Due to the query generation, we have
\begin{IEEEeqnarray}{c}
  P_{\vect{Q}_l|M}(\vect{q}_l|m)=
  \begin{cases}
    (1-p)^{\const{M}-w-1}p^w & \textnormal{if }m\in [\const{M}]\setminus\Spt{\vect{q}_l},
    \\
    (1-p)^{\const{M}-w}p^{w-1} & \textnormal{if }m\in \Spt{\vect{q}_l}.
  \end{cases}
  \nonumber\\*\IEEEeqnarraynumspace\label{eq:PQl-M_Scheme1}
\end{IEEEeqnarray}
By using the law of total probability, we obtain
\begin{IEEEeqnarray*}{rCl}  
  P_{\vect{Q}_l}(\vect{q}_l)
  & = &\sum_{m'=1}^{\const{M}}\frac{1}{\const{M}}P_{\vect{Q}_l|M}(\vect{q}_l|m')
  \nonumber\\
  % \label{eq:law_TotalProbability}\\
  % & = &\frac{1}{\const{M}}\Bigl[\Prs{\bigl\{\vect{Q}_l=\vect{q}_l\colon m\in [\const{M}]\setminus\Spt{\vect{q}_l}\bigr\}}
  % \nonumber\\
  % && \qquad +\>\Prs{\bigl\{\vect{Q}_l=\vect{q}_l\colon m\in\Spt{\vect{q}_l}\bigr\}}\Bigr]
  % \label{eq:use_query-generation_Scheme1}\\
  % & = &\frac{1}{\const{M}}\Bigl[\Prs{\bigl\{S=(\vect{q}_l)_{[\const{M}]\setminus\{m\}} \colon m\in
  %   [\const{M}]\setminus\Spt{\vect{q}_l}\bigr\}}
  % \nonumber\\
  % && \qquad +\>\Prs{\bigl\{S=(\vect{q}_l)_{[\const{M}]\setminus\{m\}} \colon m\in\Spt{\vect{q}_l}\bigr\}}\Bigr]
  % \\
  & = &\frac{1}{\const{M}}\Biggl[\binom{\const{M}-w}{1}\cdot(1-p)^{\const{M}-w-1}\cdot p^w
  \nonumber\\
  && \qquad +\>\binom{w}{1}\cdot (1-p)^{\const{M}-w}\cdot p^{w-1}\Biggr]
  % \label{eq:use_mth-query_Scheme1}
  \\
  & = &f(w,p).
\end{IEEEeqnarray*}

From the query generation  (see \cref{sec:query-generation_Scheme1}), it follows that $\Hwt{\vect{q}_1}$ must be even and $\Hwt{\vect{q}_2}$ must be odd for the case of $n=2$ servers, hence, the upload cost is equal to
\begin{IEEEeqnarray*}{rCl}
  \const{U}(\collect{C}_{\mathsf{A}})& = &\eHP{\vect{Q}_1}+\eHP{\vect{Q}_2}
  \nonumber\\
  % & = &-\sum_{\substack{\vect{q}_1\in\set{Q}_1\\ \Hwt{\vect{q}_1}\colon\textnormal{even}}}
  % P_{\vect{Q}_1}(\vect{q}_1)\log_2\bigl(P_{\vect{Q}_1}(\vect{q}_1)\bigr)
  % \nonumber\\
  % && \>-\sum_{\substack{\vect{q}_2\in\set{Q}_2\\ \Hwt{\vect{q}_2}\colon\textnormal{odd}}}
  % P_{\vect{Q}_2}(\vect{q}_2)\log_2\bigl(P_{\vect{Q}_2}(\vect{q}_2)\bigr)
  % \nonumber\\
  & = &-\sum_{w=0}^{\const{M}}\binom{\const{M}}{w}f(w,p)\log_2\bigl(f(w,p)\bigr).
  % \label{eq:use_PMF_Ql}
\end{IEEEeqnarray*}
% where \eqref{eq:use_PMF_Ql} follows directly from~\eqref{eq:Ql-PMF_BernScheme1}.
Further, by the definition in \eqref{eq:def_access}, the access complexity $\Delta(\collect{C}_{\mathsf{A}})=\Delta_{\mathsf{A}}$ follows.

Moreover, we have
\begin{IEEEeqnarray}{rCl}
  \eHPcond{\vect{Q}_l}{M}& = &% \sum_{m=1}^\const{M}\Prv{M=m}\eHPcond{\vect{Q}_l}{M=m}
  % \label{eq:use_chain-rule}
  % \\
  % & = &
  \frac{1}{\const{M}}\sum_{m=1}^\const{M}\eHP{\vect{S}}=(\const{M}-1)\Hb(p),
  \label{eq:use_IID-RVs}
\end{IEEEeqnarray}
where
% \eqref{eq:use_chain-rule} follows from the chain rule of entropy; 
\eqref{eq:use_IID-RVs} holds by the query generation, the fact that the entropy of i.i.d.\ RVs is equal to the sum of the individual entropies,  and
$S_j\sim\Bernoulli{p}$. Hence, we obtain
\begin{IEEEeqnarray*}{rCl}
  \rho^{(\mathsf{MI})}% & = &\frac{\MI{\vect{Q}_1}{M}+\MI{\vect{Q}_2}{M}}{2}
  % \\[1mm]
  & = &\frac{\eHP{\vect{Q}_1}-\eHPcond{\vect{Q}_1}{M}+\eHP{\vect{Q}_2}-\eHPcond{\vect{Q}_2}{M}}{2}
  \\
  & = &\frac{1}{2}\const{U}_{\mathsf{A}}-(\const{M}-1)\Hb(p).
\end{IEEEeqnarray*}

For the WIL metric, applying Bayes' rule, given $M=m$ and a query $\vect{q}_l$ with $\Hwt{\vect{q}_l}=w$, we get
\begin{IEEEeqnarray}{rCl}
  % \IEEEeqnarraymulticol{3}{l}{%
  P_{M|\vect{Q}_l}(m|\vect{q}_l)% }\nonumber\\*\quad%
  % & = & \frac{P_{M}(m)P_{\vect{Q}_l|M}(\vect{q}_l|m)}{P_{\vect{Q}_l}(\vect{q}_l)}
  % \nonumber\\
  % & = &\frac{P_M(m)P_{\vect{Q}_l|M}(\vect{q}_l|m)}{\sum_{m'=1}^\const{M}P_M(m')P_{\vect{Q}_l|M}(\vect{q}_l|m')}
  % \nonumber\\
  & = &\frac{P_{\vect{Q}_l|M}(\vect{q}_l|m)}{\sum_{m'=1}^\const{M}P_{\vect{Q}_l|M}(\vect{q}_l|m')},
  \label{eq:use_equally-likely}
\end{IEEEeqnarray}
where~\eqref{eq:use_equally-likely} holds since the requested file index $M$ is assumed to be uniformly distributed. Finally, since the requested index $m\in [\const{M}]$ either belongs to $[\const{M}]\setminus\Spt{\vect{q}_l}$ or $\Spt{\vect{q}_l}$, it is not too difficult to see that
\begin{IEEEeqnarray}{rCl}
  \IEEEeqnarraymulticol{3}{l}{%
    P_{M|\vect{Q}_l}(m|\vect{q}_l)}\nonumber\\*\quad%
  & = &
  \begin{cases}
    \frac{(1-p)^{\const{M}-w-1}p^w}{(\const{M}-w)(1-p)^{\const{M}-w-1}p^w+w(1-p)^{\const{M}-w}p^{w-1}}
    & \\[2mm]
    & \hspace{-2.5cm}\textnormal{if } m\in[\const{M}]\setminus\Spt{\vect{q}_l},   
    \\[1mm]
    \frac{(1-p)^{\const{M}-w}p^{w-1}}{(\const{M}-w)(1-p)^{\const{M}-w-1}p^w+w(1-p)^{\const{M}-w}p^{w-1}}
    & \\[2mm]
    & \hspace{-2.5cm}\textnormal{if } m\in\Spt{\vect{q}_l},
  \end{cases}
  \IEEEeqnarraynumspace\label{eq:use_Ql_Scheme1}
\end{IEEEeqnarray}
where \eqref{eq:use_Ql_Scheme1} follows from~\eqref{eq:PQl-M_Scheme1}. Note that to compute the entropy $\eHPcond{M}{\vect{Q}_l=\vect{q}_l}$, we only need to know the conditional PMF of $M$ given $\vect{Q}_l=\vect{q}_l$, hence, we can introduce a new RV $M_w\equiv M_{\vect{q}_l}$ with $\Hwt{\vect{q}_l}=w$ that has an equivalent PMF defined as~\eqref{eq:Mw-PMF_Scheme1}. This then gives $\rho_{\mathsf{A}}^{(\mathsf{WIL})}$.

Finally, we derive the expression for $\rho_{\mathsf{A}}^{(\mathsf{MaxL})}$. From~\eqref{eq:PQl-M_Scheme1}, it follows that
\begin{IEEEeqnarray*}{rCl}
  \IEEEeqnarraymulticol{3}{l}{%
    \max_{m\in[\const{M}]}P_{\vect{Q}_l|M}(\vect{q}_l|m)}\nonumber\\*\quad%
  & = &
  \begin{cases}
    (1-p)^{\const{M}-1} & \textnormal{if } \Hwt{\vect{q}_l}=0,
    \\
    (1-p)^{\const{M}-w}p^{w-1} & \textnormal{otherwise}.
  \end{cases}
\end{IEEEeqnarray*}
Moreover, 
\begin{IEEEeqnarray}{rCl}
  \IEEEeqnarraymulticol{3}{l}{%
    p\bigl(2^{\ML{M}{\vect{Q}_1}}-2^{\ML{M}{\vect{Q}_2}}\bigr)}\nonumber\\*\quad%
  & = &p(1-p)^{\const{M}-1}+\sum_{\substack{w\in [2:\const{M}]\\ w\colon\textnormal{even}}}\binom{\const{M}}{w}(1-p)^{\const{M}-w}p^{w}\nonumber\\
  && -\>\sum_{\substack{w\in [\const{M}]\\ w\colon\textnormal{odd}}}\binom{\const{M}}{w}(1-p)^{\const{M}-w}p^{w}
  \nonumber\\
  & = &\sum_{w\in[0:\const{M}]}\binom{\const{M}}{w}(1-p)^{\const{M}-w}(-p)^{w}\nonumber\\
  && -\>(1-p)^{\const{M}}+p(1-p)^{\const{M}-1}
  \nonumber\\
  & = &\bigl[(1-p)-p\bigr]^{\const{M}}-(1-p)^{\const{M}-1}(1-2p)<0,\label{eq:use_binomial-expansion}
\end{IEEEeqnarray}
where \eqref{eq:use_binomial-expansion} follows by binomial expansion. Since \eqref{eq:use_binomial-expansion}   is nonpositive when $0\leq p\leq\nicefrac{1}{2}$, the expression for $\rho^{(\mathsf{MaxL})}_{\mathsf{A}}$ follows immediately.

\section{Proof of Theorem~\ref{thm:SchemeA_Mn-IID-Uniform}}
\label{sec:proof_Mn-SchemeA-IID-Uniform}

Since the $(\nicefrac{\const{M}}{\eta},n)$ Scheme~A with $\{S_j\}_{j=1}^{\nicefrac{\const{M}}{\eta}-1}$ i.i.d.\ according to  $\Uniform{[0:n-1]}$ is used as a subscheme, from Section~\ref{sec:query-generation_SchemeA}, we have
\begin{IEEEeqnarray}{c}
  P_{\vect{Q}_l|M}(\vect{q}_l|(j-1)\nicefrac{\const{M}}{\eta}+m')=\inv{\bigl(n^{\nicefrac{\const{M}}{\eta}-1}\bigr)},\label{eq:Ql_SchemeA_Mn-IID-Uniform}
\end{IEEEeqnarray}
$j\in[\eta]$, $m'\in[\nicefrac{\const{M}}{\eta}]$, $\forall\,\vect{q}_l\in\set{Q}_l$, $l\in [n]$. Similar to Appendix~\ref{sec:proof_Mn2-Scheme1-IID-Bernoulli}, we obtain
\begin{IEEEeqnarray*}{rCl}
  \IEEEeqnarraymulticol{3}{l}{%
    P_{\vect{Q}_l}(\vect{q}_l) \stackrel{(a)}{=}
  \begin{cases}
    \frac{1}{\const{M}}\sum_{m=1}^\const{M} P_{\vect{Q}_l|M}(\vect{q}_l|m) & \textnormal{if } \vect{q}_l=\vect{0},
    \\[1mm]
    \frac{1}{\const{M}}\sum_{m\in\set{P}_j} P_{\vect{Q}_l|M}(\vect{q}_l|m) & \textnormal{otherwise}
  \end{cases}}\\[2mm]
  & = &
  \begin{cases}
    \frac{\const{M}}{\const{M}}\inv{\bigl(n^{\nicefrac{\const{M}}{\eta}-1}\bigr)}& \textnormal{if } \vect{q}_l=\vect{0},
    \\[1mm]
    \frac{\nicefrac{\const{M}}{\eta}}{\const{M}}\inv{\bigl(n^{\nicefrac{\const{M}}{\eta}-1}\bigr)}& \textnormal{otherwise}
  \end{cases}  
  \nonumber\\[1mm]
  & = &
  \begin{cases}
    \inv{\bigl(n^{\nicefrac{\const{M}}{\eta}-1}\bigr)}& \textnormal{if } l=1 \textnormal{ and }\vect{q}_l=\vect{0},
    \\[1mm]
    \inv{\bigl(\eta\cdot n^{\nicefrac{\const{M}}{\eta}-1}\bigr)}& \textnormal{otherwise},
  \end{cases}
  \nonumber% \label{eq:Ql-PMF_BernSchemeA}
\end{IEEEeqnarray*}
where $(a)$ holds since in Scheme~A the user can send the all-zero query to the first server to request any file in any partition group. Since $P_{\vect{Q}_1}(\vect{0})=\inv{\bigl(n^{\nicefrac{\const{M}}{\eta}-1}\bigr)}$, using~\eqref{eq:IRrate_Scheme1} gives
\begin{IEEEeqnarray*}{c}
  \const{R}(\collect{C}_{\mathsf{A}}^\mathsf{part})
  =\frac{n-1}{\Bigl[1-\inv{\bigl(n^{\nicefrac{\const{M}}{\eta}-1}\bigr)}\Bigr]+(n-1)}=\const{R}_{\mathsf{A},\mathsf{P}}.
\end{IEEEeqnarray*}

For the upload cost, since there are $\eta\bigl[n^{\nicefrac{\const{M}}{\eta}-1}-1\bigr]$ equally-likely nonzero queries in $\set{Q}_1$, it can be shown that 
\begin{IEEEeqnarray}{rCl}
  \eHP{\vect{Q}_1}
  % & = &-P_{\vect{Q}_1}(\vect{0})\log_2\bigl(P_{\vect{Q}_1}(\vect{0})\bigr)-\sum_{\vect{q}_1\neq\vect{0}}
  % P_{\vect{Q}_1}(\vect{q}_l)\log_2\bigl(P_{\vect{Q}_1}(\vect{q}_l)\bigr)
  % \nonumber\\
  & = &\frac{1}{n^{\nicefrac{\const{M}}{\eta}-1}}\log_2\bigl(n^{\nicefrac{\const{M}}{\eta}-1}\bigr)\nonumber\\
  && +\>\eta\bigl[n^{\nicefrac{\const{M}}{\eta}-1}-1\bigr]\cdot \frac{1}{\eta\cdot n^{\nicefrac{\const{M}}{\eta}-1}}
  \log_2\bigl(\eta\cdot n^{\nicefrac{\const{M}}{\eta}-1}\bigr)
  \nonumber\\
  & = &(\nicefrac{\const{M}}{\eta}-1)\log_2{n}+\log_2{\eta}-\frac{\log_2{\eta}}{n^{\nicefrac{\const{M}}{\eta}-1}}.\nonumber
\end{IEEEeqnarray}

Similarly, we have $\eHP{\vect{Q}_l}=(\nicefrac{\const{M}}{\eta}-1)\log_2{n}+\log_2{\eta}$ for all $l\in [2:n]$. This then gives the expression for $\const{U}(\collect{C}_{\mathsf{A}}^\mathsf{part})=\const{U}_{\mathsf{A},\mathsf{P}}$.

For the access complexity, recall that Scheme~A partitions all $\const{M}$ files into equally-sized $\nicefrac{\const{M}}{\eta}$ groups and Scheme~A with $\{S_j\}_{j=1}^{\nicefrac{\const{M}}{\eta}-1}$ i.i.d.~according to $\Uniform{[0:n-1]}$ is treated as a subscheme for each partition group. Thus, we have
\begin{IEEEeqnarray*}{rCl}
  \Delta(\collect{C}_{\mathsf{A}}^\mathsf{part})& = &\sum_{l=1}^n\E[\vect{Q}_l]{\delta_l(\vect{Q}_l)}% +\E[\vect{Q}_2]{\delta_2}
  \nonumber\\[1mm]
  & = &\sum_{l=1}^n\sum_{\vect{q}_l\in\set{Q}_l}\Hwt{\vect{q}_l} P_{\vect{Q}_l}(\vect{q}_l) % +\sum_{\vect{q}_2}\Hwt{\vect{q}_2}
  % P_{\vect{Q}_2}(\vect{q}_2)
  \nonumber\\
  & \stackrel{(a)}{=} &\sum_{\substack{w\in [0:\nicefrac{\const{M}}{\eta}]\\w>0}}w\cdot\eta\binom{\nicefrac{\const{M}}{\eta}}{w}(n-1)^w
  \frac{1}{\eta\cdot n^{\nicefrac{\const{M}}{\eta}-1}}
  \nonumber
  % \nonumber\\
  % & = &\sum_{\substack{w\in [0:\nicefrac{\const{M}}{\eta}]\\[1mm] w\colon\textnormal{even}}}w\cdot\eta\binom{\nicefrac{\const{M}}{\eta}}{w}
  % \frac{1}{\eta\cdot 2^{\nicefrac{\const{M}}{\eta}-1}}
  % \nonumber\\
  % && \quad +\>\sum_{\substack{w\in [0:\nicefrac{\const{M}}{\eta}]\\[1mm] w\colon\textnormal{odd}}}
  % w\cdot\eta\binom{\nicefrac{\const{M}}{\eta}}{w} \frac{1}{\eta\cdot 2^{\nicefrac{\const{M}}{\eta}-1}}
  \\
  & = &\sum_{\substack{w\in [0:\nicefrac{\const{M}}{\eta}]\\w>0}}w\binom{\nicefrac{\const{M}}{\eta}}{w}(n-1)^{w-1}\frac{n-1}{n^{\nicefrac{\const{M}}{\eta}-1}}
  \nonumber\\
  & \stackrel{(b)}{=}  &\frac{n-1}{n^{\nicefrac{\const{M}}{\eta}-1}}\cdot\bigl(\nicefrac{\const{M}}{\eta}\cdot n^{\nicefrac{\const{M}}{\eta}-1}\bigr)=(n-1)\nicefrac{\const{M}}{\eta},
\end{IEEEeqnarray*}
where $(a)$ follows since for each partition group, \eqref{eq:query-set_Scheme1} indicates that an $(\nicefrac{\const{M}}{\eta},n)$ Scheme~A consists of in total $\binom{\nicefrac{\const{M}}{\eta}}{w}(n-1)^w$ nonzero queries with Hamming weight $w$ in $\bigcup_{l=1}^n\set{Q}_l$; $(b)$ is due to the fact that
$\sum_{h=0}^{z} h\binom{z}{h}\cdot x^{h-1}=\frac{\dd}{\dd x}(1+x)^z=\frac{\dd}{\dd x}\bigl(\sum_{h=0}^{z}\binom{z}{h}x^h\bigr)=z(1+x)^{z-1}$ for some $z\in\Naturals$.

For the information leakage metric $\rho^{(\mathsf{MI})}$, similar to \eqref{eq:use_IID-RVs}, we have
\begin{IEEEeqnarray*}{rCl}
  \eHPcond{\vect{Q}_l}{M}=(\nicefrac{\const{M}}{\eta}-1)\eHP{\nicefrac{1}{n},\ldots,\nicefrac{1}{n}}=(\nicefrac{\const{M}}{\eta}-1)\log_2{n},
\end{IEEEeqnarray*}
$l\in [n]$, and hence $\rho^{(\mathsf{MI})}(\collect{C}_{\mathsf{A}}^\mathsf{part})=\nicefrac{\const{U}_{\mathsf{A},\mathsf{P}}}{n}-(\nicefrac{\const{M}}{\eta}-1)\log_2{n}=\rho^{(\mathsf{MI})}_{\mathsf{A},\mathsf{P}}$ is achievable.

Under the WIL metric $\rho^{(\mathsf{WIL})}$, if $\Hwt{\vect{q}_l}=0$, we obtain $P_{M|\vect{Q}_l}(m|\vect{q}_l)=\frac{1}{\const{M}}$, $\forall\,m\in [\const{M}]$, while $P_{M|\vect{Q}_l}((j-1)\nicefrac{\const{M}}{\eta}+m'|\vect{q}_l)=\frac{1}{\nicefrac{\const{M}}{\eta}}$ for $\Hwt{\vect{q}_l}>0$, $j\in[\eta]$, $m'\in [\nicefrac{\const{M}}{\eta}]$. Therefore, we obtain
\begin{IEEEeqnarray*}{c}
  \eHPcond{M}{\vect{Q}_l=\vect{q}_l}=
  \begin{cases}
    \log_2{\const{M}} & \textnormal{if } \vect{q}_l=\vect{0},
    \\
    \log_2{\bigl(\nicefrac{\const{M}}{\eta}\bigr)}& \textnormal{otherwise}.
  \end{cases}
\end{IEEEeqnarray*}
Because $\log_2{\bigl(\nicefrac{\const{M}}{\eta}\bigr)}\leq\log_2{\const{M}}$ for $\eta\in [\const{M}-1]$, the achievability of $\rho^{(\mathsf{WIL})}(\collect{C}_{\mathsf{A}}^\mathsf{part})=\rho^{(\mathsf{WIL})}_{\mathsf{A},\mathsf{P}}$ follows.

For the privacy metric $\rho^{(\mathsf{MaxL})}$, it follows from~\eqref{eq:Ql_SchemeA_Mn-IID-Uniform} that
\begin{IEEEeqnarray*}{rCl}
  % 2^{\rho^{(\mathsf{MaxL})}}& = &2^{\ML{M}{\vect{Q}_l}}
  % \\
  2^{\ML{M}{\vect{Q}_l}}& = &\sum_{\vect{q}_l\in\set{Q}_l}\max_{m\in[\const{M}]}P_{\vect{Q}_l|M}(\vect{q}_l|m)
  \\
  & = &\card{\set{Q}_l}\cdot\inv{\bigl(n^{\nicefrac{\const{M}}{\eta}-1}\bigr)}\\
  \IEEEeqnarraymulticol{3}{l}{
  = \begin{cases}
  \bigl(1+\eta\bigl(n^{\nicefrac{\const{M}}{\eta}-1}-1\bigr)\bigr) \cdot \inv{\bigl(n^{\nicefrac{\const{M}}{\eta}-1}\bigr)} <  \eta & \text{for $l=1$},\\
  \eta\cdot n^{\nicefrac{\const{M}}{\eta}-1} \cdot  \inv{\bigl(n^{\nicefrac{\const{M}}{\eta}-1}\bigr)} = \eta &  \text{for $l\in [2:n]$},
  \end{cases}}
\end{IEEEeqnarray*}
%for $l\neq 1$, 
which implies that $\rho^{(\mathsf{MaxL})}(\collect{C}_{\mathsf{A}}^\mathsf{part}) = \log_2{\eta} = \rho^{(\mathsf{MaxL})}_{\mathsf{A},\mathsf{P}}$.

\section{Proof of Theorem~\ref{thm:Scheme2_Mn2-UniformSphere}}
\label{sec:proof_Mn2-Scheme2-UniformShpere}

When the user requests the $M$-th file, the $(\const{M},2)$ Scheme~B sends $\vect{Q}_1=\vect{S}+\vect{v}_M$ and $\vect{Q}_2=\vect{S}$ to the respective servers, where $\vect{v}_M$ is the $M$-th $\const{M}$-dimensional unit vector. As the random strategy $\vect{S}\sim\Uniform{\set{B}_{w,\const{M}}}$ is taken, it is not too hard to see that
\begin{IEEEeqnarray*}{rCl}
  P_{\vect{Q}_1}(\vect{q}_1)=
  \begin{cases}
    \frac{w+1}{\binom{\const{M}}{w}\const{M}} & \textnormal{if } \vect{q}_1\in\set{B}_{w+1,\const{M}},
    \\[2mm]
    \frac{\const{M}-(w-1)}{\binom{\const{M}}{w}\const{M}} & \textnormal{if }\vect{q}_1\in\set{B}_{w-1,\const{M}},
  \end{cases}
\end{IEEEeqnarray*}
$\Prv{\vect{Q}_2=\vect{q}_2}=\nicefrac{1}{\binom{\const{M}}{w}}$, $\vect{q}_2\in\set{B}_{w,\const{M}}$, and $M\indep \vect{Q}_2$.

Since $\eHPcond{\vect{Q}_l}{M}=\eHP{\vect{S}}=\log_2{\binom{\const{M}}{w}}$, the results of $\const{U}(\collect{C}_{\mathsf{B}})$, $\Delta(\collect{C}_{\mathsf{B}})$, and $\rho^{(\mathsf{MI})}(\collect{C}_{\mathsf{B}})$ with $\vect{S}\sim\Uniform{\set{B}_{w,\const{M}}}$ can be determined by a simple deduction.

Moreover, one can also show that
\begin{IEEEeqnarray*}{rCl}
  \eHPcond{M}{\vect{Q}_1=\vect{q}_1}
  & = &
  \begin{cases}
    \log_2{(w+1)} & \textnormal{if } \vect{q}_1\in\set{B}_{w+1,\const{M}},
    \\
    \log_2{\bigl(\const{M}-(w-1)\bigr)} & \textnormal{if } \vect{q}_1\in\set{B}_{w-1,\const{M}}.
  \end{cases}
\end{IEEEeqnarray*}
On the other hand, it can be seen that for $\vect{q}_1\in\set{B}_{w+1,\const{M}}$,
\begin{IEEEeqnarray*}{c}
  P_{\vect{Q}_1|M}(\vect{q}_1|m)=
  \begin{cases}
    \nicefrac{1}{\binom{\const{M}}{w}} & \textnormal{if } m\in\Spt{\vect{q}_1},
    \\
    0 & \textnormal{otherwise},
  \end{cases}
\end{IEEEeqnarray*}
and for $\vect{q}_1\in\set{B}_{w-1,\const{M}}$,
\begin{IEEEeqnarray*}{c}
  P_{\vect{Q}_1|M}(\vect{q}_1|m)=
  \begin{cases}
    \nicefrac{1}{\binom{\const{M}}{w}} & \textnormal{if } m\in[\const{M}]\setminus\Spt{\vect{q}_1},
    \\
    0 & \textnormal{otherwise}.
  \end{cases}
\end{IEEEeqnarray*}
Therefore, from the above we obtain the expressions for $\rho^{(\mathsf{WIL})}_{\mathsf{B},\mathsf{U}}$ and $\rho^{(\mathsf{MaxL})}_{\mathsf{B},\mathsf{U}}$.

\section{Proof of Theorem~\ref{thm:Scheme1_Mn2-IID-Bernoulli_epsP}}
\label{sec:proof_Mn2-Scheme1-IID-Bernoulli_epsP}

From \eqref{eq:PQl-M_Scheme1} in Appendix~\ref{sec:proof_Mn2-Scheme1-IID-Bernoulli}, we know that 
\begin{IEEEeqnarray*}{rCl}
  \IEEEeqnarraymulticol{3}{l}{%
    \max_{m,m'\in[\const{M}]}\frac{P_{\vect{Q}_l|M}(\vect{q}_l|m)}{P_{\vect{Q}_l|M}(\vect{q}_l|m')}}
  \nonumber\\*[1mm]\quad%
  & = &
  \begin{cases}
    \max\left\{\frac{(1-p)^{\const{M}-w-1}p^w}{(1-p)^{\const{M}-w}p^{w-1}},
      \frac{(1-p)^{\const{M}-w}p^{w-1}}{(1-p)^{\const{M}-w-1}p^w}\right\}
    \\
    & \hspace{-3.5cm}\textnormal{if } \Hwt{\vect{q}_l}=w\in[\const{M}-1],
    \\[1mm]
    1 & \hspace{-3.5cm}\textnormal{if }\Hwt{\vect{q}_l}=0 \textnormal{ or } \const{M},
  \end{cases}  
\end{IEEEeqnarray*}
for a given query $\vect{q}_l$, for an $(\const{M},n)$ Scheme~A $\collect{C}_{\mathsf{A}}$ with $\{S_j\}_{j=1}^{\const{M}-1}$ i.i.d.\ according to $\Bernoulli{p}$, $0\leq p\leq \frac{1}{2}$. Thus, it can be easily seen that the $\eps$-privacy leakage is equal to
\begin{IEEEeqnarray*}{rCl}
  \rho^{(\eps\mhyph\mathsf{P})}{(\collect{C}_{\mathsf{A}})}& = &\ln{\left(\max\left\{\frac{p}{1-p},\frac{1-p}{p},1\right\}\right)}\nonumber\\
  & = &\ln{\left(\frac{1-p}{p}\right)}.
  % \IEEEeqnarraynumspace\label{eq:eps-privacy_Mn2-Scheme1-IID-Bernoulli}
\end{IEEEeqnarray*}
Hence, to satisfy the leakage constraint $\rho^{(\eps\mhyph\mathsf{P})}\leq\rho$, we require
\begin{IEEEeqnarray*}{c}
  \frac{1-p}{p}\leq \ee^{\rho},
\end{IEEEeqnarray*}
which gives the inequality $\inv{(1+\ee^{\rho})}\leq p\leq\nicefrac{1}{2}$ for any $\rho\geq 0$. Finally, to complete the proof, we simply pick $p_\rho\eqdef\inv{(1+\ee^{\rho})}$, and substitute it into~\eqref{eq:IRrate_Scheme1}.

\section{Proof of Lemma~\ref{lem:LB_HPm}}
\label{sec:proof_HPm-LB}

The lemma can be shown by combining Lemma~\ref{lem:entropy-diff_TV} with a similar approach to the one of the converse proofs given in the information theory literature for PIR, see, e.g., \cite{SunJafar17_1,ChenWangJafar20_1,SamyTandonLazos19_1}. To make the paper self-contained, we repeat some basic steps here.

The first objective is to find an upper bound on the absolute value of the entropy difference $\bigHPcond{\vect{A}_l^{(m)}}{\vect{Q}^{(m)}_l,\vect{X}^{\set{M}},\vect{X}^{(m)}}-\bigHPcond{\vect{A}_l^{(m')}}{\vect{Q}^{(m')}_l,\vect{X}^{\set{M}},\vect{X}^{(m)}}$ subject to $\eMI{M}{\vect{Q}_l}\leq\rho_l$, $\forall\,l\in [n]$, where $m,m'\notin\set{M}\subsetneq [\const{M}-1]$.

Observe that
\begin{IEEEeqnarray}{rCl}
  &&\Bigl |\bigHPcond{\vect{A}_l^{(m)}}{\vect{Q}^{(m)}_l,\vect{X}^{\set{M}},\vect{X}^{(m)}}\nonumber\\
  &&\; -\>\bigHPcond{\vect{A}_l^{(m')}}{\vect{Q}^{(m')}_l,\vect{X}^{\set{M}},\vect{X}^{(m)}}\Bigr |
  \nonumber\\
  & = &
  \Bigl |\bigHPcond{\vect{A}_l^{(m)},\vect{Q}^{(m)}_l}{\vect{X}^{\set{M}},\vect{X}^{(m)}}-\bigHPcond{\vect{Q}^{(m)}_l}{\vect{X}^{\set{M}},\vect{X}^{(m)}}\nonumber\\
  &&\; -\>
  \bigHPcond{\vect{A}_l^{(m')},\vect{Q}^{(m')}_l}{\vect{X}^{\set{M}},\vect{X}^{(m)}}\nonumber\\
  &&\; +\>\bigHPcond{\vect{Q}^{(m')}_l}{\vect{X}^{\set{M}},\vect{X}^{(m)}}\Bigr |
  \nonumber\\[1mm]
  & \leq &
  \Bigl |\bigHPcond{\vect{A}_l^{(m)},\vect{Q}^{(m)}_l}{\vect{X}^{\set{M}},\vect{X}^{(m)}}\nonumber\\
  &&\; -\>\bigHPcond{\vect{A}_l^{(m')},\vect{Q}^{(m')}_l}{\vect{X}^{\set{M}},\vect{X}^{(m)}}\Bigr |\nonumber\\
  &&\; +\>\abs{\bigHPcond{\vect{Q}^{(m)}_l}{\vect{X}^{\set{M}},\vect{X}^{(m)}}-\bigHPcond{\vect{Q}^{(m')}_l}{\vect{X}^{\set{M}},\vect{X}^{(m)}}}
  \nonumber\\[1mm]
  & = &  
  \Bigl |\bigHPcond{\vect{A}_l,\vect{Q}_l}{\vect{X}^{\set{M}},\vect{X}^{(m)},M=m}\nonumber\\
  &&\;\; -\>\bigHPcond{\vect{A}_l,\vect{Q}_l}{\vect{X}^{\set{M}},\vect{X}^{(m)},M=m'}\Bigr |\nonumber\\
  &&\; +\>\Bigl |\bigHPcond{\vect{Q}_l}{\vect{X}^{\set{M}},\vect{X}^{(m)},M=m}\nonumber\\
  &&\qquad -\>\bigHPcond{\vect{Q}_l}{\vect{X}^{\set{M}},\vect{X}^{(m)},M=m'}\Bigr |
  \IEEEeqnarraynumspace\label{eq:sum_entropy-differences}\\[1mm]
  & \leq & \eps^{\mathsf{MI}}(\set{Q}_l,\set{A}_l)+\eps^{\mathsf{MI}}(\set{Q}_l),\IEEEeqnarraynumspace\label{eq:UprBnd_entropy-differences}
\end{IEEEeqnarray}% QN: add \vect{X}^{[\const{M}]} in the end of the Markov chain?
where the inequality \eqref{eq:UprBnd_entropy-differences} can be justified as follows. Using~\eqref{eq:independent_files-queries} and the Markov chain $M\markov \vect{Q}_l\markov\vect{A}_l$, we have
\begin{IEEEeqnarray*}{rCl}
  \IEEEeqnarraymulticol{3}{l}{%
    \eMIcond{M}{\vect{Q}_l,\vect{A}_l}{\vect{X}^{\set{M}},\vect{X}^{(m)}}}\nonumber\\*\quad%
  & = &\eMIcond{M}{\vect{Q}_l}{\vect{X}^{\set{M}},\vect{X}^{(m)}}+\underbrace{\eMIcond{M}{\vect{A}_l}{\vect{Q}_l,\vect{X}^{\set{M}},\vect{X}^{(m)}}}_{=0}
  \\
  & = &\eMI{M}{\vect{Q}_l}\leq\rho_l.
\end{IEEEeqnarray*}
Hence, \eqref{eq:UprBnd_entropy-differences} follows from~\cref{lem:lemma_MI-TV,lem:entropy-diff_TV}. In addition, \eqref{eq:UprBnd_entropy-differences} implies that
\begin{IEEEeqnarray}{rCl}
  \IEEEeqnarraymulticol{3}{l}{%
    \bigHPcond{\vect{A}_l^{(m)}}{\vect{Q}^{(m)}_l,\vect{X}^{\set{M}},\vect{X}^{(m)}}}\nonumber\\*\quad%
  & \geq &\bigHPcond{\vect{A}_l^{(m')}}{\vect{Q}^{(m')}_l,\vect{X}^{\set{M}},\vect{X}^{(m)}}
  \nonumber\\
  && -\>\bigl[\eps^{\mathsf{MI}}(\set{Q}_l,\set{A}_l)+\eps^{\mathsf{MI}}(\set{Q}_l)\bigr],
  \IEEEeqnarraynumspace\label{eq:lwrBnd_mthFile_lthNode}
\end{IEEEeqnarray}
for all $l\in [n]$.

Now, due to \eqref{eq:retrievability} we know that
\begin{IEEEeqnarray}{rCl}
  \IEEEeqnarraymulticol{3}{l}{%
    \bigHPcond{\vect{A}_{[n]}^{(m)}}{\vect{Q}_{[n]}^{(m)},\vect{X}^{\set{M}}}  
  }\nonumber\\*\quad%
  & = &\bigHPcond{\vect{X}^{(m)},\vect{A}^{(m)}_{[n]}}{\vect{Q}^{(m)}_{[n]},\vect{X}^{\set{M}}}\nonumber\\
  && -\>\underbrace{\bigHPcond{\vect{X}^{(m)}}{\vect{A}^{(m)}_{[n]},\vect{Q}^{(m)}_{[n]},\vect{X}^{\set{M}}}}_{= 0}
  \nonumber\\
  & = &\bigHPcond{\vect{X}^{(m)}}{\vect{Q}^{(m)}_{[n]},\vect{X}^{\set{M}}}+\bigHPcond{\vect{A}^{(m)}_{[n]}}{\vect{Q}^{(m)}_{[n]},\vect{X}^{\set{M}},\vect{X}^{(m)}}
  \nonumber\\
  & = &\beta \log_2{\card{\set{X}}}+\bigHPcond{\vect{A}^{(m)}_{[n]}}{\vect{Q}^{(m)}_{[n]},\vect{X}^{\set{M}},\vect{X}^{(m)}}\nonumber\\
  &&\quad-\>\underbrace{\bigHPcond{\vect{A}^{(m)}_{[n]}}{\vect{Q}^{(m)}_{[n]},\vect{X}^{\set{M}},\vect{X}^{(m)},\vect{X}^{\cset{M}\setminus\{m\}}}}_{= 0}
  \nonumber\\
  & = &\beta\log_2{\card{\set{X}}}+\bigMIcond{\vect{A}^{(m)}_{[n]}}{\vect{X}^{\cset{M}\setminus\{m\}}}{\vect{Q}^{(m)}_{[n]},\vect{X}^{\set{M}},\vect{X}^{(m)}}
  \nonumber\\
  & = &\beta\log_2{\card{\set{X}}}\nonumber\\
  &&\qquad +\>\bigMIcond{\vect{A}^{(m)}_{[n]},\vect{Q}^{(m)}_{[n]}}{\vect{X}^{\cset{M}\setminus\{m\}}}{\vect{X}^{\set{M}},\vect{X}^{(m)}}
  \IEEEeqnarraynumspace\label{eq:use_QX-independence}\\
  & \geq &\beta\log_2{\card{\set{X}}}\nonumber\\
  &&\qquad +\>\bigMIcond{\vect{A}^{(m)}_{l},\vect{Q}^{(m)}_{[n]}}{\vect{X}^{\cset{M}\setminus\{m\}}}{\vect{X}^{\set{M}},\vect{X}^{(m)}}
  \IEEEeqnarraynumspace\label{eq:chain-rule_MI}\\
  & = &\beta\log_2{\card{\set{X}}}+\bigHPcond{\vect{A}^{(m)}_{l},\vect{Q}^{(m)}_{[n]}}{\vect{X}^{\set{M}},\vect{X}^{(m)}}\nonumber\\
  &&\quad -\>\bigHPcond{\vect{A}^{(m)}_{l},\vect{Q}^{(m)}_{[n]}}{\vect{X}^{\set{M}},\vect{X}^{(m)},\vect{X}^{\cset{M}\setminus\{m\}}}
  \nonumber\\
  & = &\beta\log_2{\card{\set{X}}}+\underbrace{\bigHPcond{\vect{Q}^{(m)}_{[n]}}{\vect{X}^{\set{M}},\vect{X}^{(m)}}}_{=\>\bigHP{\vect{Q}^{(m)}_{[n]}}}
  \nonumber\\
  &&\quad +\>\bigHPcond{\vect{A}^{(m)}_{l}}{\vect{Q}^{(m)}_{[n]},\vect{X}^{\set{M}},\vect{X}^{(m)}}
  \nonumber\\[1mm]
  &&\quad -\>\underbrace{\bigHPcond{\vect{Q}^{(m)}_{[n]}}{\vect{X}^{\set{M}},\vect{X}^{(m)},\vect{X}^{\cset{M}\setminus\{m\}}}}_{=\>\bigHP{\vect{Q}^{(m)}_{[n]}}}\nonumber\\
  &&\quad -\>\underbrace{\bigHPcond{\vect{A}^{(m)}_{l}}{\vect{Q}^{(m)}_{[n]},\vect{X}^{\set{M}},\vect{X}^{(m)},\vect{X}^{\cset{M}\setminus\{m\}}}}_{= 0}
  \nonumber\\
  & = &\beta\log_2{\card{\set{X}}}+\bigHPcond{\vect{A}^{(m)}_{l}}{\vect{Q}^{(m)}_{[n]},\vect{X}^{\set{M}},\vect{X}^{(m)}}
  \nonumber\\
  & = &\beta\log_2{\card{\set{X}}}+\bigHPcond{\vect{A}^{(m)}_{l}}{\vect{Q}^{(m)}_{l},\vect{X}^{\set{M}},\vect{X}^{(m)}}
  \IEEEeqnarraynumspace\label{eq:ineq_HPcond-m}\\
  & \geq &\beta\log_2{\card{\set{X}}}+\bigHPcond{\vect{A}^{(m')}_{l}}{\vect{Q}^{(m')}_{l},\vect{X}^{\set{M}},\vect{X}^{(m)}}\nonumber\\
  &&\quad -\>\bigl[\eps^{\mathsf{MI}}(\set{Q}_l,\set{A}_l)+\eps^{\mathsf{MI}}(\set{Q}_l)\bigr]
  \IEEEeqnarraynumspace\label{eq:ineq_HPcond_lthNode}\\
  & = &\beta \log_2{\card{\set{X}}}+\bigHPcond{\vect{A}^{(m')}_{l}}{\vect{Q}^{(m')}_{[n]},\vect{X}^{\set{M}},\vect{X}^{(m)}}\nonumber\\
  &&\quad -\>\bigl[\eps^{\mathsf{MI}}(\set{Q}_l,\set{A}_l)+\eps^{\mathsf{MI}}(\set{Q}_l)\bigr],
  \IEEEeqnarraynumspace\label{eq:ineq_HPcond}
\end{IEEEeqnarray}
for any $l\in[n]$, where \eqref{eq:use_QX-independence} follows from \eqref{eq:independent_files-queries}, \eqref{eq:chain-rule_MI} holds by the chain rule for MI, and the final inequality \eqref{eq:ineq_HPcond_lthNode} is due to \eqref{eq:lwrBnd_mthFile_lthNode}.

Thus, summing \eqref{eq:ineq_HPcond} over all possible $l\in[n]$ we have
\begin{IEEEeqnarray*}{rCl}
  \IEEEeqnarraymulticol{3}{l}{%
    n\bigHPcond{\vect{A}_{[n]}^{(m)}}{\vect{Q}_{[n]}^{(m)},\vect{X}^{\set{M}}}}\nonumber\\*\quad%
  & \geq &n\beta \log_2{\card{\set{X}}}+\sum_{l=1}^n\bigHPcond{\vect{A}^{(m')}_{l}}{\vect{Q}^{(m')}_{[n]},\vect{X}^{\set{M}},\vect{X}^{(m)}}
  \nonumber\\
  &&\hspace{0.55cm} -\>\sum_{l=1}^n\bigl[\eps^{\mathsf{MI}}(\set{Q}_l,\set{A}_l)+\eps^{\mathsf{MI}}(\set{Q}_l)\bigr]
  \\
  & \geq &n\beta\log_2{\card{\set{X}}}+\bigHPcond{\vect{A}^{(m')}_{[n]}}{\vect{Q}^{(m')}_{[n]},\vect{X}^{\set{M}},\vect{X}^{(m)}}\\
  &&\hspace{0.55cm} -\>\sum_{l=1}^n\bigl[\eps^{\mathsf{MI}}(\set{Q}_l,\set{A}_l)+\eps^{\mathsf{MI}}(\set{Q}_l)\bigr].
\end{IEEEeqnarray*}
The result of \eqref{eq:LB_HPm} then follows by dividing both sides by $n$.

On the other hand, following a similar derivation as \eqref{eq:ineq_HPcond-m}, we can obtain 
\begin{IEEEeqnarray*}{rCl}
  \IEEEeqnarraymulticol{3}{l}{%
    \bigHPcond{\vect{A}_{[n]}^{(\const{M})}}{\vect{Q}_{[n]}^{(\const{M})},\vect{X}^{[\const{M}-1]}}}\nonumber\\*\quad%
  & \geq &\beta \log_2{\card{\set{X}}}+
  \underbrace{\bigHPcond{\vect{A}^{(\const{M})}_{l}}{\vect{Q}^{(\const{M})}_{l},\vect{X}^{(\const{M})},\vect{X}^{[\const{M}-1]}}}_{=0}.
\end{IEEEeqnarray*}
This completes the proof of \eqref{eq:LB_HP-Mth}.

\section{Proof of Lemma~\ref{lem:lemma_MaxL-TV}}
\label{sec:proof_MaxL-TV}
We start the proof by defining a set $\set{B}_{x,x'}\eqdef\{x,x'\}$ with two arbitrary elements $x\neq x'$, $x,x'\in\set{X}$, and a subset $\set{Z}_{x,x'}\subseteq\set{Y}$ as $\{y\in\set{Y}\colon P_{Y|X}(y|x)\geq P_{Y|X}(y|x')\}$. Next, we introduce a new RV $Z_{x,x'}$ as
\begin{IEEEeqnarray*}{c}
  Z_{x,x'}(y)=
  \begin{cases}
    1 & \textnormal{if }y\in\set{Z}_{x,x'},
    \\
    0 & \textnormal{otherwise}.
  \end{cases}
\end{IEEEeqnarray*}
By~\eqref{eq:expression_MaxL} we have
\begin{IEEEeqnarray*}{rCl}
  \IEEEeqnarraymulticol{3}{l}{%
    \sum_{z\in\{0,1\}}\max_{b\in\set{B}_{x,x'}}P_{Z_{x,x'}|X}(z|b)}\nonumber\\*\quad%
  & = &\sum_{y\in\set{Z}_{x,x'}}P_{Y|X}(y|x)+\sum_{y\in\cset{Z}_{x,x'}}P_{Y|X}(y|x')
  \nonumber\\
  &  \stackrel{(a)}{=} &\sum_{y\in\set{Y}}\max_{b\in\set{B}_{x,x'}}P_{Y|X}(y|b)
\\
  &  \stackrel{(b)}{\leq} &\sum_{y\in\set{Y}}\max_{x\in\set{X}}P_{Y|X}(y|x)\leq 2^{\rho},
  \IEEEeqnarraynumspace\label{eq:use_maximizing-subset}
\end{IEEEeqnarray*}
where~$(a)$ follows by the definition of the subset $\set{Z}_{x,x'}$, and~$(b)$ holds simply because maximizing over a subset leads to a smaller value. Moreover, by using the relation between TV distance and variational distance~\cite[Lem.~3.12]{Moser19_v47}, it follows that
\begin{IEEEeqnarray*}{rCl}
  \IEEEeqnarraymulticol{3}{l}{%
    \norm{P_{Z_{x,x'}|X=x}-P_{Z_{x,x'}|X=x'}}_\mathsf{TV}}\nonumber\\*\quad%
  & = &\frac{1}{2}\biggl(\abs{P_{Z_{x,x'}|X}(1|x)-P_{Z_{x,x'}|X}(1|x')}\nonumber\\
    &&\quad\quad +\>\abs{P_{Z_{x,x'}|X}(0|x)-P_{Z_{x,x'}|X}(0|x')}\biggr)
  \\
  & = &\frac{1}{2}\Biggl(\sum_{y\in\set{Z}_{x,x'}}\bigl[P_{Y|X=x}(y)-P_{Y|X=x'}(y)\bigr]\nonumber\\
  &&\quad\quad +\>\sum_{y\in\cset{Z}_{x,x'}}\bigl[P_{Y|X=x'}(y)-P_{Y|X=x}(y)\bigr]\Biggr)
  \\
  & = &\frac{1}{2}\sum_{y\in\set{Y}}\abs{P_{Y|X=x}(y)-P_{Y|X=x'}(y)}
  \\
  & = &\norm{P_{Y|X=x}-P_{Y|X=x'}}_\mathsf{TV}.%
\end{IEEEeqnarray*}
Therefore, since $x,x'$ are chosen arbitrarily, \cref{lem:lemma_MaxL-TV} holds for any alphabets $\set{X}$ and $\set{Y}$ if we can show that the assertion is true when $X$ and $Y$ are binary.

To complete the proof, we show that Lemma~\ref{lem:lemma_MaxL-TV} holds for any binary RVs $X$ and $Y$. The proof is quite straightforward: Since $\set{X}=\set{Y}=\{0,1\}$, we can define $P_{Y|X}$ by $a\eqdef P_{Y|X}(1|0)$ and $b\eqdef P_{Y|X}(1|1)$ with $0\leq a, b\leq 1$. Thus, by definition $\enorm{P_{Y|X=0}-P_{Y|X=1}}_\mathsf{TV}=\abs{a-b}$ and $\ML{X}{Y}=\log_2\bigl(\max\{1-a,1-b\}+\max\{a,b\}\bigr)=\log_2\bigl(1+\abs{a-b}\bigr)\leq\rho$. As $\log_2(\cdot)$ is a strictly increasing function,  $\abs{a-b}\leq 2^\rho-1$.

% \IEEEtriggeratref{3}

% \balance

%%%%%%%%%%%%%%%%%%%%%%%%%%%%%%%%%%%%%%%%%%%%%%%%%%%%%%%%%%%%%%%%%%%%%%%%%%%%% 
% \bibliographystyle{IEEEtran}
% \bibliography{defshort1,biblioHY}


%%%%%%%%%%%%%%%%%%%%%%%%%%%%%%%%%%%%%%%%%%%%%%%%%%%%%%%%%%%%%%%%%%%%%%%%%%%%% 

\begin{IEEEbiographynophoto}{Hsuan-Yin~Lin}
  (S'09--M'13--SM'19) received his B.S.~from National Tsing-Hua University (NTHU), Taiwan, in 2007, and his M.S. degree and Ph.D. degree from National Chiao Tung University (NCTU), Taiwan, in 2008 and 2013, respectively. He was a visiting scholar at Universitat Pompeu Fabra, Barcelona, Spain, and TU Darmstadt, Germany, and a postdoctoral at Simula UiB, Norway. Currently, he is a research scientist at Simula UiB. His current interests include privacy-preserving technologies, information-theoretic cryptography, coding in distributed storage systems, finite blocklength information theory, scheduling in millimeter-wave cellular networks, and distributed detection and estimation. In 2014, Dr.~Hsuan-Yin Lin was awarded the Honor Membership of the Phi Tau Phi Scholastic Honor Society of the Republic of China (Taiwan) and the New Partnership Program for the Connection to the Top Labs in the World (subsidized by the Ministry of Science and Technology, Taiwan). Dr. Lin was the Publicity Chair of the 9th International Workshop on Signal Design and its Applications in Communications, Dongguan, China. He is currently serving as Guest Editor of Entropy: Special Issue ``Information-Theoretic Approach to Privacy and Security,'' and the TPC member of 2022 International Zurich Seminar on Information and Communication (IZS).
\end{IEEEbiographynophoto}

\begin{IEEEbiographynophoto}{Siddhartha Kumar}
  (S'15--M'18) received the M.Sc. degree in communications engineering from Chalmers University of Technology, Gothenburg, Sweden, in 2015, and his Ph.D. in informatics from the University of Bergen, Bergen, Norway, in 2018. He is currently a postdoctoral fellow at Simula UiB. He has been a TPC member for the IEEE Global Communications Conference (GLOBECOM), 2020 and 2021, and the IEEE Wireless Communications and Networking Conference (WCNC), 2021. His research interests are geared towards information-theoretic privacy and security in distributed storage and computing. In 2013, his bachelor's thesis was funded by the Karnataka government, India, under the KSCST student project program.
\end{IEEEbiographynophoto}

\begin{IEEEbiographynophoto}{Eirik Rosnes}
  (S'01--M'04--SM'11) was born in Stavanger, Norway, in 1975. He received the Cand. Scient. degree in physics and the Dr. Scient. degree in informatics from the University of Bergen, Bergen, Norway, in 1999 and 2003, respectively. From 2003 to 2011, he was with the Department of Informatics, University of Bergen, first as a Postdoctoral Researcher and then as a Senior Researcher. From September 2001 to March 2002 and from March 2005 to September 2005, he was a Visiting Scholar with the Center for Magnetic Recording Research, University of California at San Diego, La Jolla, CA, USA. From October 2011 to August 2013, he was a Senior Engineer with Ceragon Networks, and since August 2013, he has been with both Simula Research Laboratory / Simula UiB as a Senior Researcher (from 2020 as a Chief Research Scientist) and Section Leader and the University of Bergen (until August 2017) as an Adjunct Associate Professor. His research interests are in the areas of communication theory and information theory and include classical error-control coding, codes on graphs, codes for distributed storage systems, and coding for privacy and security. Dr.~Rosnes was a Technical Program Co-Chair for the 2009 International Workshop on Coding and Cryptography, Ullensvang, Norway, and served as Area Editor for the AEU INTERNATIONAL JOURNAL OF ELECTRONICS AND COMMUNICATIONS from October 2014 to January 2020. He is currently serving as Associate Editor for IEEE TRANSACTIONS ON COMMUNICATIONS.
\end{IEEEbiographynophoto}

\begin{IEEEbiographynophoto}{Alexandre Graell~i~Amat} (S'01--M'05--SM'10) is a Professor at the Department of Electrical Engineering, Chalmers University of Technology, Gothenburg, Sweden. He
received the M.Sc. degree in Telecommunications Engineering from the Universitat Polit\`ecnica de Catalunya, Barcelona, Catalonia, Spain, in 2001, and the M.Sc. and Ph.D. degrees in Electrical Engineering from the Politecnico di Torino, Turin, Italy, in 2000 and 2004, respectively. From 2001 to 2002, he was a Visiting Scholar with the University of California San Diego, La Jolla, CA, USA. From 2002 to 2003, he held a visiting appointment at Universitat Pompeu Fabra, Barcelona, and the Telecommunications Technological Center of Catalonia, Barcelona. From 2001 to 2004, he held a part-time appointment at the STMicroelectronics Data Storage Division, Milan, Italy, as a consultant on coding for magnetic recording channels. From 2004 to 2005, he was a Visiting Professor with Universitat Pompeu Fabra, Barcelona. From 2006 to 2010, he was with the Department of Electronics, IMT Atlantique (formerly ENST Bretagne), Brest, France. Since 2019 he is also Adjunct Research Scientist at Simula UiB, Bergen, Norway. His research interests are in the field of coding theory with application to distributed computing, privacy and security, random access, and optical communications. 

Prof. Graell i Amat received the Marie Sk\l{}odowska-Curie  Fellowship from the European Commission and the Juan de la Cierva Fellowship from the Spanish Ministry of Education and Science. He received the IEEE Communications Society 2010 Europe, Middle East, and Africa Region Outstanding Young Researcher Award. He was the General Co-Chair of the 7th International Symposium on Turbo Codes and Iterative Information Processing, Sweden, 2012, and the TPC Co-Chair of the 11th International Symposium on Topics in Coding, Canada, 2021. He was an Associate Editor of the IEEE COMMUNICATIONS LETTERS from 2011 to 2013. He was Associate Editor and Editor-at-Large of the IEEE TRANSACTIONS ON COMMUNICATIONS from 2011 to 2016 and 2017 to 2020, respectively. He is currently Area Editor of the IEEE TRANSACTIONS ON COMMUNICATIONS.
  
\end{IEEEbiographynophoto}

\begin{IEEEbiographynophoto}{Eitan Yaakobi}
  (S'07--M'12--SM'17) is an Associate Professor at the Computer Science Department at the Technion --- Israel Institute of Technology. He received the B.A. degrees in computer science and mathematics, and the M.Sc. degree in computer science from the Technion --- Israel Institute of Technology, Haifa, Israel, in 2005 and 2007, respectively, and the Ph.D. degree in electrical engineering from the University of California, San Diego, in 2011. Between 2011-2013, he was a postdoctoral researcher in the department of Electrical Engineering at the California Institute of Technology and at the Center for Memory and Recording Research at the University of California, San Diego. His research interests include information and coding theory with applications to non-volatile memories, associative memories, DNA storage, data storage and retrieval, and private information retrieval. He received the Marconi Society Young Scholar in 2009 and the Intel Ph.D. Fellowship in 2010-2011.
\end{IEEEbiographynophoto}

\end{document}